\documentclass{aa}  
\usepackage{graphicx}
\pdfoutput=1
\usepackage{txfonts}
\usepackage[font=small, labelfont=bf]{caption}
\usepackage{subcaption}
\usepackage{booktabs,siunitx}
\usepackage{babel,blindtext}
\usepackage{natbib}
\bibpunct{(}{)}{;}{a}{}{,} % to follow the A&A style
\usepackage{enumitem}
\usepackage{amsmath}
\usepackage{breqn}
\usepackage{subcaption}
\usepackage{graphicx} 
\usepackage[colorlinks=true, linkcolor = blue, urlcolor=blue, citecolor = blue]{hyperref}
\usepackage{longtable}
\begin{document} 

   %\titlerunning{HeViCS Red Galaxies}% Part of RIGHT running header
   \title{Knocking on giants' doors:}

   \subtitle{I. The evolution of the dust-to-stellar mass ratio in distant dusty galaxies} %Blinded by Colours?}

   \author{D. Donevski\inst{1,2}
   \and A. Lapi\inst{1,3}
    \and K. Ma\l{}ek\inst{4,5}
     \and D. Liu\inst{6}
      \and {C. G\'omez-Guijarro} \inst{7}
        \and R. Dav\'e\inst{8,9,10}
     \and K. Kraljic\inst{8}
     \and L. Pantoni\inst{1,3}
        \and A. Man\inst{2}
       \and S. Fujimoto\inst{11,12}
       \and A. Feltre\inst{13}
   \and W. Pearson\inst{4}
   \and Q. Li\inst{14}
    \and D. Narayanan\inst{11,14,15}
   %\and C. Pappalardo\inst{13,14}
          %\fnmsep\thanks{Fill other institutions later...}
          }

   \institute{SISSA, Via Bonomea 265, Trieste, Italy,\\
              \email{darko.donevski@sissa.it}
              \and
              Dunlap Institute for Astronomy \& Astrophysics, 50 St. George Street, Toronto, ON M5S 3H4, Canada
               \and
               IFPU - Institute for fundamental physics of the Universe, Via Beirut 2,
               34014 Trieste, Italy
               \and
              National Centre for Nuclear Research, ul. Pasteura 7, 02-093 Warsaw, Poland
              \and
              Aix Marseille Univ. CNRS, CNES, LAM, Marseille, France
                	\and
              Max-Planck-Institut f{\"u}r Astronomie, K{\"o}nigstuhl 17, D-69117 Heidelberg, Germany
            	\and
            	AIM, CEA, CNRS, Universit\'e Paris-Saclay, Universit\'e Paris Diderot, Sorbonne Paris Cit\'e, F-91191 Gif-sur-Yvette, France
            		\and
            	Institute for Astronomy, Royal Observatory, Edinburgh EH9 3HJ, United Kingdom 
            	\and
            	University of the Western Cape, Bellville, Cape Town, 7535, South Africa
            	\and
            	South African Astronomical Observatories, Observatory, Cape Town, 7925, South Africa
            	\and
            	Cosmic Dawn Center (DAWN), Copenhagen, Denmark
            	\and
            	Niels Bohr Institute, University of Copenhagen, Juliane Mariesvej, 30-2100 Copenhagen, Denmark
                \and
                INAF – Osservatorio di Astrofisica e Scienza dello Spazio di Bologna, Via P. Gobetti 93/3, 40129 Bologna, Italy
                \and
             Department of Astronomy, University of Florida, 211 Bryant Space Sciences Center, Gainesville, FL, USA
             \and
             University of Florida Informatics Institute, 432 Newell Drive, CISE Bldg E251, Gainesville, FL 32611
          %  \and
           %Centro de Astronomia e Astrofísica da Universidade de Lisboa, Observatório Astronómico de Lisboa, Tapada da Ajuda,
            %1349-018 Lisboa, Portugal
            %and 
            %Instituto de Astrofísica e Ciencias do Espaço, Universidade de Lisboa, OAL, Tapada da Ajuda, 1349-018 Lisboa, Portugal
             }

   \date{Received ; accepted }

% \abstract{}{}{}{}{} 
% 5 {} token are mandatory
 
  \abstract
  % context heading (optional)
  % {} leave it empty if necessary  
 {

The dust-to-stellar mass ratio ($M_{\rm dust}$/$M_{\rm \star}$) is a crucial yet poorly constrained quantity to understand the complex physical processes involved in the production of dust, metals and stars in galaxy evolution. In this work we explore trends of $M_{\rm dust}$/$M_{\rm \star}$ with different physical parameters using observations of 300 massive, dusty star-forming galaxies detected with ALMA up to $z\approx5$. Additionally, we interpret our findings with different models of dusty galaxy formation. We find that $M_{\rm dust}$/$M_{\rm \star}$ evolves with redshift, stellar mass, specific star formation rate and integrated dust size, differently for main sequence and starburst galaxies. In both galaxy populations $M_{\rm dust}$/$M_{\rm \star}$ increases until $z\sim2$ followed by a roughly flat trend towards higher redshifts, suggesting efficient dust growth in the distant universe. We confirm that the inverse relation between $M_{\rm dust}$/$M_{\rm \star}$ and $M_{\star}$ holds up to $z\approx5$ and can be interpreted as an evolutionary transition from early to late starburst phases. We demonstrate that $M_{\rm dust}$/$M_{\rm \star}$ in starbursts reflects the increase in molecular gas fraction with redshift, and attains the highest values for sources with the most compact dusty star-formation.
%a combination of higher star formation efficiency and molecular gas fraction in compact objects, while in main sequence galaxies $M_{\rm dust}$/$M_{\rm \star}$ is predominantly regulated by molecular gas fraction.

The state-of-the-art cosmological simulations that include self-consistent dust growth, broadly reproduce the evolution of $M_{\rm dust}$/$M_{\rm \star}$ in main sequence galaxies, but underestimate it in starbursts. The latter is found to be linked to lower gas-phase metallicities and longer dust growth timescales relative to observations. Phenomenological models based on the main-sequence/starburst dichotomy and analytical models that include recipes for rapid metal enrichment are consistent with our observations. Therefore, our results strongly suggest that high $M_{\rm dust}$/$M_{\rm \star}$ is due to rapid dust grain growth in metal enriched interstellar medium. This work highlights multifold benefits of using $M_{\rm dust}$/$M_{\rm \star}$ as a diagnostic tool for: (1) disentangling main sequence and starburst galaxies up to $z\sim5$; (2) probing the evolutionary phase of massive objects; and (3) refining the treatment of the dust life cycle in simulations.

}


\maketitle

\section{Introduction}

Recent advent of infrared (IR) instruments such are \textit{Herschel} and ALMA, allowed us to identify long, high-redshift tail ($2<z<7$) of individual dusty star-forming galaxies (DSFGs,  e.g. \citealt{weiss13}, \citealt{riechers13}, \citealt{oteo17b}, \citealt{zavala17}, \citealt{strandet17}, \citealt{jin19}, \citealt{casey18}, for comprehensive reviews see \citealt{casey14} and \citealt{hodge20}). The nature of these sources is critical to our understanding of how massive galaxies assemble and how could their large dust reservoirs have been formed at early cosmic times (e.g. \citealt{dwek14},  \citealt{zhukovska16}, \citealt{popping17}, \citealt{aoyama19}, \citealt{nanni20}). Along with the progress in increasing the statistics of DSFGs, many observational works revisited the correlation between the star-formation rate (SFR) and stellar mass ($M_{\star}$) in star-forming galaxies, showing that in the vast majority of known DSFGs these two quantities are expected to form almost linear relation so-called "main-sequence" (MS hereafter, \citealt{brinchmann04}, \citealt{noeske07}, \citealt{elbaz10}, \citealt{daddi10}, \citealt{rodighiero11}, \citealt{sargent14}, \citealt{speagle14}, \citealt{whitaker15}, \citealt{schreiber15}, \citealt{pearson18}). The prominent positive outliers of this sequence with boosted specific SFR ($\rm sSFR=SFR/M_{\star})$ are so-called starbursts (SB). Knowing the physical properties of these two populations of objects allows us to understand the heterogeneous characteristics of distant DSFGs concerning their evolutionary stage within the main-sequence paradigm (e.g. \citealt{sargent14}, \citealt{scoville17b}, \citealt{silverman18}). %The intrinsic MS scatter could reflect the physical evolution of galaxies undergoing the phases of gas compaction, depletion, replenishment, and quenching (Tacchella et al. 2016). Also, if the SFR-M⋆ scatter is (partly) a result of bursty star for- mation phases (Sparre et al. 2017), it could be linked to the fluc- tuations of the baryon accretion rate in the dark matter host halo%.

It is generally perceived that multiwavelength observations (e.g. from ultraviolet (UV) to sub-mm) are key to providing a complete picture of the stellar mass ($M_{\star}$) and $\rm SFR$ of dusty galaxies. On one hand, dust affects the spectral energy distributions (SED) of galaxies such that at shorter wavelengths stellar light is more absorbed by dust and re-emitted in far-infrared (FIR). On the other hand, %dust grains increase the rate of forming molecules by two orders of magnitude comparing to the cases without dust. 
along with molecular and atomic lines, dust is one of major coolants of the interstellar medium (ISM) and prevents gas heating up from the general interstellar radiation field, thus playing an important role in the process of star formation (\citealt{cuppen17}). As a consequence, the ratio between the dust and stellar mass ($M_{\rm dust}$/$M_{\rm \star}$) stands as a key parameter for understanding the physical processes involved in producing the dust, metals, and stars in DSFGs. It has been suggested that $M_{\rm dust}/M_{\star}$ can be a useful marker of the galaxy ISM and survival capacity of dust grains against the multiple destruction processes (\citealt{dunne11}, \citealt{rowlands15}, \citealt{tan14}, \citealt{bethermin15}, \citealt{calura17}, \citealt{deVis17}, \citealt{michalowski19}, \citealt{burgarella20}). 

 % It is now generally regarded that observing IR emission in combination with ulrtraviolet (UV) emission is key to providing a complete picture of the stellar mass ($M_{\star}$) and star-formation rate (SFR) of dusty galaxies. FIR emission is a key component for accurately determining the SFR of and object, since part of the UV radiation from young stars heats galaxy dust, which then re-radiates in the IR. In the case of DSFGs, observing only in the UV would provide a direct measure of SFR, but could significantly underestimate the total SFR as this absorption by dust obscures the UV emission (e.g. \citealt{williams18}, \cite{buat14}). 
In spite of its importance, the cosmic evolution of $M_{\rm dust}/M_{\star}$ has not yet been fully understood. %Despite the rapidly flourishing statistics of DSFGs at high-$z$,
Linking the cosmic evolution of dust-to-stellar properties in massive galaxies is extremely challenging task due to various reasons: (1) A proper constraint of dust quantities such as dust luminosity ($L_{\rm IR}$) and $M_{\rm dust}$ requires exquisite IR SEDs with rich wavelength sampling towards Rayleigh-Jeans (RJ) tail. Until recently, the limited depth of FIR surveys restricts statistical studies of high-$z$ DSFGs either to the most luminous objects (e.g. \citealt{riechers13}, \citealt{dowell14}, \citealt{oteopc}, \citealt{ddrisers}, \citealt{miller18}, \citealt{pavesi18}), or strongly lensed galaxies (e.g. \citealt{negrello10}, \citealt{wardlow14}, \citealt{strandet17}, \citealt{ciesla20}); (2) %The widest coverage of galaxies' IR SEDs come from blind \textit{Herschel} surveys. 
Considering the limiting beam size of single-dish FIR instruments, extensive follow-ups with high spatial resolution are required to better constraint sources' IR SEDs and redshifts (\citealt{cox11}, \citealt{hodge13}, \citealt{simpson15}, \citealt{dacunha15}, \citealt{miettinen15}, \citealt{oteo17b}, \citealt{dunlop17}, \citealt{miettinen17}, \citealt{fudamoto17}, \citealt{stach19}, \citealt{an19}, \citealt{as2uds19}, \citealt{gomezguijarro19a}, \citealt{simpson20}); (3) The DSFGs are usually highly dust-obscured, meaning that their continuum emission is very faint at rest-frame UV/optical wavelengths. Due to above stated reasons, there are still considerable uncertainties in the derived physical properties of DSFGs. Consequently, some vital quantities, such as $M_{\star}$, $M_{\rm dust}$ or AGN-fraction ($f_{\rm AGN}$), are poorly constrained, which often prevent us from knowing the position of DSFGs in the $\rm SFR$-$M_{\star}$ plane with respect to the "main-sequence". 
%In order to deal with DSFGs that lie close to the confusion limit, 

To partially overcome those issues, statistical methods based on stacking are often applied to infer the average properties of DSFGs that lie close to the confusion limit (e.g., \citealt{schreiber15}, \citealt{bethermin15}). More recently, a new generation of source extraction methods based on positional, redshift and/or SED priors, were tested in order to directly resolve individual galaxies from confused IR images (e.g. \citealt{pearson17}, \citealt{xid+17}, \citealt{liu17}). These techniques enabled observational constraints of DSFGs, e.g. the scaling between the dust mass and gas mass (\citealt{leroy11}, \citealt{magdis12}, \citealt{zahid14}, \citealt{scoville17b}, \citealt{tacconi18},  \citealt{liu19b}), and the evolution of dust temperature (\citealt{magdis12}, \citealt{bethermin15},  \citealt{bethermin15}, \citealt{liang19}). Such discoveries have raised important questions about the dust mass content in the early universe, in particular what are the observational imprints of the main sources of dust mass production/destruction at high-$z$'s. %suggest that the measurement of Mdust/M⋆ could be a very powerful mean for distinguishing starbursts from normal galaxies at high-$z$, during the phase of early metal enrichment in starburst galaxies.

Along with observational efforts, throughout the last decade an extensive attention has been given to the theoretical  studies of the formation of DSFGs and their dark matter (DM) halos by applying different classes of cosmological simulations (\citealt{hayward13}, \citealt{narayanan15}, \citealt{mckinon17}, \citealt{simba}, \citealt{aoyama19}) or semi-analytic and analytic methods (\citealt{lacey16}, \citealt{popping17}, \citealt{imara19}, \citealt{cousin19}, \citealt{vijayan19}, \citealt{lagos19}, \citealt{pantoni19}). To investigate the evolution of the dust content of high-$z$ DSFGs, the models have made significant progress by replacing the simplified scaling relations with the physical recipes for self-consistent dust formation, growth and destruction in evolving galaxies (e.g. \citealt{mckinon17}, \citealt{aoyama19}, \citealt{hou19}, \citealt{graziani19}, \citealt{simba}). While this empowers to study diverse samples of DSFGs from the statistical point of view, it remains challenging to interpret the key contributors to their dust-to-stellar mass ratio. The main reason is the existing tension between the modelled and observed high number density of the most massive DSFGs ($M_{\star}>10^{10}\:M_{\odot},\:M_{\rm dust}>10^{9}\:M_{\odot}$, \citealt{mckinon17}).

We consider that is timely to link the methods described above and inspect the nature of $M_{\rm dust}$/$M_{\rm \star}$ in a large sample of individually detected high-$z$ DSFGs. There are two main questions we address in this work. The first one is: \textit{How the $M_{\rm dust}$/$M_{\rm \star}$ evolves with cosmic time and position of the galaxy with respect to the main-sequence?} Properly answering to this question requires careful examination of all observational challenges outlined above. For this goal, we assemble large statistical data set that contains MS and SB DSFGs identified over a wide redshift range in the COSMOS field with ALMA. We complement deep multi-wavelength catalogue with the carefully de-blended sources' IR fluxes, and apply physically motivated SED modelling to self-consistently derive physical properties of DSFGs. We study different trends with $M_{\rm dust}$/$M_{\rm \star}$ for galaxies within and above the MS. %This approach allows us to achieve the large statistics, homogeneity and high robustness over SED derived quantities. Having a well-constrained understanding of the observed physical properties, such as the cosmic evolution of $M_{\rm dust}$/$M_{\rm \star}$ in DSFGs, 
We then address the second question: \textit{How can the $M_{\rm dust}$/$M_{\rm \star}$ be understood within the framework of dusty galaxy formation and evolution?} We employ state-of-the-art galaxy models, with the aim in achieving comprehensive understanding of the nature of rapid dust evolution in our sources. %We estimate galaxy compactness in order to examine which mechanism is the main driver of observed galaxy properties. %in order to complement the models with detailed treatment of dust production, we also examine phenomenological models in order to better understand statistical properties of 

%The main aim of this work is in studying the relative dependence of how the dust-to-stellar content of DSFGs scales with the galaxy star formation rate and size. 
The paper is organised as follows: in \hyperref[sec:2]{Section 2} we describe the data analysed in this work. In \hyperref[sec:3]{Section 3} we explain the SED fitting methodology and provide average statistical properties of our sample. In \hyperref[sec:4]{Section 4} we present the main results that show how the dust-to-stellar mass ratio of ALMA detected DSFGs scales with the galaxy redshift, sSFR and $M_{\star}$. We provide the recipe to model the observed data based on simple empirical prescriptions. In \hyperref[sec:5]{Section 5} we compare our results to different models of galaxy formation and evolution. %We perform simulations to review all selection biases and highlight the necessity of a further refinement of selection criteria in searching for $z\gtrsim4$ sources. 
We discuss the role of compact dusty star-formation on observed $M_{\rm dust}$/$M_{\rm \star}$ in \hyperref[sec:6]{Section 6}, while our main conclusions are outlined in \hyperref[sec:6]{Section 7}. Throughout the paper we assume a \cite{planck16} cosmology and Chabrier IMF (\citealt{chabrier03}). 

\section{Data and sample selection}
   \label{sec:2}

To build the statistically significant sample of DSFGs suitable for our analysis, we adopt homogeneously calibrated multiwavelength catalogues released by the \textit{Herschel} Extragalactic Legacy Project (HELP , \citealt{malek19}, \citealt{shirley19}, Oliver et al., in preparation). The HELP catalogues offer observational information across the well-known and well-studied extragalactic fields that were targeted by $\mathit{Herschel}$. We chose the COSMOS field (\citealt{scoville07}) due to the wealth of multi-wavelength data complemented to several hundreds of galaxies that exist in public ALMA archive. The main advantage of panchromatic catalogue provided by HELP is its homogeneous calibration and implementation of state-of-the-art source extracting/de-blending tool (XID+, \citealt{xid+17}) that allows us to overcome the confusion limit in FIR observations made with the \textit{Herschel} telescope. %Namely, the final master list contain more than 170 million of sources selected at the deepest IRAC band 1 ($3.4 \mu$m) priors. 

In order to extract the fluxes beyond the conventional \textit{Spitzer} and $\mathit{Herschel}$ confusion limit, the  MIPS ($24\mu$m), PACS (100, 160$\mu$m) and SPIRE fluxes (250, 350 and 500 $\mu$m) are assigned to each source with use of probabilistic de-blending method \texttt{XID+} (\citealt{xid+17}). The code \texttt{XID+} de-blends confusion limited maps with use of positional and redshift information from the deepest IRAC priors (up to 23.4 $\mathrm{mag}$ at $3.6\: \mu$m). In this way, we take the advantage of fluctuations within the confused maps and place strong constraints on the peak of sources' FIR SEDs, improving upon dust embedded star formation and identify the main contributors to the flux detected with higher-resolution instruments that operate at longer wavelengths (i.e. ALMA). Given the positional prior from the HELP catalogue, we identify counterparts to ALMA detected galaxies, either in Band 6 (1.1mm, 121 source) or Band 7 (870 $\mu$m, 207 sources). We adopt fluxes available within the ALMA archive ($\mathrm{A^{3}COSMOS}$, see \citealt{liu19a} for more details).

For the final galaxy sample we require source detections with $\rm S/N\geq3$ in at least 5 photometric bands in the mid-IR-to-FIR/sub-mm range ($8\:\mu$m < $\lambda$ <$1100\:\mu$m), and with $\rm S/N\geq5$ in at least 10 photometric bands covering the optical-NIR range ($0.3\: \mu$m < $\lambda$ <$8\: \mu$m). These requirements are particularly important for achieving the robustness to physical parameters estimated from SED fitting (see e.g. \citealt{malek19}). When multiple measurements are available in similar optical-NIR pass-bands we take the deepest one to reduce the measurement uncertainties. The optical to NIR data come from Subaru Suprime-Cam (6 bands), HSC ($Y$-band), VISTA ($J, H, K_{s}$ bands) and IRAC (4 bands).  We assemble the final list of sources (329 in total), out of which 73 galaxies have known publicly available spectroscopic redshifts ($z_{\rm spec}$), while the rest have photometric redshifts ($z_{\rm phot}$) generated using a Bayesian combination approach (the precision of $z_{\rm phot}$ is estimated to be $\delta z/1+z_{\rm spec}<0.005$, see \citealt{duncan18} for details). The full redshift distribution extends over a wide range ($0.5<z<5.25$) as shown in top panel of \hyperref[fig:Fig.2]{Fig. 1}.	

\section{Panchromatic SED modelling of the data}
\label{sec:3}

\subsection{Tools: CIGALE}

We make use of very dense panchromatic data coverage and apply full SED (UV+IR) modelling of our DSFGs. As a main tool we adopt the newest release of Code Investigating GALaxy Emission (CIGALE; \citealt{cigale19}\footnote{\url{https://gitlab.lam.fr/cigale/cigale}}, \citealt{noll09})  CIGALE is a state-of-the-art SED modelling and fitting code, which combines UV-optical stellar SED with IR component. The code entirely conserves the energy between dust absorption in the UV-to-NIR domain and emission in the mid-IR and FIR. CIGALE is designed for estimating the wide range of physical parameters by comparing modelled galaxy SEDs to observed ones. For each parameter CIGALE makes a probability distribution function (PDF) analysis, and the output value is the likelihood-weighted mean of the PDF (and consequently, the associated error is likelihood-weighted standard deviation). %While the number of options provided by CIGALE is large, the code operates in a modular way, which allows the user to manually select desired ones. 
In this work we carefully chose the model parameters following some of the most recent prescriptions that are extensively tested on large multi-band datasets with available deep IR observations, thus being optimised for a wide range of DSFGs (e.g. \citealt{lofaro17},  \citealt{ciesla17}, \citealt{pearson18}, \citealt{malek19}, \citealt{buat19}). In the following we briefly summarise the choice of modules and parameters presented in \hyperref[tab:3.1]{Table 1}. %For the purpose of our analysis, here we use newest release of CIGALE  (\citealt{cigale19})\footnote{\url{https://gitlab.lam.fr/cigale/cigale}}. %rom COSMOS2015 to estimate the SPIRE  flux densities for use as a flux density prior in XID+ (see Sec. 3.1.2 and 3.2). After the SPIRE band flux densities had been extracted, 

\subsubsection*{Stellar component}

To construct the stellar component of our SED model we use Bruzual $\&$ Charlot stellar population synthesis model (\citealt{bruzual03}, BC03) together with a \citealp{chabrier03} IMF. We fix metallicity to the solar value, which is usually seen as good assumption because the more recent star formation events are using more metallic gas (\citealt{asano13}). Our assumption is additionally motivated by recent spectroscopy studies of DSFGs in HUDF field, for which metallicities consistent with solar are inferred at $1<z<3$ (\citealt{boogaard19}, see also \citealt{nagao12}, \citealt{kriek16}, \citealt{debreuck19})\footnote{Additionally, in the next Section we also compare our results to studies that explore grid of metallicities and star-formation histories.}. We adopt the flexible star-formation historiy (SFH) which is composed of a delayed component with additional burst. The functional form is given as: 

\begin{equation}
\rm SFR(\mathit{t}) = \rm SFR_{\rm delayed}(\mathit{t})+\rm SFR_{\rm burst}(\mathit{t}),
\end{equation}
where $\rm SFR_{\rm delayed}(\mathit{t})\propto \mathit{t} e^{-t/\tau_{main}}$, and $\rm SFR_{\rm burst}(\mathit{t}) \propto e^{-(t-t_{0})/\tau_{burst}}$. Here $\tau_{main}$ represents e-folding time of the main stellar population, while $\tau_{main}$ represents e-folding time of the late starburst. The e-folding time of the two stellar populations (old and young) in the SFH was roughly matched to that of \cite{malek19}. 

Our choice of SFH is motivated by study of \cite{ciesla17} (see also \citealt{forrest18}) who investigate how accurate are different choices of SFHs in reproducing the IR observations with respect to the $\mathrm{SFR-M_{\star}}$ plane. \cite{ciesla17} show that exponentially declining and delayed SFH struggle to model high SFRs in $z > 2$ DSFGs, while exponentially rising and log-normal SFHs has the ability to reach highest SFRs, but show some inconsistency with observed data of massive galaxies at intermediate and lower redshifts. 
%%%%%%%%%%%%%%%%%%%%%%%%%%%%
\begin{table*}
	\caption{Parameters used for modelling the SEDs with CIGALE. All ages/times are given in Gyr.}
	\label{tab:3.1} 
	\centering   
	%       \vspace{0.03cm}   
	\scalebox{0.73} {\begin{tabular}{c c c c} 
			%\hline\hline       
			% To combine 4 columns into a single one 
			%       %HJD & $E$ & Method\#2 & \multicolumn{4}{c}{Method\#3}\\ 
			\hline
			\toprule 
			\\[-12pt]\\
			Parameter&Values&&Description\\[-7pt]\\\\
			\hline
			%	\\[-7pt]\\
			& & Star Formation History & \\[-7pt]\\
			
			$\tau_{main}$&1.0, 1.8, 3.0, 5.0, 7.0& & e-folding time (main)\\
			$\tau_{burst}$&0.01& & e-folding time (burst)\\
			\centering
			$f_{burst}$&0.001, 0.1, 0.20, 0.30& & Mass fraction of the late burst\\
			Age & 0.5, 1.0, 2.0, 3.0, 4.5, 6.0, 7.5, 9.0, 10.0, 11.0, 12.0 & & Population age (main)\\
			Burst age & 0.001, 0.05, 0.08, 0.11, 0.3 && Age of the late burst\\
			\hline
			%	\\[-7pt]\\
			& & Stellar emission & \\[-7pt]\\
			
			IMF& Chabrier 2003& & Initial mass function\\
			Z&0.02& & Metallicity (0.02) in Solar\\	
			Separation age& 0.01 && Age difference between old and young population\\
			\hline
			%	\\[-7pt]\\
			& & Dust attenuation & \\[-7pt]\\
			$A^{BC}_{v}$&0.3, 0.8, 1.2, 3.3, 3.8& & V-band attenuation\\	
			slope BC&-0.7& & Power law slope of BC attenuation\\	
			BC to ISM factor&0.3, 0.5, 0.8, 1.0 & & Ratio of the BC-to-ISM attenuation\\		
			slope ISM&-0.7& & ISM attenuation power law slope\\
			\hline
			& & Dust emission & \\[-7pt]\\
			$q_{PAH}$&0.47, 1.12, 3.9& & Mass fraction of PAH\\
			$U_{\min}$&5.0, 10, 25.0, 40. & & Minimum radiation field\\
			$\alpha$&2.0& & Dust emission power law slope\\
			$\gamma$&0.02& & Illuminated fraction\\
			\hline
			& & AGN emission & \\[-7pt]\\
			$r_{ratio}$&60.& &Maximum to minimum radii of the dust torus\\
			$\tau$&1.0, 6.0& &Optical depth at 9.7$\mu$m\\
			$\beta$&-0.5& &Radial dust distribution within the torus\\
			$\gamma$&0.0& &Angular dust distribution within the torus\\
			Opening angle&$100^{\circ}$& &$\gamma$ Opening angle of the torus\\
			$\psi$&0.001, 89.99& &Angle between eq.axis and line of sight\\
			$f_{\rm AGN}$&0.0, 0.1, 0.25, 0.5, 0.8&&  AGN fraction\\[-3pt]\\
			%	\centering
			%30-40 mJy & 1.67$\pm$0.2 & 1.74$\pm$ 0.15 & 0.07\\[-3pt]\\
			%\centering
			
			\hline    
			\bottomrule               
		\end{tabular}
	}\\
	\vspace{0.5cm} 
	
\end{table*}
%%%%%%%%%%%%%%%%%%%%%
\subsubsection*{Dust attenuation}

In order to model the effects of dust on the integrated spectral properties for the large variety of galaxies, we adopt a double power-law recipe for dust attenuation initially described in \cite{charlott00}. The \cite{charlott00} attenuation law (CF00) assumes that birth clouds (BCs) and the ISM each attenuate light according to fixed power-law attenuation curves. The formalism is based on age-dependent attenuation, meaning that differential attenuation between young (age $<10^7\:\rm yr$) and old (age $>10^7\:\rm yr$) stars is assumed. Both attenuation laws are modelled by a power law function, with the amount of attenuation quantified by the attenuation in the $V$ band. We chose to keep both power law slopes (BC and ISM) of the attenuation fixed at -0.7. The parameters we adopt for CF00 are already used for the fitting of a large sample of DSFGs (\citealt{malek19}).

The choice of dust attenuation laws can significantly impact estimated stellar masses of massive, dusty galaxies and our motivation to chose CF00 is strengthen by two recent findings: (a) it has been shown  that hydrodynamical galaxy models require the inclusion of a birth cloud component to properly match the observed optical depth-attenuation curve slope relation in galaxies (\citealt{trayford20}, see \citealt{salim20} for the review); (b) it has been found that the widely used Calzetti attenuation law (\citealt{calzetti00}) sometimes tends to underestimate $M_{\star}$ by $\rm 0.3-0.5\:dex$ in massive high-$z$ DSFGs (see discussions in \citealt{lofaro17}, \citealt{williams18} and \citealt{buat19}). 
%birthcloud model has allowed for additional degrees of freedom in SED fittingsoftware, including both age-selective attenuation, as well as the ability to model birthcloudswith a range of parameterized curves,
\subsubsection*{Dust emission model}
%The robust estimation of the dust mass ($M_{\rm dust}$) is in the limelight of this paper, and we take particular care choosing the dust component in our SED modelling. 
In general $M_{\rm dust}$ can be estimated either with more simplified methods such as single SED template fitting (\citealt{schreiber18}) and modified blackbody (MBB) fitting (\citealt{pozzi19}, \citealt{clements19}), or with more complex and physically motivated dust emission models (\citealt{dl07}, \citealt{galliano11}, \citealt{draine14}, \citealt{themis17}, see \citealt{galliano18} for an extensive review). We choose to perform the modelling of galaxies'  IR SEDs with the physically motivated, dust emission library of \citealp{draine14} (DL14 hereafter).

 %Therefore we choose to perform the modelling of galaxies'  IR SEDs with the physically motivated, dust emission library of \cite{draine14} (DL14 hereafter). %UBACI KASNIJE footnote{Recently, several studies found (i.e. \cite{smithhayward18}) that there is a good match between the dust parameters of massive galaxies recovered with SED fitting codes from hydro simulations, where dust is treated with radiative transfer.} 

The DL14 is a multi-parameter library which describes the interstellar dust as a mixture of carbonaceous and amorphous silicate grains. The grain size distributions are chosen to realistically "mimic" the observed extinction in the Milky Way (MW), the Large Magellanic Cloud (LMC), and the Small Magellanic Cloud (SMC). The properties of dust grains are parametrised by the so-called polycyclic aromatic hydrocarbon (PAH) index ($q_{\mathrm{PAH}}$), defined as the fraction of the dust mass in the form of PAH grains. The IR SEDs are calculated for dust grains heated by starlight for various distributions of intensities. The majority of the dust is heated by a radiation field with constant intensity from the diffuse ISM, while much smaller fraction of dust (defined as a fraction $\gamma$) is exposed to starlight with interstellar radiation field (ISRF) intensity in a range comprised between $U_{\rm min}$ to $U_{\rm max}$ following a power-law distribution. We sample different $U_{\rm min}$, keeping the dust emission slope fixed at $\beta=2$ along with illumination fraction fixed at $\gamma=0.02$ (see \citealt{magdis12}, \citealt{malek19}, \citealt{buat19}). In our modelling, $L_{\rm IR}$ is an integral of a SED over the rest-frame wavelength range of $\lambda=8-1000\:\mu$m, while the dust masses are derived by fitting and normalising the IR photometry to the DL14 library.

It has been shown that modelling of broadband SEDs with physically motivated models increases the robustness of dust mass estimates (e.g. \citealt{dl07}, \citealt{berta16}, \citealt{schreiber18}). Furthermore, it is demonstrated that single $T_{\rm dust}$ MBB fitting tends to significantly underestimate the $M_{\rm dust}$ by a factor of $\sim2$ as compared to those derived from physically based libraries (\citealt{dale12}, \citealt{magdis12}, \citealt{deVis17}). The discrepancies in estimated $M_{\rm dust}$ could be larger for galaxies with colder dust and higher vertical distance to the galaxy MS (\citealt{berta16}). On top of this, the consistency has been found between the dust properties ($M_{\rm dust}$ and $L_{\rm IR}$) derived with CIGALE DL14 library to those modelled in hydro simulations where dust is treated with radiative transfer (\citealt{smithhayward18}, \citealt{trcka20}).

\subsubsection*{AGN model}

AGN activity is known to be present in DSFGs (\citealt{simeon13}, \citealt{brown19}) and can significantly impact derived physical properties, particularly stellar mass (\citealt{ciesla15}, \citealt{salim16}, \citealt{leja18}). Thus, to improve the derived galaxy properties we chose to quantify the contribution of AGNs to the total predicted $L_{\rm IR}$. We derive the fractional contribution of the AGN, defined as the relative impact of the dusty torus of the AGN to the $L_{\rm IR}$ ("AGN fraction"). We adopt AGN templates presented in \citealp{fritz06} (see also \citealt{feltre12}). The templates are computed at different lines of sight with respect to the torus equatorial plane and account for both (Type 1 and Type 2) AGN emission, from $0^{\circ}$ to $90^{\circ}$ respectively. The parameters in the AGN model were matched to those from \cite{ciesla15}. Due to computational reasons we somewhat reduce the number of input options, and model the two extreme values for inclination angle ($0^{\circ}$ and $90^{\circ}$).

%%%%%%%%%%%%%%%%%%

\subsection{Statistical properties of our sample}

\begin{figure*}[ht]
	\vspace{-0.2cm}
	\centering
	%\hspace{-1.0cm}
	%\includegraphics [width=9.19cm]{Plot_16.pdf}
	\includegraphics [width=18.33cm]{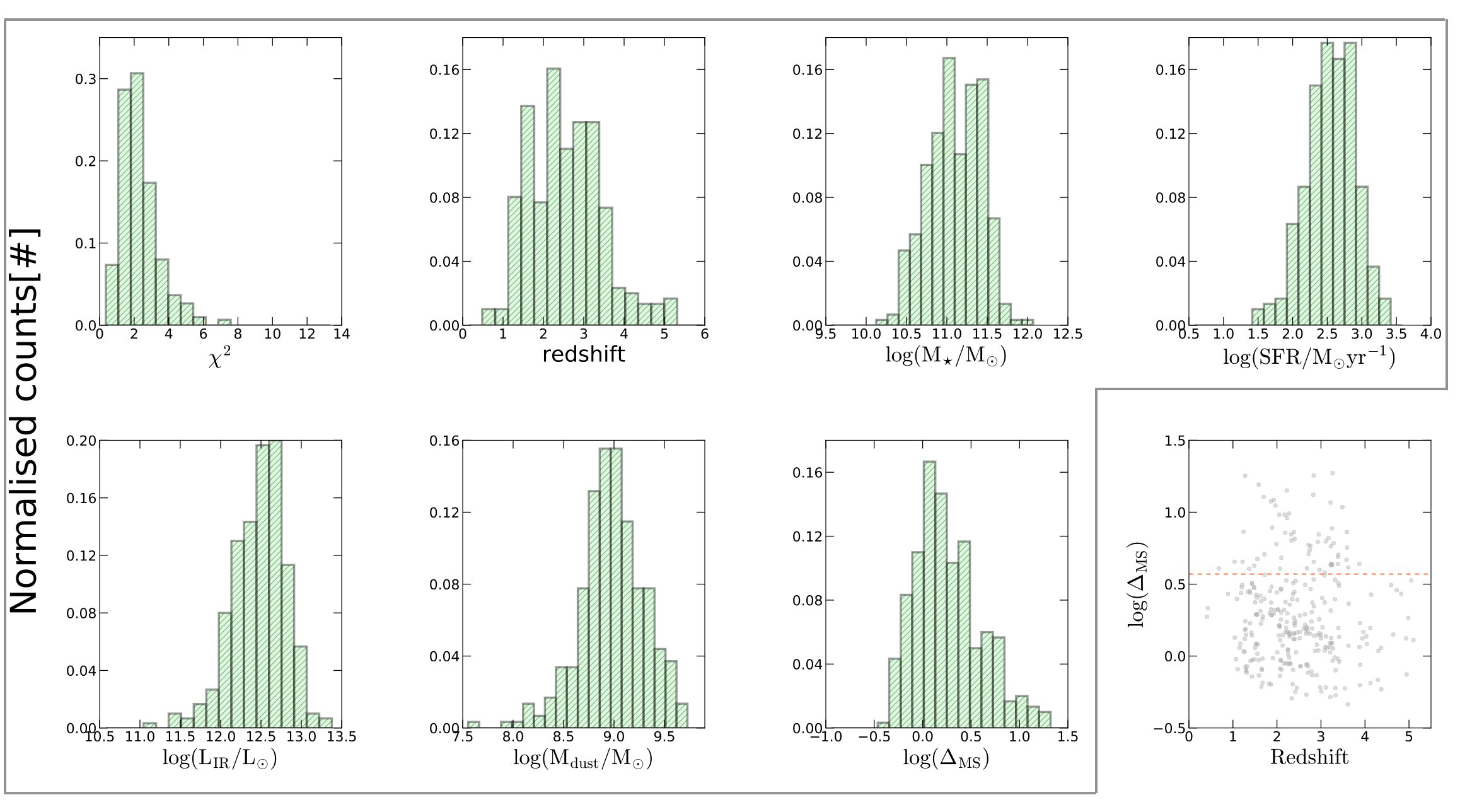}
	\caption{For all panels inside the grey box: distributions of the physical properties estimated for our DSFGs from the SED fitting with CIGALE. \texttt{From top left to bottom right}: the goodness of fit expressed as reduced $\chi^{2}$; galaxy redshift; stellar mass; star-formation rate; IR luminosity; dust mass and galaxy linear offset from the MS (in $\log$ scale). \texttt{Bottom right panel:} linear offset of galaxy’s observed SFR to the SFR expected from the modelled MS ($\Delta_{\rm MS}$ in $\log$ scale), as a function of redshift. A border between the sources considered as MS and SB DSFGs is indicated by a horizontal, dashed line.}
	\label{fig:Fig.2}
\end{figure*}

We now fit the full datasets with the models defined in previous section. Before using our SED derived quantities for the science analysis, we confirm that fitted SEDs have a good quality, quantified  with
the reduced value of $\chi^{2}$<10 (top left panel of \hyperref[fig:Fig.2]{Fig. 2}). We additionally assign the modelling option available within CIGALE to produce mock catalogues and then follow the approach implemented by \cite{malek19} to ensure that our SED fitting procedure does not introduce significant systematics to our measurements (see \hyperref[sec:appA]{Appendix A}). %test the reliability of derived quantities. We show the corresponding results in \hyperref[sec:appA]{Appendix A} and here just note that we find no significant offset and dispersion from the distribution of observed and simulated values of $M_{\rm dust}$ and $M_{\star}$. We ensure that our SED fitting procedure does not introduce significant systematics to our measurements. 
We discard from the further analysis all objects for which the $f_{\rm AGN}$ from our full SED is higher than 20$\%$ (29/329 sources, or $9\%$ of the total sample). We also double-check for additional X-ray-bright AGNs in the COSMOS (\citealt{civano16}) and find none. After this step, the remaining 300 sources are used for our final analysis. The full list of sources and their main properties are presented in \hyperref[tab:4]{Table 3}. %As an example, in \hyperref[fig:Fig.1]{Fig.1} we show one of our distant sources identified through different pass-bands and its corresponding SED.

In \hyperref[fig:Fig.2]{Fig. 2} we show the distribution of SED derived properties, while in  \hyperref[tab:3]{Table 2} we tabulate median physical values confronted to similar ALMA studies (\citealt{dacunha15}, \citealt{as2uds19}). We infer the high median redshift of $z=2.39$ with corresponding 16-84th percentile range ($z=1.66-3.31$). These are also IR luminous, with the median of $L_{\rm IR}=2.93\times10^{12}L_{\odot}$, and $\sim80\%$ of sources above $L_{\mathrm IR}=10^{12}L_{\odot}$. %The minimal and maximal values are $L_{\mathrm IR}=8\times10^{11}L_{\odot}$ and $L_{\mathrm IR}=3.1\times10^{13}L_{\odot}$, respectively. 
The two studies we used for statistical comparison applied the \texttt{MAGPHYS} code (\citealt{daCunha10}) and derived physical parameters fitting the UV-to-sub-mm data of ALMA 870 $\mu$m selected galaxies in the ALESS field (\citealt{dacunha15}) and UDS field (\citealt{as2uds19}). From \hyperref[tab:3]{Table 2} we see there is a consistency among the studies over SED-derived physical quantities ($z$, $L_{\mathrm IR}$, $\rm SFR$ and $M_{\rm dust}$). We note that \cite{as2uds19} analysed a large statistical sample (707 objects in total) but considered only 4 photometric bands in optical-NIR part of SED. A similar criterion is imposed by \cite{dacunha15} who analysed 99 sources out of which 22 have less than 4 photometric detections in optical-NIR range. It is thus likely that inclusion of "optically fainter" sources is responsible for marginally higher median redshifts and wider corresponding range inferred for these DSFGs relative to ours. %However, as evident from \hyperref[tab:3]{Table2}, the median values of our SED-derived physical quantities ($z$, $L_{\mathrm IR}$, $\rm SFR$ and $M_{\rm dust}$) are fairly consistent with the two aforementioned studies, thus being suitable for a direct comparison. %: \cite{dacunha15} report the median value $L_{\mathrm IR}=3.14\times10^{12}L_{\odot}$, while \cite{as2uds19} find $L_{\mathrm IR}=2.88\times10^{12}L_{\odot}$. We reach the similar agreement for other statistical parameters such as $M_{\rm dust}$ or SFR, as evident from \hyperref[tab:3]{Table2}. The distribution of redshifts, $L_{\rm IR}$ and $M_{\star}$ is very similar to ours, thus being suitable for a direct comparison.  

To model the position with respect to the MS for each source in our sample, we apply the functional form of the MS defined by \citealp{speagle14} (their "best-fit", provided with Eq. 28). For each object from our final catalogue we assign $\Delta_{\rm MS}$ defined as an linear offset of galaxy’s observed SFR to the SFR expected from the modelled MS.  We assume the galaxy is a starburst if $\Delta_{\rm MS}\geq4$ (e.g. \citealt{elbaz17}), while the galaxy having $\Delta_{\rm MS}\leq4$ is considered a MS DSFG \footnote{In principle, our results would depend on how well the evolution of the MS with redshift is constrained. We thus also test the MS relation of \cite{schreiber15} but find that the choice of adopted MS does not significantly impact the statistics of our MS and SB DSFGs.We thus kept \citealp{speagle14} relation for easier comparison to studies that build gas-scaling relations upon the same MS modelling method (see Section 4.4).}. %However, we find that the number of sources that can be considered as starbursts insignificantly changes from 58 to 60 galaxies. We thus kept \citealp{speagle14} relation for easier comparison to studies that build gas-scaling relations upon the same MS modelling method (see Section 4.4)}. 
We infer that our sample contains 242 MS DSFGs ($81\%$ of the total) and 58 SB DSFGs ($19\%$ of the total).

%%%%%%%%%%%%%%%%%%
%%%%%%%%%%%%%%%%%%%

\begin{table*}
	\caption{The statistics of our SED derived physical properties with \texttt{CIGALE}, compared to those from known statistical ALMA studies (\citealt{dacunha15} and \citealt{as2uds19}). The physical parameters in these two studies are estimated via multi-band SED fitting with the code \texttt{MAGPHYS}. The range indicated with each median corresponds to the 16th-84th percentile of the likelihood distribution.}
	\label{tab:3} 
	\centering   
	%	\vspace{0.03cm}   
	\scalebox{0.84} {\begin{tabular}{c c c c } 
			%\hline\hline       
			% To combine 4 columns into a single one 
			%       %HJD & $E$ & Method\#2 & \multicolumn{4}{c}{Method\#3}\\ 
			\hline
			\toprule 
			& This work & ALESS & AS2UDS\\
			&&\cite{dacunha15}&\cite{as2uds19}\\[-7pt]\\
			%& $T_{\mathrm{d}}=38\pm10$K &&&&&&&& \\[-3pt]\\
			\hline 
			\\
			\centering
			%$\#$ & 301 & 99 & 707\\[-3pt]\\
			$\langle z \rangle$ & $2.39^{+0.92}_{-0.73}$ &
			$2.7^{+1.39}_{-1.1}$ & 
			$2.61^{+0.79}_{-0.81}$\\[-3pt]\\
			\centering
			%	$\hat{z}$ & $(2.49\pm0.73)$ & $(2.82\pm0.97)$ & $(2.61\pm0.07)$\\[-3pt]\\
			$\langle L_{\rm IR}\rangle$ & $2.93^{+2.17}_{-1.31}\times10^{12}L_{\odot}$  &  $3.24^{+2.4}_{-1.8}\times10^{12}L_{\odot}$  & 
			$2.88^{+2.52}_{-1.3}\times10^{12}L_{\odot}$ \\[-3pt]\\
			\centering
			$\langle\rm SFR\rangle$ & $270^{+255}_{-170}\:\rm M_{\odot}yr^{-1}$  & 
			$281^{+420}_{-190}\:\rm M_{\odot}yr^{-1}$ & 
			$236^{+240}_{-150}\:\rm M_{\odot}yr^{-1}$\\[-3pt]\\
			\centering
			$\langle M_{\star}\rangle$ & $1.02^{+0.7}_{-0.4}\times10^{11}M_{\odot}$ &
			$0.89^{+0.7}_{-0.4}\times10^{11}M_{\odot}$ & 
			$1.26^{+0.5}_{-0.5}\times10^{11}M_{\odot}$\\[-3pt]\\
			\centering
				$\langle M_{\rm dust}\rangle$ & $7.33^{+5.2}_{-3.4}\times10^{8}M_{\odot}$ &
			$6.01^{+5.6}_{-3.8}\times10^{8}M_{\odot}$ & 
			$6.8^{+5.1}_{-3.6}\times10^{8}M_{\odot}$\\[-3pt]\\
			
			\hline    
			\bottomrule               
	\end{tabular} }\\
	\vspace{0.5cm} 
	
\end{table*}
%%%%%%%%%%%%%%%%%%%%%

\section{The evolution of dust-to-stellar properties over cosmic time}
\label{sec:4}

\subsection{The evolution of $M_{\rm dust}$ with redshift}

\begin{figure}[h]
	%	\vspace{-0.3cm}
	\centering
	\hspace{-0.5cm}
	\includegraphics [width=9.03cm]{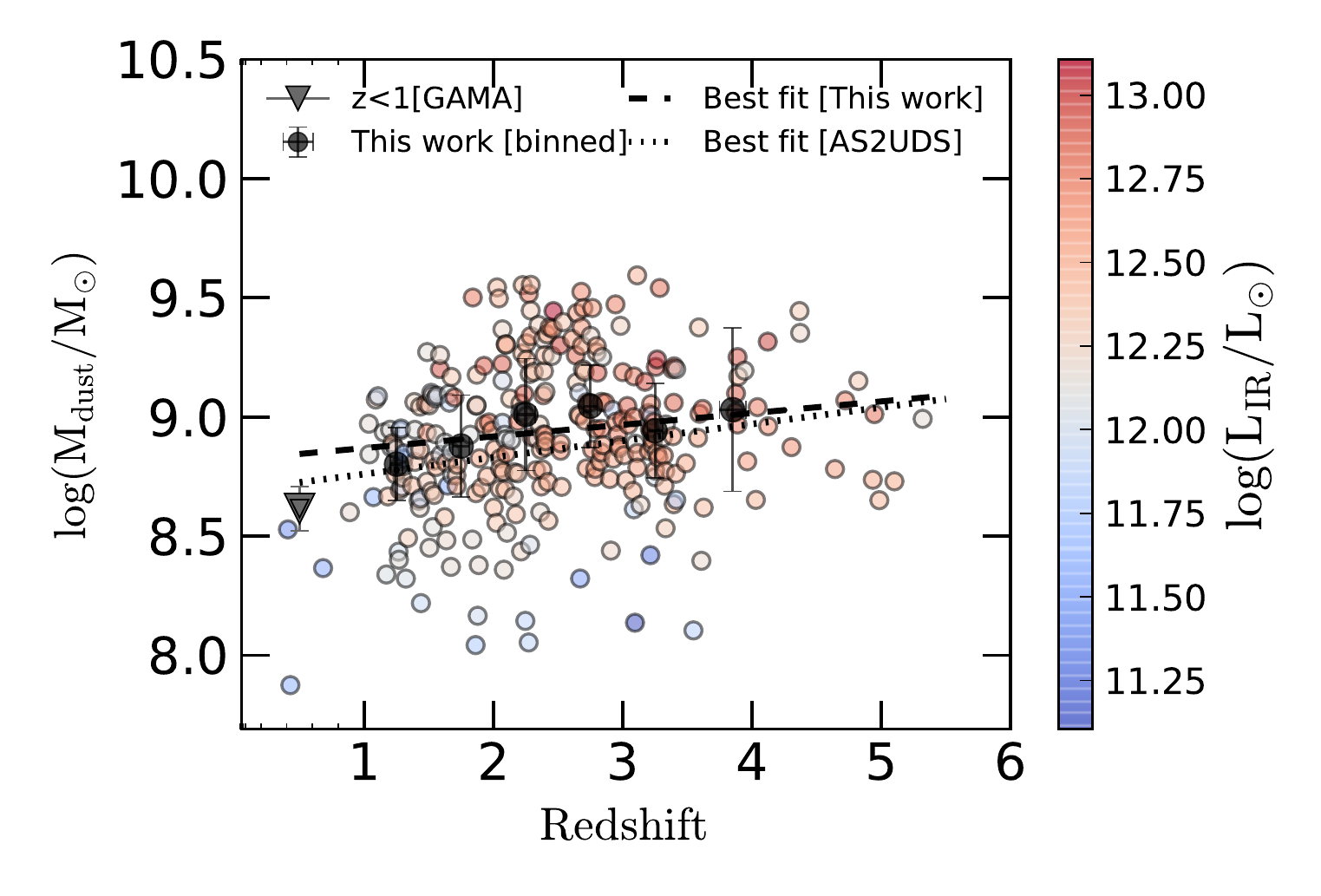}
	
	\caption{The observed redshift evolution of $M_{\rm dust}$. Individual values are displayed with circles, coloured with corresponding $L_{\rm IR}$. Binned means are shown with black circles and associated $1\sigma$ errors. For comparison, we also show the mean	 $M_{\rm dust}$ for a large sample of dusty galaxies up to $z\sim0.5$ (\citealt{driver18}, black inverted triangle). The black dashed and dotted lines are best regression fits from this work and \cite{as2uds19} respectively.}
	\label{fig:Fig.4}
\end{figure}

In \hyperref[fig:Fig.4]{Fig. 2} we plot the redshift evolution of the $M_{\rm dust}$ for the full sample of our DSFGs. From the multi-band SED fitting we find the median of $M_{\rm dust}=7.33^{5.2}_{-3.4}\times10^{8}M_{\odot}$. The relative contribution to the total sample of DSFGs with $M_{\rm dust}>10^{9}M_{\odot}$ is $29\%$, which places one third of our DSFGs towards the most extreme tail of the dust mass function (DMF). As we see from \hyperref[tab:3]{Table 2}, the median $M_{\rm dust}$ from this work is in consistency with findings from \cite{dacunha15} and \cite{as2uds19}. We note that \cite{dacunha15} applied slightly different prescription in their dust SED model and explore the grid of SFHs and stellar metallicities in order to derive physical properties of ALMA observed galaxies. This strengthens the conclusion that high $M_{\rm dust}$ in ALMA detected DSFGs is not an observational artefact due to adopted SED fitting procedure. %Although not surprising, \hyperref[fig:Fig.4]{Fig. 2} clearly shows that more luminous sources have $M_{\rm dust}$ higher in average than the less luminous ones (e.g. those below $10^{12}L_{\odot}$). %are very high irrespective of the SED fitting methods. %We see a great overall agreement between the two fits, with the difference that our function imposes somewhat shallower slope and higher normalisation, pronouncing the small difference up to $z\sim2$. 
The inferred $M_{\rm dust}$ of our DSFGs are in average $\sim0.2\:\rm dex$ larger than the values measured through a $\textit{Herschel}$ stacking analysis of galaxies at $z<2.5$ (\citealt{santini14}. The similar difference is seen if we compare to the median $M_{\rm dust}$ of a large sample of dusty sources in the local Universe  ($z<0.5$) detected within the GAMA survey (\citealt{driver18}).

%with black square and triangle in \hyperref[fig:Fig.4]{Fig.3}\cite{driver18} analysed a large sample of dusty sources up to $z\sim0.5$ within the GAMA survey and found that $M_{\rm dust}$ is order of magnitude higher than what has been measured in H-ATLAS galaxies at $z<0.1$ (presented with black square and triangle in \hyperref[fig:Fig.4]{Fig.3}, respectively).

%It has already been pointed in the literature (see e.g. \citealt{rowlands15}, \citealt{pozzi19}) that the most massive dusty galaxies show increase of their dust content with redshift. For example, for 29 bright SPIRE galaxies at $z>1$, \cite{rowlands15} derive the median dust mass $\hat{M_{\mathrm dust}}=(1.9\pm0.54)\times10^{9}M_{\odot}$, very similar to our estimates. However,
Despite the fact that SFRs of our sources range over almost three orders of magnitude (namely, from $40\:M_{\odot}\rm yr^{-1}$ to $1640\:M_{\odot}\rm yr^{-1}$), their $M_{\rm dust}$ exhibits milder variation, in average $\sim25\%$ across the observed redshift range. We quantify the observed cosmic evoluton of $M_{\rm dust}$ using linear regression fit of the form:  
\begin{equation}
\log(M_{\rm dust}) = (0.052\pm0.04)\times z + (8.80\pm0.09)
\end{equation}
The slow rise with redshift is qualitatively consistent with findings from \citet[their Fig. 11] {as2uds19}. 
As pointed out by \cite{as2uds19}, since $M_{\rm dust}$ is strongly correlated to ALMA 870 $\mu$m flux, the broad agreement amongst the different ALMA studies likely reflects the similar flux limits of the single-dish surveys followed-up with ALMA. It has been found  that sub-mm flux of extremely luminous DSFGs selected from single dish SCUBA2 camera (at $850\:\mu$m), strongly correlates with redshift (\citealt{stach19}, \citealt{simpson20}). Due to tight connection between dust and gas (young stars are predominantly formed in dense molecular clouds, while dust catalyses transformation from atomic hydrogen into molecular), it has been proposed that steady range of $M_{\rm dust}$ should correspond to a similarly uniform selection in terms of $M_{\rm gas}$ (e.g \citealt{swinbank14}, \citealt{simpson20}).

%%%%%%%%%%%%%%%%%%

%%%%%%%%%%%%%%%%%%%
\subsection{The evolution of $M_{\rm dust}-\rm SFR$ with respect to the main-sequence}

In order to achieve a closer insight to the ISM of our DSFGs, we now explore how the dust masses relate to their star-formation rates. The $M_{\rm dust}$ and $\rm SFR$ are expected to be correlated  in galaxies (\citealt{daCunha10}, \citealt{dunne11}, \citealt{bourne12}, \citealt{santini14}, \citealt{rowlands15}, \citealt{kirkpatrick17}, \citealt{aoyama19}). Such a relationship can naturally be understood due to dust mass being a good tracer of the $M_{\rm gas}$ in DSFGs (\citealt{scoville17b}), while $M_{\rm gas}$ and SFR are linked through the known Kennicutt–Schmidt relation (KS, \citealt{schmidt59}, \citealt{kennicutt98}, \citealt{sargent14}). However, it is less known whether the expected relation holds towards the highest redshifts and higher SFRs. 

\begin{figure}[h]
	%\vspace{-0.2cm}
	\centering
	\hspace{-0.4cm}
	\includegraphics [width=9.03cm]{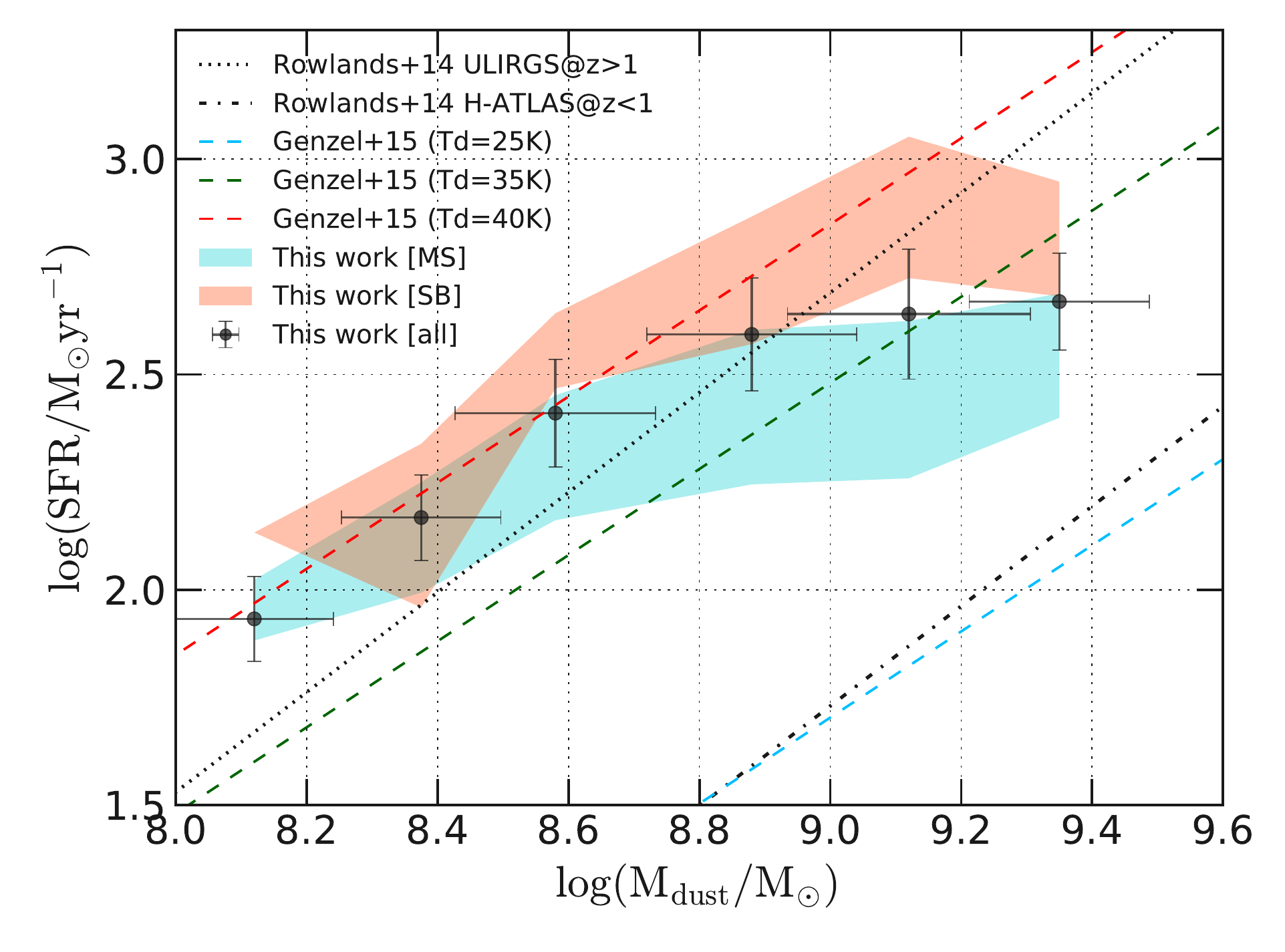}
	\caption{The observed relation between $M_{\rm dust}$ and $\rm SFR$ in our DSFGs, shown for the whole sample (binned means, shown as black circles), and divided on SB and MS galaxies (shaded in dark cyan and orange, respectively). The known, empirically based scaling relation between $T_{\rm dust}-\rm SFR-\it M_{\rm dust}$ (\citealt{genzel15}) are overlaid with dashed lines. Different colours correspond to their fixed $T_{\rm dust}$, as indicated in the legend. The best fit for local and intermediate redshift ULIRGs (\citealt{rowlands15}) are displayed with dotted and dot-dashed line, respectively.}
	\label{fig:Fig. 4}
\end{figure}

In \hyperref[fig:Fig. 4]{Fig. 3} we display how the $M_{\rm dust}$ relates to SFR. We show the median values of the binned data for the full sample, and for MS and SB DSFGs, separately. We find positive evolutionary trend of SFR with $M_{\rm dust}$. which holds for both populations of galaxies, while at the fixed $M_{\rm dust}$, SB DSFGs have on average higher SFR than the MS sample. Interestingly, we see that the linear trend between $M_{\rm dust}$ and SFR starts to flatten towards the upper right part of the diagram. Up to the $M_{\rm dust}\lesssim10^{9}M_{\odot}$, the flattening of the relation is mainly caused by MS galaxies. Considering the relation between $M_{\rm dust}$ and SFR as a consequence of KS law \cite{miettinen17} argued that a shallower slope towards higher $M_{\rm dust}$ could mean that DSFGs deviate from a traditional KS law (see also \citealt{santini14}).
 %The relation presented in \cite{rowlands15} is the best fit to their data -  local and intermediate ULIRGs selected in H-ATLAS field from \textit{Herschel} SPIRE 250$\mu$m band. The scaling relation from \cite{genzel15} was built under assumption that the dust re-emits the absorbed energy of stellar photons in the optically thin IR regime, at an average constant $T_{\rm dust}$ with the fixed emissivity assumed to be $\beta = 1.5$. 
To better understand this finding, we overlaid our data with the best scaling relations between $M_{\rm dust}$ and $\rm SFR$ derived by \cite{genzel15} and \cite{rowlands15}. The scaling relations are built upon the approximation that dust SED can be represented with an average constant $T_{\rm dust}$ and dust emissivity which is assumed to be $\beta = 1.5$. The scaling relations we show in \hyperref[fig:Fig. 4]{Fig. 3} imply that $M_{\rm dust}$ and $L_{\rm IR}$ are correlated as $L_{\rm IR}/M_{\rm dust}\propto T_{\rm dust}^{4+\beta}$ (\citealt{blain03}). From this we would expect fully linear trend between $M_{\rm dust}$ and SFR. Our binned data differ from this expectation, and can be better described by (logistic) function that saturates at $M_{\rm dust}\sim10^{9}M_{\odot}$. Therefore, the sub-linear relation deduced from our sample could also reflect the possible change in ISM conditions (e.g. wide distribution of the radiation field intensities, different optical depths and source geometry). This is in line with results from studies which explored the cosmic evolution of interstellar radiation fields and its complex link to galaxy stellar mass (\citealt{bethermin15}, \citealt{schreiber18}) or gas mass (\citealt{kirkpatrick17}, \citealt{mckinney20}).

%%%%%%%%%%%%%%%%%%%

%%%%%%%%%%%%%%% 

%However, for the case of massive galaxies such relation is far from being trivial, as already pointed by seminal works by \cite{daCunha10}, \cite{dunne11}, see also \cite{tan14} and \cite{bethermin15}. We come to this point again in the next Chapter, where we analyse the evolutionary picture related to our sources. 

 %This implies shallower relation between $M_{\rm dust}$ and SFR at high-$z$, as compared to what is usually found in their local "analogues". 
Our DSFGs are probing the highest end of the SFR-$M_{\rm dust}$ plane, which sparsely overlap with local DSFGs (see \citealt{daCunha10}). This is evidence for extremely efficient and rapid dust formation process at earlier cosmic epochs (\citealt{hjorth14}, \citealt{rowlands15}, \citealt{lesniewska19}, \citealt{dwek19}). There are several possible explanations why $M_{\rm dust}$–SFR relation of our DSFGs lie above the locally inferred values. Studying the DSFGs at $z\sim2$, \cite{kirkpatrick17} conclude that high-$z$ DSFGs have larger than average molecular gas reservoir than galaxies with similar $M_{\rm dust}$ at lower redshifts. Other works argued towards much higher efficiency of converting gas to stars. \cite{magdis12} demonstrate that dust luminosity emitted per unit of dust mass could also serve as a good indicator of star formation efficiency ($\rm SFE=\rm SFR/M_{\rm gas}\propto L_{\rm IR}/M_{\rm dust}$). Such approximation is valid if $L_{\rm IR}\propto \rm SFR \propto M_{\rm gas}$ and under the assumption that ratio between the dust and gas mass (hereafter $\delta_{\rm DGR}$) is roughly constant. By examining this formalism, \cite{schreiber17} conclude that physical changes in the ISM could be responsible for enhanced SFE, such that most massive galaxies at $z<1$ have reduced interstellar radiation fields, and correspondingly reduced SFEs.  %An interesting point to stress is that the difference between MS and SB DSFGs seen in \hyperref[fig:Fig. 4]{Fig. 3} could originate if $M_{\rm dust}$ grows on timescales faster than $M_{\star}$. Therefore an object which shows stronger positive offset relative to main sequence can have accumulated at early times almost all the dust of a MS object. We will return to this point in \hyperref[sec:6]{Section 6}.

The cause of the flattening of $M_{\rm dust}$-SFR relation is interesting to discuss. From our data we see that shallower rise is mostly driven by MS DSFGs, while SB DSFGs are more compatible with linear scaling from \cite{genzel15}. At the first glance, the flattening could be a consequence of our sample being incomplete at a fixed stellar mass. Nevertheless, the similar departure from the linear trend between $M_{\rm dust}$ and SFR has been found in the complete AzTEC survey of the brightest DSFGs selected as $S_{1.1\rm mm}>3.5\rm\:mJy$ (\citealt{miettinen17}). \cite{hjorth14} investigate simple analytical limiting cases for early dust production, being the first that propose the bending of the SFR-$M_{\rm dust}$ relation. They postulate that a maximum attainable $M_{\rm dust}$ is in early starburst phase in which the rapid dust build-up in very massive systems at early cosmic times is the cause of the observed bend-over of the $\rm SFR$-$M_{\rm dust}$ relation. However, to reproduce high dust yields, the scenario proposed by \cite{hjorth14} imposes extreme dust-formation efficiency by SNe under the galaxy closed-box solution which is found to be unrepresentative for most of known DSFGs (see e.g discussion in \citealt{pantoni19}). Therefore, the fact that we see plateau rather than a linear rise of SFR towards the $M_{\rm dust}$ can be explained if the dust mass build-up is related to additional dust production source, e.g. the grain growth in the ejecta/remnant or the ISM. The process is believed to be very fast with a timescale of a few tens of million years (\citealt{asano13}, \citealt{hirashita17}, \citealt{popping17}, \citealt{pantoni19}). In the next Section, we will closely investigate this possibility through different scaling relations that link dust, gas and metal content in our DSFGs.

It is also possible that the dust emission in compact DSFGs is affected by opacity effects. As we show in \hyperref[sec:6]{Section 6}, some of our sources have extreme surface densities of dusty star-formation, which would make the gaseous ISM highly optically thick even in the IR regime (\citealt{cortzen20}). The fact that the $\rm SFR$-$M_{\rm dust}$ relation becomes flat for our DSFGs at the high $M_{\rm dust}$ end, further supports this possibility. To check how the opacity assumption affects our results, we use prescription of a thick dust model from \cite{dowell14} and fit the IR SEDs to the sources from the highest dust mass bin ($M_{\rm dust}>2\times10^{9}\:M_{\odot}$). We find that use of a thick dust model returns $\sim2-3\times$ lower $M_{\rm dust}$ due to increase in $T_{\rm dust}$ at a given $L_{\rm IR}$. This is in line with \cite{cortzen20}, who studied GN20, known starburst at $z=4$, and reported $\sim2\times$ discrepancy between the dust masses derived from optically thin and optically thick dust model. However, as pointed by \cite{cortzen20}, it is difficult to properly quantify these effects because optically thin or thick solutions are heavily degenerate, and require independent proxy for $T_{\rm dust}$ to discriminate between the two. For the sake of consistency, we thus keep our DL14-based $M_{\rm dust}$ for the rest of the paper.

\subsection{The evolution of $M_{\rm dust}/M_{\star}$ with respect to the main-sequence}
\label{sec:4.3}

\begin{figure*}[ht]
	\vspace{-0.2cm}
	\centering
	%\hspace{-1.0cm}
	%	\includegraphics [width=13.89cm]{dsfg-tracks.pdf}
	\includegraphics [width=18.69cm]{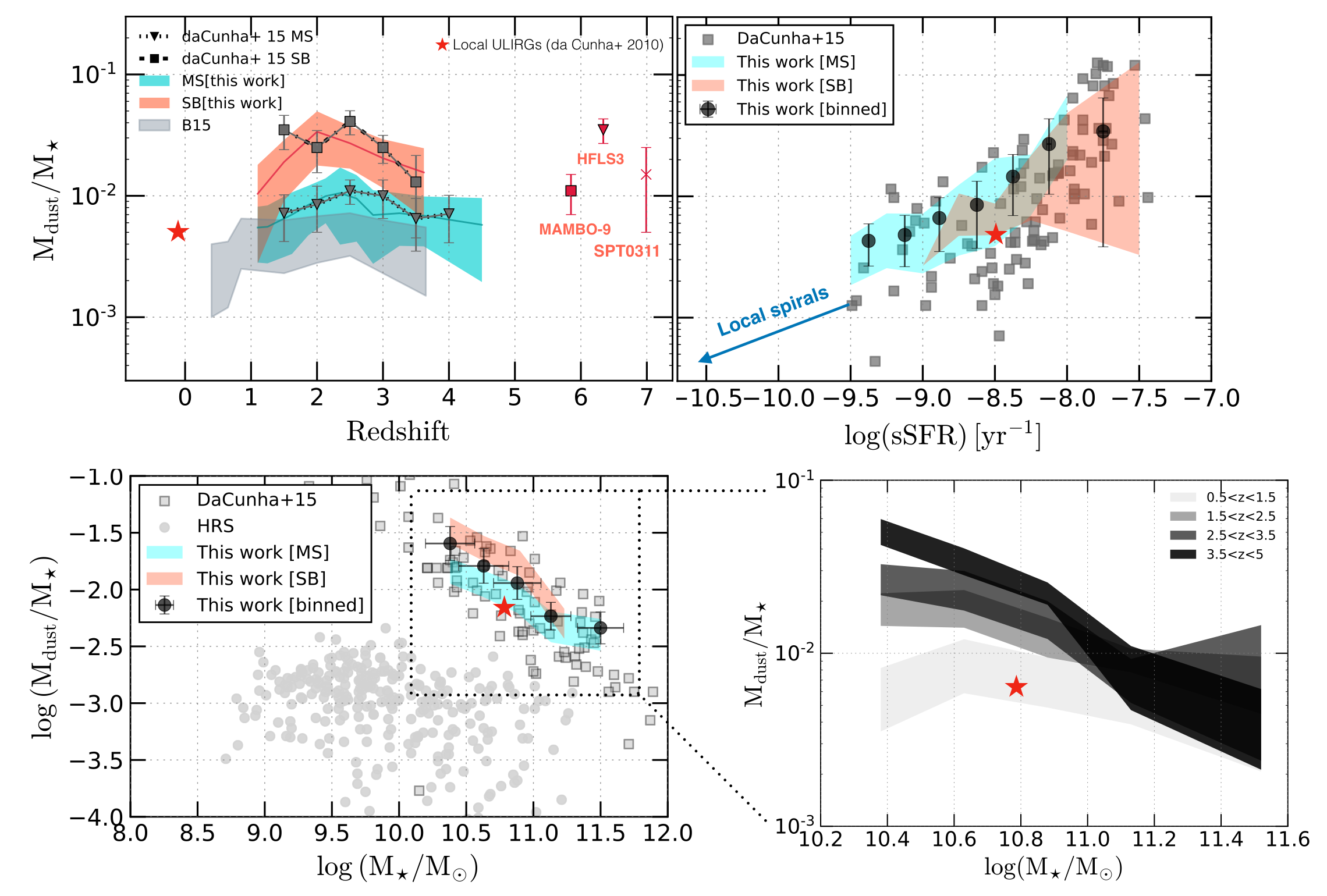}
	\caption{\texttt{Top panel:} The $M_{\rm dust}/M_{\star}$ versus redshfit (left) and sSFR (right), of our MS and SB DSFGs.The binned averages and corresponding standard errors for MS and SB subsample are shown as shaded dark cyan and orange area, respectively. The grey, shaded area is the observed trend obtained via stacking analysis by \cite{bethermin15}. The red star in each panel indicate the median value of most extreme local ULIRGs (\citealt{daCunha10}). Binned values for MS and SB from high-redshift ALMA sample analysed by \citealt{dacunha15} are shown with triangles and squares, respectively. Also displayed with red symbols are individual detections of the most distant DSFGs at $z>5$ (MAMBO-9	\citealt{casey18}, \citealt{jin19}; HFLS3, \citealt{riechers13}; and SPT0311, \citealt{strandet17}). For consistency, we use public data and recalculate $M_{\rm dust}$ of latter three objects following our method, finding a good agreement with their archival estimates. In the right panel, the binned mean values of the full sample are presented with black circles and the results from \cite{dacunha15} with filled squares. The error bars represent the dispersion ($1\sigma)$ associated to the mean. \texttt{Bottom panel:} The $M_{\rm dust}/M_{\star}$ versus galaxy stellar mass.The left panel shows our estimates compared to those that span the similar stellar mass range. Points are colour-coded in the same way as in the upper plot.  Grey circles indicate the local galaxy sample composed of \textit{Herschel} detected galaxies (both passive and active) from the Virgo cluster (HRS, \citealt{andreani18}). On the right side, we resolve the overall trend of $M_{\rm dust}/M_{\star}$ vs. $M_{\star}$ from this work per different redshift bins and plot the trend with corresponding $1\sigma$ uncertainty. Redshift bins are colour-coded as in the legend. } 
	\label{fig:Fig.6}
\end{figure*}

We now explore how various physical quantities of our DSFGs relate to $M_{\rm dust}/M_{\star}$ in MS and SB DSFGs. Our goal here is to use the $M_{\rm dust}/M_{\star}$ as a tool to assess the efficiency of the specific dust production and destruction mechanisms in galaxies. In \hyperref[fig:Fig.6]{Fig. 4} we present different evolutionary trends of $M_{\rm dust}/M_{\star}$ for MS and SB DSFGs against the redshift, sSFR and stellar mass. %In order to better understand how our data relate to other works of DSFGs, we also compile different data sets, as described in the legend of \hyperref[fig:Fig.6]{Fig. 4}.

The upper left panel of \hyperref[fig:Fig.6]{Fig. 4} illustrates the evolution of $M_{\rm dust}/M_{\star}$ ratio with redshift. We placed estimated values for MS (SB DSFGs) in 11 (6) redshift bins of 0.3 (0.5). We did not analyse the highest redshift bins ($z>5$ for MS and $z>3.5$ for SB DSFGs) due to lack of statistical significance (they contain only 2 and 1 objects, respectively). The binned means and their standard errors are shown as cyan and orange regions for MS and SB DSFGs, respectively. The medians of our DSFGs are found to be $M_{\rm dust}/M_{\star}=0.006^{+0.004}_{-0.003}$ for MS DSFGs and $M_{\rm dust}/M_{\star}=0.017^{+0.010}_{-0.006}$ for SB DSFGs. We find that for both populations $M_{\rm dust}/M_{\star}$ rises up to the certain redshift ($z\sim2-2.25$) and flattens/bend towards earlier epochs. It is worth noting that no SB DSFGs are observed at $z>4$. This can be a consequence of our source selection, but also an indication that SB DSFGs at high-$z$ lie systematically below the central relation for starbursts predicted from KS law (\citealt{santini14}, \citealt{bethermin15}, \citealt{silverman18}, \citealt{liu19b}). Nonetheless, the SB DSFGs typically have $3-4$ times higher $M_{\rm dust}/M_{\star}$ as compared to MS DSFGs, regardless of the observed redshift. The variation with redshift amongst the binned values is mild, about 0.3 dex. All these imply that the carefully estimated $M_{\rm dust}/M_{\star}$ can be applied as a useful tool for distinguishing SB and MS dusty galaxies over wide redshift range.

The different evolution of $M_{\rm dust}/M_{\star}$ with redshift for MS and SB DSFGs has already been reported in \cite{bethermin15}  (see also \citealt{tan14}). They construct the average SED of MS and SB DSFGs detected from stacking analysis. By deducing the mean intensity of the radiation field, they estimate the $M_{\rm dust}$ from \cite{dl07} dust SEDs. Their result for MS DSFGs is displayed with grey shaded region in \hyperref[fig:Fig.6]{Fig. 4}.  Considering the MS DSFGs, we can see the similarity between the trends, such that values inferred by \cite{bethermin15} suggest an increase in $M_{\rm dust}/M_{\star}$ until $z\sim1.5$, and slight decline towards higher-$z$'s. This is very similar to the overall evolutionary shape we infer from our data, although average values from \cite{bethermin15} are slightly lower than ours, due to the stacking technique they adopt in order to reach lower $L_{\rm IR}$, thus inferring somewhat lower normalisation for $M_{\rm dust}/M_{\star}$. In any case, our estimates are within $1\sigma$ uncertainty from those of \cite{bethermin15} over the whole redshift range. The same authors also analyse extreme starbursts ($\Delta_{\rm MS}>10$), obtaining very steep slope for SB DSFGs at $0<z<2$ (see their figure 8). To compare observed trends with other studies of individual ALMA galaxies, we bin the data from \cite{dacunha15} and select objects as MS and SB DSFGs in the exact same way as for our sample. As evident from \hyperref[fig:Fig.6]{Fig. 4}, coherence between our data and those from \cite{dacunha15} is present over full redshift range. For the sake of clarity, we also show the median value obtained for the sample of the most extreme local ULIRGs (\citealt{daCunha10}), along with some of the most distant individual DSFGs confirmed to date (\citealt{riechers13}, \citealt{strandet17}, \citealt{casey18}). Despite the fact that we yet have to reach the census of observed DSFGs at $z>4-7$, it is clear that even the most distant DSFGs have very high $M_{\rm dust}/M_{\star}$, hinting there could be a wide spread in the dust-to-stellar ratio of star-forming galaxies. \cite{jin19} recently showed that some number of the most distant sources with high $M_{\rm dust}/M_{\star}$ could be a rare population of cold starbursts ($T_{\rm dust}<30$K). They argue on observed cold dust temperatures being a result of either low star formation efficiency with rapid metal enrichment, or evidence for optically thick dust continuum in the FIR.  \citealt{cortzen20}).%As recently proposed by \cite{jin19}, some number of these sources could be rare population of very cold starbursts, with $T_{\rm dust}$<30 K, and RJ values steepened by CMB. lowe Considering the although it is uncertain whether or not they can be considered starbursts, due to currently unknown "main-sequence" relation for such distant objects. 

The top-right panel of \hyperref[fig:Fig.6]{Fig. 4} discloses a strong mutual correlation between $M_{\rm dust}/M_{\star}$ and $\rm sSFR$. %slightly higher in average in high-$z$ DSFGs than in local ULIRGs. 
This is in agreement to what has been reported in the literature for a different statistical samples of DSFGs (\citealt{hunt14}, \citealt{martis19}) and Lyman Break Galaxies (LBGs, \citealt{burgarella20}).  We find that correlation shows a substantial scatter but extends over two orders of magnitude and saturates at the highest sSFR which is in consistency with \cite{dacunha15}. The scatter could be due to cosmic evolution of the relation between $T_{\rm dust}$ and $M_{\star}$, which is found to be non-monotonic and strongly dependent on galaxy ISM (\citealt{kirkpatrick17}, \citealt{imara19}). %} found complex, non-monotonic dependence of the redshift evolution of $T_{\rm dust}$ with $M_{\star}$ is non-monotonic, and its exact shape depends on the galaxy ISM.%Considering the position of our galaxies in the $M_{\rm dust}/M_{\star}$-sSFR diagram, we can see the clear difference between distant DSFGs and the most extreme, local ULIRGs. Despite having very high $M_{\rm dust}/M_{\star}$ ratio, they do not enter the upper right corner of the plane. We observe a large scatter of our data in all sSFR bins, which can be a consequence of evolving interstellar radiation field (ISRF). 

The relation between $M_{\rm dust}/M_{\star}$ and $\rm sSFR$ can be interpreted as age-evolutionary sequence. That said, the difference between objects populating the opposite corners of the $M_{\rm dust}/M_{\star}-\rm sSFR$ plane could originate if $M_{\rm dust}$ grows on timescales faster than $M_{\star}$. The important outcome of this interpretation is that DSFGs from the upper-right side of the diagram could be objects dominated by young stellar populations that could have accumulated at early times almost all the dust of a normal "main-sequence" object. These young DSFGs are expected to own large amounts of molecular gas relative to stars which would place them in the uppermost part of the $M_{\rm dust}/M_{\star}-\rm sSFR$ diagram, in line with the picture where the sSFRs in more massive DSFGs peak earlier in the Universe than those of less massive objects (\citealt{lefloch05}, \citealt{behroozi13}). The subsequent decrease of sSFR is due to exhaustion of their gas reservoirs and reflects the efficiency of dust removal. Such interpretation would be consistent with the scenario proposed by \cite{burgarella20}, who studied LBGs at $z>5$ and found that sources with the youngest stellar populations have the highest sSFRs (see also \citealt{calura17}). We will return to this point in \hyperref[sec:5.3]{Section 5.3} and \hyperref[sec:6]{Section 6}. %but rather peaks at characteristic values of stellar mass. The latter can also be an indication of temporal changes of the amount of large dust grains relative to small ones, which is required to induce changes in $T_{\rm dust}$ if dust is optically thin. 

In the lower panel of \hyperref[fig:Fig.6]{Fig. 4} we show $M_{\rm dust}/M_{\star}$ as a function of $M_{\star}$. For our sources we show the median values  computed in bins of stellar mass, along with trends inferred for MS and SB DSFGs, separately. We observe a clear anti-correlation between $M_{\rm dust}/M_{\star}$ and $M_{\star}$, with the normalisation being higher in SB DSFGs than in MS DSFGs. We confirm that such distinction holds until $\log(M_{\star}/M_{\odot})=11.2$, while above this $M_{\star}$ there are no sources considered as starbursts. The anti-correlation of $M_{\rm dust}/M_{\star}$ with $M_{\star}$ is known to exists in the locally observed galaxies (\citealt{bourne12}, \citealt{deVis17}, \citealt{casasola20}). We see that our DSFGs tend to have in average slightly higher median $M_{\rm dust}/M_{\star}$ per fixed stellar mass, than of the most extreme local ULIRGs (marked with the red star).  The difference is much larger and exceeds an order of magnitude if we compare to locally detected early and late-type galaxies from the  \textit{Herschel} Reference Survey (HRS, \citealt{andreani18}). %The inferred trend seems more complex than what is predicted from simple empirical prescriptions, perhaps due to non-negligible contribution of early starbursts which cause broken power law dependence between $M_{\rm dust}/M_{\star}$ and stellar mass, rather than a strong linearity. 

We further inspect the evolution of this inverse relation with redshift, by dividing the full sample in four redshift bins, as denoted in the lower-right side of \hyperref[fig:Fig.6]{Fig. 4}. We unveil several interesting features. Firstly, we provide for the first time the strong observational evidence that anti-correlation of $M_{\rm dust}/M_{\star}$ with $M_{\star}$, continues up to $z\sim5$ in massive DSFGs. %This result implies that in low mass galaxies, the specific production of dust is particularly efficient and that the balance between dust pro- duction and destruction is somehow dependent on the mass of the galaxy (see Sect. 5).%
We find a systematic shift towards higher $M_{\rm dust}/M_{\star}$ with increasing redshift. This seems valid at least until $M_{\star}\sim10^{11}M_{\odot}$, after which the difference in normalisation becomes less prominent, coincidental with the stellar mass range mostly unpopulated with SB DSFGs. Secondly, there is a tentative evidence for a change of slope of the inverse relation with redshift. Lurking at the lowest redshift bin we see that our data indicate a slight turnover of $M_{\rm dust}/M_{\star}$ at characteristic $M_{\star}$, which is followed by a mild overall change of the amplitude. Towards  higher-$z$'s the inverse relation becomes steeper, and can be roughly quantified as a simple power law evolving from $M_{\rm dust}/M_{\star}\propto M^{-0.21}_{\star}$ to $M_{\rm dust}/M_{\star}\propto M^{-0.57}_{\star}$ at $1.5<z<5$. We caution that less biased sample of spectroscopically confirmed candidates at $z>3-5$ is necessary for confirming this claim. 

The observed anti-correlation of $M_{\rm dust}/M_{\star}$ with $M_{\star}$ seems a natural reflection of the dust life-cycle: $M_{\star}$ grows with time as galaxy evolving, while dust grains (altogether with ISM metals) decrease from the budget being incorporated into the stellar mass. \cite{calura17} applied the chemical galaxy model on proto-spheroidal galaxies, suggesting that observed trend of $M_{\rm dust}/M_{\star}$ with $M_{\star}$ is strongly dependent on galaxy star-formation history. They demonstrate that galaxies characterised by a prolonged (bursty) star formation activity, shows a rather flat (steep) behaviour of $M_{\rm dust}/M_{\star}$ with respect to $M_{\star}$. They concluded that the observed inverse relation is due to the time evolution of $M_{\rm dust}/M_{\star}$ in the late starburst phase of DSFGs. During this evolutionary phase, $M_{\star}$ is still increasing, but the the galaxy SFR and the dust production rate decrease resulting in a downhill of $M_{\rm dust}/M_{\star}$ towards the point which characterises the end of star formation. 

\cite{imara19} developed analytical solution based on simplified empirical prescriptions and found that the evolution of  $M_{\rm dust}/M_{\star}$ with $M_{\star}$ can be parametrised as a broken power law, where the breaking point is controlled by $\delta_{\rm DGR}$. They highlight that the evolution of galaxy molecular gas mass ratio (defined as $\mu_{\rm gas}=M_{\rm gas}/M_{\star}$) is crucial in regulating the observed $M_{\rm dust}/M_{\star}$ per fixed $M_{\star}$. In this regard, the decreasing trend with higher $M_{\star}$ could be due to the deficiency of galaxies with high $\mu_{\rm gas}$ above the critical stellar mass ($M_{\star}\simeq10^{11}M_{\odot}$). Above this value, the gas infall and condensation towards the central regions would become less efficient, while feedback caused by black-holes (BH) would suppress star-formation (e.g. \citealt{mancuso16}). 

%The large relative contribution of objects with high $M_{\rm dust}/M_{\star}$ at high-$z$'s is difficult to model. It is worth emphasizing that \cite{calura17} and  \cite{imara19} claimed difficulty in reproducing the dustiest objects under the standard IMF. Many chemical models predict that without accounting for dust growth in collisions with interstellar gas, the maximal dust-to-stellar mass ratio would be limited to values of $M_{\rm dust}/M_{\star}\leq10^{-3}$ (\citealt{dunne11}, \citealt{rowlands15}, \citealt{hirashita17}, \citealt{deVis17}). This is an order of magnitude lower than the median $M_{\rm dust}/M_{\star}$ derived for our SB DSFGs which bolsters the need of a significant dust growth in order to explain the high $M_{\rm dust}/M_{\star}$. %Even in the extreme scenario where SNe produce more dust than observed, and condense nearly all ejected metals, this would only result in $M_{\rm dust}/M_{\star}\sim10^{-2}$ (\citealt{dunne11}). 

\subsection{Modelling the observed evolution of $M_{\rm dust}/M_{\star}$}
\label{sec:4.4}

%In \hyperreff_{\rm gas}]{Fig. 4} we show how the $M_{\rm dust}/M_{\star}$ evolves as a function of redshift, stellar mass, sSFR and IR size as measured with ALMA. 

To better understand what drives the cosmic evolution of $M_{\rm dust}/M_{\star}$, we further model our data based on simplified empirical prescriptions. We follow the approach presented in seminal works of \cite{tan14} and \cite{bethermin15} by rewriting the $\delta_{\mathrm{DGR}}$ as:

%\begin{equation}
%\frac{M_{\rm dust}}{M_{\star}}\propto \frac{M_{\rm gas}}{M_{\star}} \times \delta_{\rm DGR},
%\end{equation}

% It has been shown that for massive galaxies ($M_{\star}>10^{10} M_{\odot}$), it is reasonable to assume $\log \delta_{\rm DGR}=\log (\frac{Z}{Z_{\odot}})$ (\citealt{schreiber18}, \citealt{remyruyer14}, \citealt{magdis12}, \citealt{leroy11}. In this case, the previous equation becomes:

\begin{equation}
\label{eqn:Eq.1}
\frac{M_{\rm dust}}{M_{\star}}\propto\frac{M_{\rm gas}}{M_{\star}} \times Z_{\rm gas},
\end{equation}

The equation unveils that the evolution of $M_{\rm dust}/M_{\star}$ depends on the evolution of molecular gas mass ratio and gas-phase metallicity\footnote{It has been shown that for massive galaxies ($M_{\star}>10^{10} M_{\odot}$), it is reasonable to assume $\log \delta_{\rm DGR}\propto \log (\frac{Z}{Z_{\odot}})$ (\citealt{leroy11}, \citealt{magdis12}, \citealt{remyruyer14}, \citealt{schreiber18}).}. To solve \hyperref[eqn:Eq.1]{Eq. 3} %we apply $M_{\star}$ derived from our SED fittinig procedure, while for %
we model the redshift evolution of $M_{\rm gas}/M_{\star}$ and $Z_{\rm gas}$ relying on scaling relations from the literature. In the following, we briefly describe our choice of parameters entering the right side of \hyperref[eqn:Eq.1]{Eq. 3}.

%A number of studies have explored the correlation between galactic gas mass and stellar mass or SFR (e.g., \citealt{zahid14}, \cite{sargent14}, \citealt{magdis12}, \citealt{scoville16}). 
To model the redshift evolution of $M_{\rm gas} / M_{\star}$ we apply the gas scaling relations that are based on IR/sub-mm data. Namely, we consider Eq. 9 from \cite{scoville17b}, Eq. 6 from \cite{tacconi18}, and Eq. 11 from \cite{liu19b}. The scaling relations provide empirical recipes for connecting galaxy-integrated properties ($M_{\rm gas}$, $M_{\star}$, and SFR) in the framework of the star-formation main sequence. These are mostly valid in tracing the molecular mass component. For relatively high $M_{\star}$ of our sample, this is a fair assumption if one considers that rising ISM pressure to high-$z$'s would induce negligible contribution of atomic hydrogen to the total gas mass (\citealt{combes18}, \citealt{tacconi20}). We instruct the reader to \cite{scoville17b}, \cite{tacconi18} and \cite{liu19b} for detailed descriptions and briefly outline the main points below: (1) \cite{scoville17b} derive $M_{\rm gas}$ from the optically thin RJ tail of dust emission, assuming that IR SED can be well described with the constant mass-weighted $T_{\rm dust}$, which is assumed to be 25 K; (2) \cite{tacconi18} determine $M_{\rm gas}$ combining three independent methods based on CO line fluxes, FIR SEDs, and single sub-mm flux (1 mm) photometry; (3) \cite{liu19b} provide a new functional form for $M_{\rm gas}$ by re-analysing different systematics and photometric bands' conversions for a large sample of $\sim700$ DSFGs detected with ALMA. All methods are based upon investigating statistically significant number of star-forming galaxies whose stellar masses spanning three orders of magnitude at $0<z<4$. %Their analysis was complemented with more than 1000 CO-detected sources from the literature (although vast majority of those are detected at $z<1$). 

There are different ways to parametrise $Z_{\rm gas}$ as a function of $M_{\star}$ and/or redshift and SFR, either through Fundamental Metallicity Relation (FMR; \citealt{mannucci10}, \citealt{mannucci11}, \citealt{curti19}, \citealt{chruslinska19}), or Mass-Metallicity Relation ($\rm MZR$, e.g. \citealt{kewley08}, \citealt{maiolino08}, \citealt{zahid14},\citealt{genzel15}, \citealt{hunt16}). %There are also alternative prescriptions, e.g. those suggesting the sharp break of FMR at $z>1$ (e.g. \citealt{bethermin15} and \citealt{steidel14}). 
%It is a matter of active debate whether or not these relations are valid for massive DSFGs at $z>3-4$. On one hand, optical/near-IR spectroscopy suffers from high dust attenuation, and on the other hand, the statistics of sources that have been spectroscopically studied through fine structure lines with ALMA is still limited (\citealt{boogaard19}). 

We apply three different prescriptions known in the literature: \textbf{(1)} The $\rm MZR$ from \cite{hunt16}. It is based on compiled observations of almost 1000 galaxies observed up to $z=3.7$. The $Z_{\rm gas}$ from their sample stretch over two orders of magnitude, while SFRs and stellar masses span five orders of magnitude. %The relation is given as redshift independent, so-called "Fundamental plane of metallcity".
They quantify the metalicity as: $Z_{\rm gas} = -0.14\log(\rm SFR)+0.37\log(\mathit{M}_{\star})+4.82$; \textbf{(2)} The $\rm MZR$ from \citet[their Eq.12a] {genzel15}. They analyse a large sample of galaxies with either CO line measurements or well-sampled dust SEDs. The galaxies studied by \cite{genzel15} span wide redshift range ($0<z<3$), and contain significant fraction of DSFGs. \textbf{(3)}  Broken metallicity relation (BMR) proposed by \cite{bethermin15}. The relation is in principle FMR with a correction of $0.30\times(1.7-z)$ dex at $z>1.7$. %Utilizing the RJ-tail dust continuum method (at rest-frame 850 $\mu$m), \cite{scoville17b} studied the gas-scaling relations with a large sample of 700 FIR-selected galaxies up to $z=4.5$. 

\begin{figure}[h]
	%	\vspace{-0.2cm}
	\centering
	\hspace{-1.0cm}
	\includegraphics [width=8.69cm]{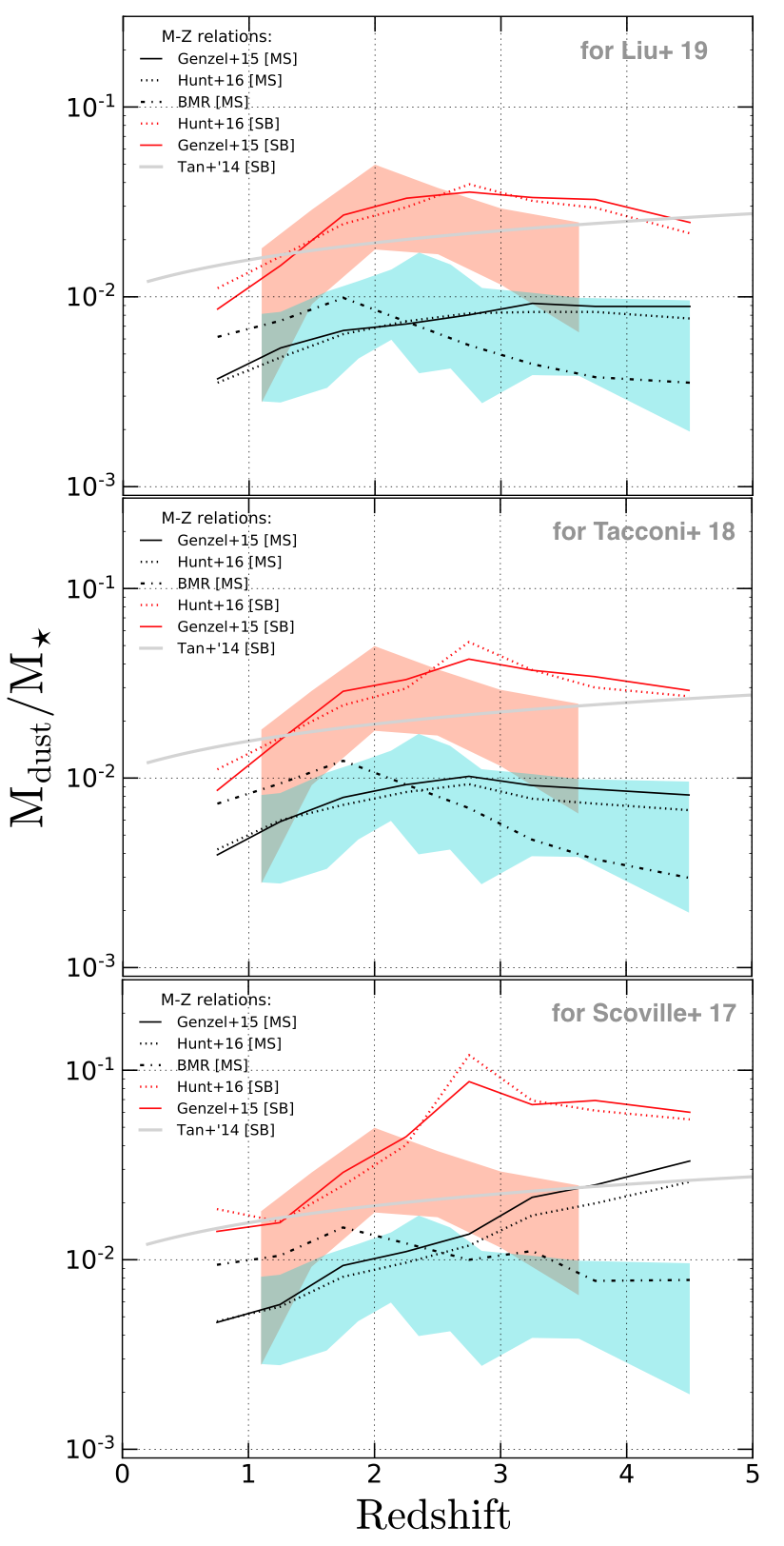}
	\caption{\texttt{From up to bottom:} The cosmic evolution of $M_{\rm dust}$/$M_{\star}$ modelled as a combination of gas mass scaling relations and $\rm MZRs$. We test the gas scaling relations from \cite{liu19b}, \cite{tacconi18} and \cite{scoville17b}. To model the evolution of gas-phase metallicity, we test different $\rm MZRs$ from \cite{hunt16}, \cite{genzel15}, along with the broken fundamental metallicity relation (BMR, \citealt{bethermin15}). The overplotted shaded regions are the same observed data (MS and SB DSFGs) presented in the upper left panel of \hyperref[fig:Fig.6]{Fig. 4}.} %We note that fundamental metallicity relation from \cite{mannucci10} overestimates observed trend by factor of 3, so we exclude it from our further consideration.}
	\label{fig:Fig.06}
\end{figure}

We now substitute different prescriptions for $Z_{\rm gas}$ and $M_{\rm gas}/M_{\star}$ into \hyperref[eqn:Eq.1]{Eq. 3}. For parameters entering $Z_{\rm gas}$ and $M_{\rm gas}/M_{\star}$, we use our SED derived  $M_{\star}, \rm SFR, \mathit{z}$, along with $\Delta_{\rm MS}$. By doing this, from \hyperref[eqn:Eq.1]{Eq. 3} we infer related cosmic evolution of the $M_{\rm dust}/M_{\star}$.  In \hyperref[fig:Fig.06]{Fig.5 } we display the modelled evolutionary tracks for MS and SB DSFGs compared to the observed data. We see that up to $z\sim2-2.5$, irrespective of their $\Delta_{\rm MS}$, the observed dust-to-stellar mass evolution can be well described by any of adopted gas scaling relations, along with the evolution of $Z_{\rm gas}$ derived from \cite{hunt16} or \cite{genzel15}. At $z>2.5$, predictions significantly differ, and we find that our data favour the best-fit function from \cite{liu19b} and \cite{tacconi18} rather than that of \cite{scoville17b} which overestimates our values both for MS and SB DSFGs. We also find larger dispersion of the residuals from the model fits in SB DSFGs than in MS DSFGs, which could imply a wider range of intrinsic physical properties (e.g.$Z_{\rm gas}$) in our starbursts. Our results for SB DSFGs broadly agree with that of \cite{tan14} who fit a compilation of individual starbursts with mildly rising trend of $M_{\rm dust}/M_{\star}$ with redshift, quantified as $M_{\rm dust}/M_{\star}\propto(1+z)^{0.51}$. While we see a broad agreement with \cite{tan14} at $z\gtrsim2$, we find that at $z\lesssim2$ the evolution of $M_{\rm dust}/M_{\star}$ in our SB DSFGs can be best modelled as $M_{\rm dust}/M_{\star}\propto(1+z)^{1.13}$, suggesting much steeper rise.
%That observed $M_{\rm dust}/M_{\star}$ in starbursts is compatible within $1\sigma$ with the trend found by \cite{tan14}, although we find much steeper rise up to $z\sim2$. 
%\footnote{Alternative, but less likely possibility, could be that observed deviance at high-$z$ is a consequence of underestimated dust masses from our side. This cannot be neglected if majority of our DSFGs are optically thick even at longest wavelengths (see e.g. \citealt{scoville15}, \citealt{spilker14}).}. 

%The other two relations provide much better fit, with the difference that  $M_{\rm dust}/M_{\star}$  starts to flatten (drops) if we assume \cite{liu19b} (\citealt{tacconi18}). 

We apply the relation from \cite{liu19b} and compute the median $M_{\rm gas}$, obtaining $M_{\rm gas}=9.1\times10^{10}M_{\odot}$ for MS DSFGs, and $M_{\rm gas}=1.1\times10^{11}M_{\odot}$ for SB DSFGs. The relation of \cite{scoville17b} tends to overpredict these values by factor of 1.5-2 relative to \cite{liu19b} and \cite{tacconi18}. Interestingly, the same difference has been reported by \cite{zavadsky20} who applied $\rm [CII]$ as a tracer of  molecular gas content in a large sample of MS galaxies with a median stellar mass of $10^{9.7}M_{\odot}$. The differences between the gas scaling relations have already been investigated in the literature (see discussions in \citealt{liu19b} and \citealt{millard20}). As mentioned in \cite{miettinen17}, a potential caveat of \cite{scoville17b} approach could be assumption of a constant $T_{\rm dust}$. Intensity of the radiation field is expected to evolve and we find that increasing the $T_{\rm dust}$ (e.g. from 25 to 45K) the inferred $M_{\rm gas}$ decreases by a factor of $\sim1.3$, which would partially explain the offset to our data. An additional reason why the $M_{\rm dust}/M_{\star}$ computed from \cite{scoville17b} best-fit overestimates our data could also be due to the way they assign stellar masses to their galaxies. The scaling relation is calibrated based on the $\Delta_{\rm MS}$ of IR-bright DSFGs for which SFRs were computed from $L_{\rm IR}$, but assigned $M_{\star}$ were derived separately from optical-NIR SEDs. This approach has a risk of underestimating $M_{\star}$ to those computed from self-consistent SED fitting from UV to sub-mm (\citealt{mitchell13}, \citealt{buat14}). On top of this, the gas scaling relations from \cite{tacconi18} and \cite{liu19b} account for metalliciity correction, which is not the case for \cite{scoville17b}. %I

%UBACI KASNIIJE: While it is generally agreed that high-redshift galaxies are gas-dominated, the exact details of the redshift evolution of $f_{\rm gas}$ are uncertain (see \citealt{popping19a}, \citealt{popping19b}, \citealt{narayanan15}). The formulation of the gas fraction can be re-expressed to the form $1/(1+(\tau_{\rm dep}\times \rm sSFR)^{-1})$, where both $\tau_{\rm dep}$ and sSFR are redshift dependent. For the rest of the paper, we adopt relation provided by \cite{liu19b} to calculate $\tau_{\rm dep}$ from estimated gas masses. In galaxy formation theory, the baryonic gas fraction is set by a balance between gas accretion and the removal of gas by star formation and galactic outflows (\citealt{moster13}). The baryonic inflow rate ensures that large gas reservoirs are built up in massive galaxies and it is expected that such high inflow rate is strongly connected to initial DM halo mass (\citealt{lapi18}).

We apply $\rm MZR$ from \cite{genzel15} and infer median $Z_{\rm gas}$ expressed as $12+\log(\rm O/H)$, obtaining $12+\log(\rm O/H)=8.64\pm0.05$ and $12+\log(\rm O/H)=8.52\pm0.09$, for MS and SB DSFGs, respectively. %Imposing the relation of \cite{hunt16} we find $12+\log(\rm O/H)=8.6\pm0.04$ and $12+\log(\rm O/H)=8.37\pm0.09$ respectively. 
These values let us characterise both MS and SB DSFGs as metal rich objects, since the estimated $Z_{\rm gas}$ are close to solar ($12+\log(\rm O/H)=8.69$, \citealt{allende01}). %and appear to saturate at a certain $M_{\rm dust}/M_{\star}$ value which is unlikely to be coincidental. 

%While Calibrations of these correlations are widely studied in the literature, however, their validity for different types of galaxies  are rarely studied. 
It is important to stress that $Z_{\rm gas}$ of high-$z$ SB DSFGs is a matter of active debate (see e.g. discussions in \citealt{tan14}, \citealt{liu19b} and \citealt{tacconi20}). %It is a matter of active debate whether or not these relations are valid for massive DSFGs at $z>3-4$. 
On one hand, optical/near-IR spectroscopy suffers from high dust attenuation, and on the other hand, the statistics of sources that have been spectroscopically studied through fine structure lines with ALMA is still limited (\citealt{boogaard19}). %Both the conversion factor of CO-to-$\rm H_{2}$ and $\delta_{\mathrm{DGR}}$ depends on $Z_{\rm gas}$ which. 
While many studies suggest that at fixed $M_{\star}$ objects with higher $\Delta_{\rm MS}$ are more gaseous and less metallic, there are recent, opposite claims suggesting super-solar metallicities that imply lower $M_{\rm gas}$, but higher SFE and higher $\delta_{\mathrm{DGR}}$ of SB DSFGs, much like in local ULIRGs (\citealt{downes98}, \citealt{magdis12}, \citealt{puglisi17}, \citealt{silverman18}, \citealt{valentino}). For example, using the prescription for $M_{\rm gas}$ given by \citealp{sargent14}, \cite{bethermin15} found that, in order to match the observed $M_{\rm dust}/M_{\star}$, their extreme starbursts require $\delta_{\mathrm{DGR}}\approx1/50$, appropriate for $Z_{\rm gas}$ twice as high as solar ($12+\log(\rm O/H)\approx9$). The fact that modelled curves for SB DSFGs at $z>2.5$ are slightly above the data also supports this hypothesis. Our data cannot fully solve this issue, and we caution the reader that our conclusions rely on the assumption that our DSFGs do not deviate strongly from adopted scaling relations. Nevertheless, even with large uncertainties in $Z_{\rm gas}$, the high $M_{\rm dust}/M_{\star}$ and its very slow decline towards high-$z$ suggest that SB DSFGs were substantially metal abundant even in the distant Universe. This strongly implies the need of rapid metal enrichment in early star-formation phase. Furthermore, our general conclusion from this modelling exercise is that high $M_{\rm dust}/M_{\star}$ are originating from massive DSFGs being metal rich. Such rapid metal enrichment at the high-$z$ would lead to the solar (or even few times solar) $Z_{\rm gas}$ of very massive, quiescent objects into which these DSFGs might evolve (Man et al., in prep.).%therefore likely to deviate from these relations

%The large relative contribution of distant objects with high $M_{\rm dust}/M_{\star}$ is hard to model. 
The chemical models that do not include grain growth in the ISM have difficulty in matching the dustiest objects under the standard IMF (\citealt{dunne11}, \citealt{rowlands15}, \citealt{deVis17}, \citealt{calura17}). For example, \cite{burgarella20} proposed the dust formation scenario assuming the high dust condensation efficiencies from stellar ejecta and non-standard "top-heavy" IMF, but found the maximum values limited to $M_{\rm dust}/M_{\star}\leq10^{-2}$. Consequently, it would be hard to fully  reproduce the significant number of observed DSFGs that populate the top-right corner of $M_{\rm dust}/M_{\star}-\rm sSFR$ plane. %On top of this, \cite{mckinon17} have found evidence that top-heavy IMF does not alleviate the tension in distant ($z>2$) DSFGs. %This is an order of magnitude lower than the median $M_{\rm dust}/M_{\star}$ derived for our SB DSFGs. 

It has been postulated that the timing when the $M_{\rm dust}$ growth in the ISM becomes effective is determined by $Z_{\rm gas}$ (see e.g. \citealt{asano13}). If $Z_{\rm gas}$ in a galaxy exceeds a certain critical value, the grain growth becomes active and the $M_{\rm dust}$ rapidly increases until metals are depleted from the ISM. This critical value of $Z_{\rm gas}$ is larger for a shorter star formation timescales, which is well supported by our data, since the typical star-formation timescale ($M_{\star}/\rm SFR$) of our DSFGs is less than 1 Gyr, with the median of $8.3\times10^{8}$yr. In addition, 47 sources ($15\%$ of the total sample) form their stellar masses at very short timescales of $\lesssim 100\:\rm Myr$. Studying the evolution of galaxies in the SAGE semi-analytical model, \cite{triani20} have found that the grain growth starts to dominate overall dust production if  $12+\log(\rm O/H)\gtrsim8.5$ and $\log (M_{\star}/M_{\odot})\gtrsim9.2$. The $84\%$ of our sources fulfil both of these criteria, which implies that the dust grain growth in ISM would be the dominant source of dust production in the vast majority of observed DSFGs. However, to better understand the evolution of $M_{\rm dust}/M_{\star}$ within the framework of dusty galaxy formation, in the next Section we inspect models, along with the state-of-the-art cosmological simulations that track dust life cycle in a self-consistent way.

%Even in the extreme scenario where SNe produce more dust than observed, and condense nearly all ejected metals, this would only result in $M_{\rm dust}/M_{\star}\sim10^{-2}$ (\citealt{dunne11}). 

%\cite{deVis17} and \cite{rowlands15} find that inclusion of a grain growth in massive galaxies, in addition to their bursty SFHs, can compensate the half of the shortfall in the predicted $M_{\rm dust}$. However, these studies failed to reproduce the high $M_{\rm dust}/M_{\star}$ and reasonable $Z_{\rm gas}$ at the same time.  suggest a fast removal (destruction + outflow) of grains 

%While it is generally agreed that high-redshift galaxies are gas-dominated, the exact details of the redshift evolution of $f_{\rm gas}$ are uncertain (see \citealt{popping19a}, \citealt{popping19b}, \citealt{narayanan15}). 

%\subsection{Comparison to the results from the literature}

 \section{Comparison to the models of dusty galaxy formation and evolution}
  \label{sec:5}
  %Dust evolution is the modification of the constitution of a grain mixture under the effect of environmental processing. Most of the models links the dust evolution to star formation, consequently describing it as three-phase process. These are: (1) Dust grain growth (2) Dust processing; and (3) Dust grain destruction. %the formation of grain cores associated to star formation; (b) stellar ejecta; (c) SN shock waves; and (d) UV and high-energy radiation.

Theoretical works that aim at investigating the evolution of dust content in galaxies can broadly be separated into analytic and semi-analytical solutions (e.g. \citealt{lacey11}, \citealt{gioannini17}, \cite{popping17}, \citealt{imara19}, \citealt{pantoni19}, \citealt{triani20}) and hydrodynamical simulations (\citealt{mckinon17}, \citealt{aoyama18}, \citealt{hou19}, \citealt{vijayan19}, \citealt{simba}). On top of this, there are also phenomenological models (e.g. \citealt{cai13}, \citealt{s16}, \citealt{bethermin17} ). The latter group of models are not \textit{"ab initio"}, but they could be very useful for complementing our knowledge about specific galaxy population. 

\subsection{Models}
	\label{sec:5.1}

In this work we consider all three classes of models outlined above. Namely, we analyse the predictions from the \textbf{(I)} Cosmological galaxy formation simulation with self-consistent dust growth and feedback (SIMBA, \citealt{simba}); \textbf{(II)} Analytical model of \cite{pantoni19}; and the \textbf{(III)} Phenomenological model based on multi-band surveys (\citealt{bethermin17}).

\subsubsection{SIMBA cosmological simulation (\citealt{simba})}

The cosmological galaxy formation simulation SIMBA utilizes mesh-free finite mass hydrodynamics (\citealt{hopkins15}, \citealt{mufasa}). We refer the reader to \cite{simba} for extensive description of the simulation, and here we summarize the most important points. The primary SIMBA simulation has $1024^{3}$  dark matter particles and $1024^{3}$ gas elements in a cube of $100\:{\rm Mpc} h^{-1}$ side length. The simulation preserves the mass within each fluid element during the evolution, thereby enabling detailed tracking of gas flows. Star formation is modelled using a molecular $\rm H_{2}$ gas relation from \citealt{schmidt59}, with the abundance of $\rm H_{2}$  computed from sub-grid prescription that connects $Z_{\rm gas}$ and gas column density in the local Universe (\citealt{krumholz11}). SIMBA applies fully physically-motivated black hole growth following the work of \cite{alcasar17}. The novel sub-grid prescriptions for AGN feedback and X-ray feedback are also included. The implementation of dust life cycle is introduced in \cite{li19}. It is broadly based on the seminal work by \cite{dwek98} and its updated version by \cite{popping17} and \cite{mckinon17}. 

The net dust production/destruction rate in SIMBA can be generalised as:
\begin{equation}
\label{eqn:4}
\Sigma \dot{M}_{\rm dust}\propto \dot{M}_{\rm dust}^{\rm SNe}+\dot{M}_{\rm dust}^{\rm ISM}
- \dot{M}_{\rm dust}^{\rm destr}- \dot{M}_{\rm dust}^{\rm SF}+ \dot{M}_{\rm dust}^{\rm inf}- \dot{M}_{\rm dust}^{\rm out}
\end{equation}

The first term in the right side of \hyperref[eqn:4]{Eq. 4} describes the dust produced by condensation of a fraction of metals from the ejecta of $\rm SNe$ and asymptotic giant branch $\rm AGB$) stars; the second term describes the dust by accretion in the ISM; the third term describes the dust destructed by $\rm SNe$ shock waves; the fourth term is the destruction of dust by astration and stellar feedback; the fifth term is an additional dust production by gas infall; the sixth term describes the expelled dust mass, due to $\rm SNe$ and AGN. The latter two mechanisms are responsible for heating up/removal of gas from the ISM into the DM halo (or even further out).

The full treatment of dust is explained in details by Eqs. (11)-(31) in \cite{simba} and Eqs. (1)-(11) in \cite{li19}. In general, the dust model makes the explicit assumption that dust can grow only in the dense regions of the ISM. The production of dust by condensation of metals from $\rm SNe$ and AGB ejecta is estimated by Eq. (4)–(7) in \cite{popping17}. The dust model within SIMBA does not include contribution from $\rm Ia \:SNe$ which is opposite to some models that proposed the same condensation efficiency between Type $\rm Ia\:SNe$ and Type II SNe (\citealt{dwek98}, \citealt{mckinon17}, \citealt{popping17}). %SIMBA follows suggestion by more recent theoretical works (e.g. \citealt{gioannini17}) which disfavour such a high contribution from $\rm Ia\:SNe$.

Overall, SIMBA accounts for dust produced from ageing, stellar populations, grain growth, destruction in SN shocks, and the advection and transport of dust in galactic winds. Dust is injected into the ISM as stars evolve off the MS, with $M_{\rm dust}$ calculated using stellar nucleosynthetic yields and grain condensation efficiencies. The timescale for grain growth through collisions depends on local gas density and temperature, while the timescale for dust destruction through SN sputtering scales inversely with the local SNe rate. The dust grains are assumed to all have the same radius and density ($a=0.1\:\mu$m and $\sigma=2.4\:g/cm^{3}$, respectively (see e.g. \citealt{draine14}). The condensation efficiencies for AGB and core-collapse supernovae ($\rm CC\:SNe$) are constant (0.2 and 0.15, respectively). These values are tuned in order to match the observed $\delta_{\mathrm{GDR}}-Z_{\rm gas}$ relation by \cite{remyruyer14}. 

\subsubsection{\cite{pantoni19} model}

%The model of \cite{pantoni19} (hereafter P19) is a new set of analytic solutions aimed at self-consistently describing the spatially-averaged time evolution of the gas, stellar, metal, and dust content in individual DSFGs hosted within a DM halo of a given mass and formation redshift. The basic framework of the model is described in \citealt{pantoni19}. The model presumes a galaxy as an open (one-zone) system comprising three interlinked mass components: a reservoir of warm gas, molecular gas (fed by gas infall and depleted by star formation and stellar feedback), and stellar mass (partially restored to the cold phase by stars during their evolution). The model is built under assumption that DSFGs at high-$z$ are progenitors of local elliptical galaxies. The P19 model relies on \textit{"in-situ"} co-evolution scenario for star formation and BH accretion meaning that the star formation in DSFGs at high-$z$ is regulated by internal processes, while marginally influenced by galaxy merging (see e.g. \citealt{moster13}, \citealt{lapi18}). The P19 assumes a spatially-averaged star formation law. The corresponding analytic solutions for the metal enrichment and stellar mass are self-consistently derived in P19, using as input the solutions for the evolution of the mass components. The evolution of $M_{\rm dust}$ takes into consideration all physical processes contained in \hyperref[eqn:4]{Eq.4}. For exact details about the gas and dust treatment, see equations (33)-(41) in \cite{pantoni19}.

\citet[hereafter P19] {pantoni19} has presented a new set of analytic solutions that self-consistently describes the spatially-averaged time evolution of gas, stellar, metal, and dust content in individual galaxies hosted within a DM halo of a given  mass and formation redshift. In particular, the solutions have been applied to the description of high-$z$ DSFGs as the progenitors of local ellipticals. The basic framework is described in P19. It presumes the galaxy as an open (one-zone) system comprising three inter-linked mass components: a reservoir of infalling  gas (subject to cooling and condensation), cold star-forming gas (fed by gas infall and depleted by star formation and feedback), and stellar mass (partially restored to the cold phase by stars during their evolution). The corresponding metal and dust enrichment history of the cold gas is self-consistently computed using as input the solutions for the evolution of the mass components. The evolution of $M_{\rm dust}$ takes into consideration all the relevant physical processes contained in \hyperref[eqn:4]{Eq. 4}. For exact details about the gas metallicity and dust treatment, see Eqs.(9)-(14) and (33)-(39), respectively, in \cite{pantoni19}. The main parameters entering the solutions have been set by relying on an\textit{"in-situ"} evolution framework, implying that the star formation in DSFGs at high-$z$ is mainly regulated by internal processes (e.g., \citealt{moster13}, \citealt{lapi18}). Coupling the outcome for individual galaxies with merger rates based on the state-of-the-art numerical simulations, the P19 model show success in reproducing the main statistical relationships followed by high-$z$ DSFGs (e.g., galaxy MS, $M_{\rm gas}$, $M_{\rm dust}$ etc.) and by their local descendants (e.g., mass-metallicity relation, alpha-enhancement, etc.).

\subsubsection{\cite{bethermin17} model}
\label{sec:5.1.2}
%\vspace{-0.23cm}
The phenomenological, model of \citet[hereafter B17] {bethermin17} relies on the combination of observed dust SED templates of galaxies and IR luminosity functions. %This method is advantageous over other similar methods since it is initiated in observational data, but the main uncertainty is whether calibrated "aver" SED templates can be applicable to variety of galaxies at the same luminosity at higher redshifts. 
The B17 is built on IR/sub-mm data and it is one of few models that are able to simultaneously match the total IR number counts and the evolution of sSFR.  The model applies the abundance matching procedure to populate the DM halos of a light cone constructed from the Bolshoi-Planck simulation (\citealt{darkmatter16}. The halo catalogues are matched to the observed galaxy stellar mass function (SMF) described by a double Schechter function (\citealt{davidzon17}). Physical properties ($\rm SFR, M_{\star}$) are assigned to each object based on the dichotomy model which decomposes bolometric IR-luminosity function with MS and SB dusty galaxies. B17 assume that the scatter on the MS is constant with $M_{\star}$ and redshift. Shape of the SEDs is controlled by the galaxy type (MS or SB) and with the mean intensity of the radiation field $\langle{U}\rangle$, which couples with the $T_{\rm dust}$. Contribution of AGNs and strong lensing are also included following the recipe presented in \cite{bethermin12}.

%\\
%%%%%%%%%%%%%%%%%% 	

\subsection{Confronting observed results to models}
\label{sec:5.2}
%%%%%%%%%%%%%%%%%%
\begin{figure*}[h]
	\vspace{-0.2cm}
	\centering
	%\hspace{-1.0cm}
	\includegraphics [width=17.69cm]{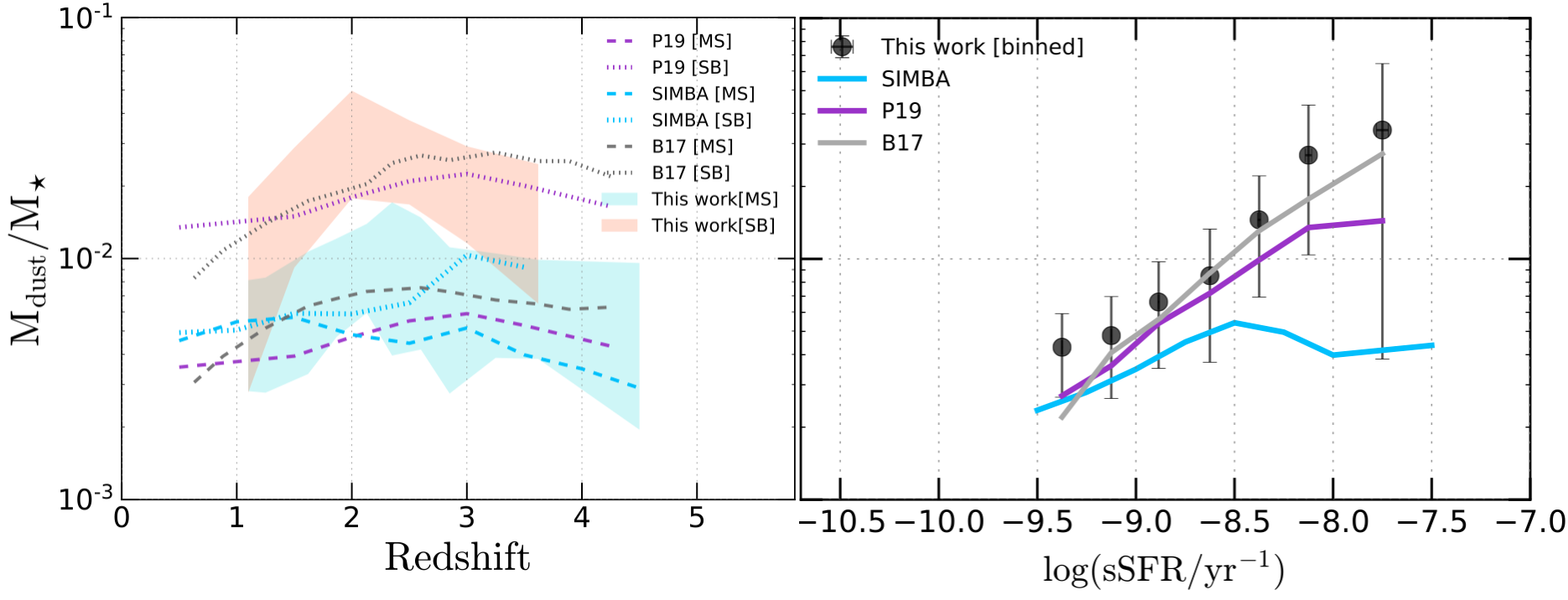}
	\caption{\texttt{Left:}The redshift evolution of $M_{\rm dust}/M_{\star}$ predicted by different models (see \hyperref[sec:5.1]{Section 5.1}). The observed data are overplotted with same colours as in previous figures. The grey, blue and purple dashed and dotted lines represent predictions  for galaxies selected as MS and SB DSFGs from \citealt{bethermin17}, \citealt{simba} and \citealt{pantoni19}, respectively. %For the analytic model of \cite{pantoni19}, we show the punctual predictions for the typical highly star-forming MS galaxy with $M_{\rm DM}\approx10^{12}\:M_{\odot}$ (orchid line). The dashed and dotted blue lines show the predictions from SIMBA (\citealt{simba}). In order to properly exemplify models, we consider the same modelled stellar mass and sSFR range as we observed for our DSFGs ($\log M_{\star}/M_{\odot}>10^{9.8}$ and $\log(\rm sSFR/yr^{-1})>-9.5)$. 
     \texttt{Right:} The evolution of $M_{\rm dust}/M_{\star}$ as a function of $\rm sSFR$ for the full sample of observed and modelled galaxies. The binned values from this work are shown with black circles with corresponding $1\sigma$ vertical error bars. The model predictions are denoted with same colours as in the left panel.} %A comparison between the estimated dust-to-stellar evolution, and model predictions. In the lower panel, we show the same observed trend against the predictions from the phenomenological model of \cite{bethermin17}}
	\label{fig:Fig.7}
\end{figure*}
%%%%%%%%%%%%%%%%%%%

We now confront our observational findings to the models described above. In order to achieve this goal, we analyse the simulated catalogues. To ensure consistency between observed and simulated data, we impose the same range of modelled $M_{\star}$ and $\rm sSFR$ as in our observations ($\log (M_{\star}/M_{\odot})>10$ and $\log(\rm sSFR/\rm yr^{-1})>-9.5$). We note that same IMF (\citealt{chabrier03}) is adopted both in observed and simulated data. We separate modelled galaxies into MS and SB DSFGs following the exact same method we apply over our real data. %To fully evaluate the data to models, it would be ideal to simulate the full extraction process and SED modelling as for the real data (see \citealt{bethermin17}, \citealt{ddrisers}). Since the sources followed-up with ALMA are selected in different ways in the literature, is not possible to produce the fully realistic "end-to-end" selection. Nonetheless, by selecting the same range of $M_{\star}$ and $\rm sSFR$, we ensure consistency between modelled and observed data.

We illustrate our findings in \hyperref[fig:Fig.7]{Fig. 6}, where we show how the $M_{\rm dust}/M_{\star}$ changes as a function of redshift (left panel) and sSFR (right panel). Considering the B17 model, we see that the model predictions are in a good agreement with our data both for MS and SB DSFGs. Despite the fact that B17 is based on averaged observed statistical properties of galaxies and very simplified physical prescriptions, it is successful in matching both the observed evolution of $M_{\rm dust}/M_{\star}$ versus $z$ and $\rm sSFR$ within the $1\sigma$ uncertainties. From the left panel we see that the $M_{\rm dust}/M_{\star}$ modelled for SB DSFGs has a small positive offset of 0.05 dex to our data at the highest redshift bins ($z>3$). This indicates that due to our selection criteria we likely miss some rare and prodigious  starbursts at high-$z$. These sources are usually barely detected  even in very deep NIR data, and their existence at $3<z\leq5$ is confirmed by recent blind ALMA surveys (\citealt{williams18}, \citealt{franco18}, \citealt{wang19}). 
%\footnote{At higher redshifts, and in more massive galaxies, the ISM is on average denser and the condensation/star formation timescales are shorter. Thus the star formation in a galaxy of given stellar mass is higher, causing the main sequence locus to shift upwards. We stress that, moving toward higher redshift, the fraction of starbursting objects decreases appreciably. This is because, given the evolution of the SFR function and the shorter age of the universe, it becomes more difficult to spot galaxies of appreciably different ages and featuring very high SFRs.} 
It is worth noting that in B17 the mean interstellar radiation field $\langle U\rangle$ steadily evolves in MS DSFGs, but is tuned to be constant in SB DSFGs over $0<z<3$. This implies the existence of starbursts that could be slightly colder than MS DSFGs at the same redshift, having very high dust masses ($M_{\rm dust}\gtrsim10^{9}M_{\odot}$). The latter could be an additional reason why at $z\sim3$ the B17 predicts slightly higher $M_{\rm dust}/M_{\star}$ than in our observations. Nonetheless, the agreement we see between our data and B17 model, entails that our empirical knowledge of how the $M_{\rm dust}/M_{\star}$ evolves within the MS paradigm, is moving towards the comprehensible picture. Such conclusion is strongly supported with the observations of the cosmic evolution of $M_{\rm gas}$ and $\rm sSFR$ (\citealt{liu19b}). 

%In that case, we significantly reduce uncertainties of our overdensity estimations.% who adopt the stellar mass function and luminosity evolution modelled in B17, and successfully reproduce the observed cosmic evolution of $M_{\rm gas}$ in a large statistical sample of galaxies up to $z\sim6$.

The theoretical predictions of P19 are also broadly consistent with the observed evolution of $M_{\rm dust}/M_{\star}$ with redshift and $\rm sSFR$, and the overall agreement is valid both for MS and SB DSFGs. One of the major forecasts of P19 is very rapid evolution of $Z_{\rm gas}$, which attains high values in a quite short timescale ($\lesssim10^{8}\rm yr$) while being mainly related to "\textit{in-situ}" processes. Such a rapid evolution becomes particularly important for reproducing the $Z_{\rm gas}$ in $z>3$ DSFGs. The P19 predicts that $Z_{\rm gas}$ in massive  ($\sim10^{12}M_{\odot}$) DM halos saturates close to slightly super-solar values for the case of standard \citealp{chabrier03} IMF. This is important finding, since many chemical and semi-analytical models propose the use of "top-heavy" IMF as the only solution for assuring very high $M_{\rm dust}$ and rapid metal enrichment in massive galaxies (e.g. \citealt{lacey16}, \citealt{calura17}). The P19 model also predicts that $M_{\rm gas}$ increases monotonically up to $M_{\star}\sim10^{11}M_{\odot}$, above which the gas infall and condensation become less efficient causing the subsequent decline in $M_{\rm gas}$. This can be a cause of a rapid downfall of $M_{\rm dust}/M_{\star}$ towards the lower sSFR. The good agreement with P19 model provides a strong support to the scenario where significant dust growth in the metal-rich ISM is needed to explain the high $M_{\rm dust}/M_{\star}$. 

%UBACI KASNIJE: For example, \cite{lacey16} predict the gas metallicities of MS galaxies to be supersolar on average, resulting in higher dust masses. Even such model can explain slightly higher dust masses we estimate, there are no clear evidences for top-heavy IMF assumed in this model, so we did not include it in our analysis in the present paper. 

%We now turn our attention to the hydrodynamical cosmological simulation SIMBA. We find that SIMBA is successful in reproducing DSFGs with $M_{\rm dust}/M_{\star}>0.01$, which remains extremely challenging task for most of actual cosmological simulations (see e.g. discussions in \citealt{mckinon17} and \citealt{graziani19}). However, from the left panel of \hyperref[fig:Fig.7]{Fig.7} we see that there are still existing tensions relative to data, such that the observed evolution of $M_{\rm dust}/M_{\star}$ is underestimated by factor of 2-5, depending on the redshift range and modelled galaxy's distance with respect to the chosen main-sequence.

Compared to the observations of MS DSFGs, the cosmological simulation SIMBA reproduces $M_{\rm dust}/M_{\star}$ well up to $z=1.5$, while at $z>1.5$ the modelled values are lower but still compatible with the data within $1\sigma$. The modelled $M_{\rm dust}/M_{\star}$ remains as a weakly decreasing function of $z$, pronounced with an overall change of amplitude by roughly 0.25 dex. This is another success of SIMBA and indicates that the simulation is able capturing the massive dust production ($M_{\rm dust}>10^{9}M_{\odot}$) towards earlier cosmic times, witch is hardly reproduced by most of cosmological simulations (see e.g. discussions in \citealt{mckinon17} and \citealt{graziani19}). SIMBA is less successful in reproducing the observed $M_{\rm dust}/M_{\star}$ in SB DSFGs and underpredicts this quantity by factor of 3-6 depending on the redshift. The discrepancy between the observed and modelled $M_{\rm dust}/M_{\star}$ towards the higher $\Delta_{\rm MS}$ is well illustrated in the right panel of \hyperref[fig:Fig.7]{Fig. 6}. 

We see that at $\log(\rm sSFR/\rm yr^{-1})\gtrsim-8.5$ the SIMBA predicts much flatter trend with $\rm sSFR$ relative to data, which implies the deficiency of simulated objects with $M_{\rm dust}/M_{\star}\gtrsim10^{-2}$. We find that the relative contribution of sources  fulfilling the criterion $M_{\rm dust}/M_{\star}>10^{-2}$ is $20\%$ in B17 and only $2\%$ in SIMBA. The underestimation of modelled DSFGs with the highest dust masses ($M_{\rm dust}>10^{9}M_{\odot}$) has already been discussed by \cite{li19}. They compared simulated dust mass functions from SIMBA and the observed ones at $0<z<2$, inferring $\sim2-4$ underestimation of model to data.  We note that if $M_{\rm dust}$ and $z_{\rm phot}$ are derived from FIR data only, they could suffer from large uncertainties, and the high-$z$ DMFs are very uncertain constraint on cosmological models. On the contrary, due to the wealth of multiwavelength data coverage and de-blended IR photometry complemented with ALMA observations, estimated $M_{\rm dust}$ and $M_{\star}$ have significantly smaller uncertainties. Therefore, it seems unlikely that our technique led to significant underestimation (overestimation) of derived $M_{\star}$ ($M_{\rm dust}$). If the latter is true, this could indicate that some model ingredients in SIMBA need to be refined (such as amount of molecular gas relative to stars, or dust destruction mechanisms in the sub-grid model).

We note that our sample is incomplete at fixed $M_{\star}$, since most of sources were preselected for the purpose of ALMA follow-ups. This would cause difference in galaxy SEDs and starburst fractions as compared to complete samples within the same range of $M_{\star}$. We instruct the reader to \cite{liu19a} for detailed discussion of ALMA selection biases. Since ALMA Band 6/7 is sensitive to the galaxies with the $T_{\rm dust}$ colder than that of \textit{Herschel} at a fixed $L_{\rm IR}$, %Other sources ended up within the observed ALMA FoV without any preselection. While for the latter we may be biased towards colder sources, this is less true for the former, which may be biased towards different selections. To better quantify the 
we further approximate what would be the $M_{\rm dust}/M_{\star}$ of a mass complete sample of modelled DSFGs. We use the full catalogue based on B17 model which is perfectly suitable for our goal since it is based on $2\:\mathrm{deg}^{2}$ simulation. We relaxed selection criteria in order to inspect the average $M_{\rm dust}/M_{\star}$ of all unlensed sources with $M_{\star}>10^{10}M_{\odot}$ below the detection limit. %By using phenomenological simulations over analysing the real data, our conclusions which come from incomplete sample can be complemented with other modelled sources beyond the completeness limit. We obtained that our ALMA detections catch around $\sim15\%$ of all sources at $M_{\star}>10^{10}M_{\odot}$. 
The "missed" DSFGs peak at $z\approx3$ and are warmer than ALMA selected sample due to higher average $\langle U\rangle$ (thus $T_{\rm dust}$). However, inclusion of these sources does not significantly impact our results since the median $M_{\rm dust}/M_{\star}$ of "missed" objects is find to be 0.004 for MS and 0.008 for SB DSFGs. %Because our sample accounts for some of the brightest DSFGs at the particular redshift, our sample is clearly biased towards very dust-rich systems. 
%The lower than observed $M_{\rm dust}/M_{\star}$  in SB DSFGs could indicate that some model ingredients in SIMBA need to be refined. Thus, in order to properly reproduce the upper locus of $M_{\rm dust}/M_{\star}$-sSFR plane, the models would require either a higher $f_{\rm gas}$, more rapid metal enrichment, or refinement of dust destruction mechanisms in the sub-grid model. %The careful treatment of all these factor is out of the scope of this paper, but it will be discussed in our accompanying paper (Donevski et al., in prep). In the following we briefly discuss physical quantities underlying beneath the observed tension. %} 

\subsection{What lies behind the tension between cosmological simulations and observations?}
\label{sec:5.3}
In \hyperref[sec:4.4]{Section 4.4} we give a sense of how the $M_{\rm dust}/M_{\star}$ is influenced by different evolutions of molecular gas mass ratio and $Z_{\rm gas}$.
We now turn our attention in investigating the trends with latter two quantities in models. The P19 model predicts the average $Z_{\rm gas}$ that is $\rm 0.1\:dex$ and $\rm 0.02$ dex above the solar value for MS and SB DSFGs, respectively. The median $Z_{\rm gas}$ in MS (SB DSFGs) modelled in SIMBA are $\rm 0.24\:dex$ ($\rm 0.22\:dex$) below the solar. The values are reasonable high for both galaxy populations, but still lower by a factor of $\sim1.5$ to the medians derived from our data (see \hyperref[sec:4.4]{Section 4.4}). This implies that lower $M_{\rm dust}/M_{\star}$ predicted by SIMBA would be partly resulting from lower modelled $Z_{\rm gas}$. %Comparing their observations to then available semi-analytical models, \cite{bethermin15} found much larger discrepancy ($\sim0.4-0.5 \rm dex)$ between the observed and modelled $Z_{\rm gas}$ in starbursts.  %To a certain extent, this may be a source of observed deficit of modelled galaxies with $M_{\rm dust}/M_{\star}\geq0.01$. 

\begin{figure}[h]
	\vspace{-0.2cm}
	\centering
	%	\hspace{-1.0cm}
	%	\includegraphics [width=13.89cm]{dsfg-tracks.pdf}
	
	\includegraphics [width=9.33cm]{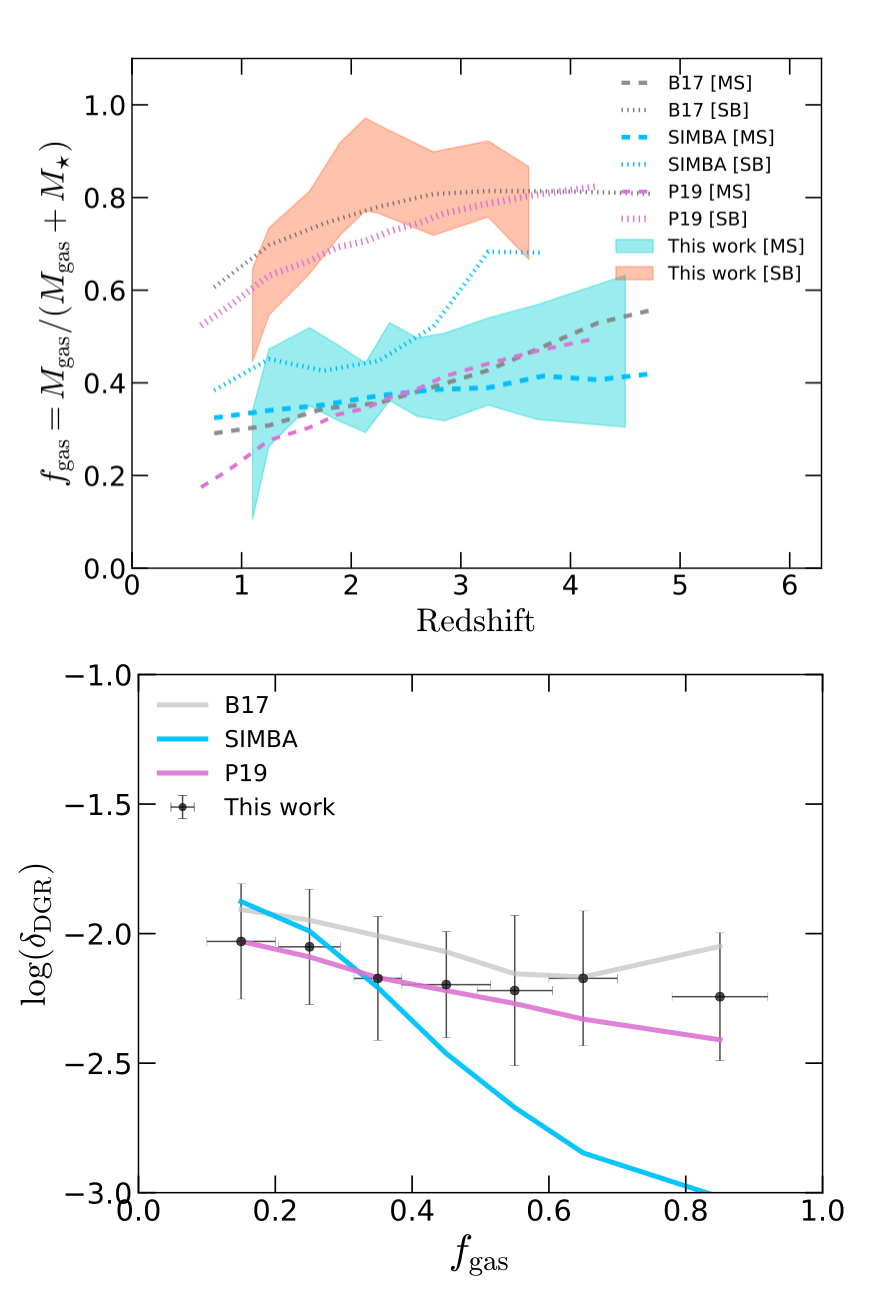}
	\caption{\texttt{Upper panel:} The cosmic evolution of molecular gas fraction ($f_{\rm gas}$) in our DSFGs, estimated from the functional form of \cite{liu19b} and illustrated with shaded areas. The model predictions for MS and SB DSFGs are overplotted with dashed and dotted lines, respectively. Purple, blue and grey lines correspond to the P19, SIMBA and B17 respectively. For P19 and SIMBA, $f_{\rm gas}$ is derived self-consistently, while  for B17 we test the same scaling relation we apply in our observations. .\texttt{Lower panel:} The $\delta_{\rm DGR}$ as a function of $f_{\rm gas}$ for the full sample of observed and modelled galaxies. Observed mean ratios with corresponding $1\sigma$ uncertainty are presented with black circles. The meaning of colours that correspond to modelled values is the same as in the upper panel.}
	\label{fig:Fig.17}
\end{figure}

%We now analyse how the stellar mass, dust mass and molecular gas mass relate in models. 
In order to unveil how the dust, stellar and molecular gas budget are interlinked in models, we further analyse the simulated $\delta_{\mathrm{DGR}}$ and molecular gas fraction, defined as $f_{\rm gas}$=$M_{\rm gas}/(M_{\rm gas}+M_{\star})$.  In the upper panel of \hyperref[fig:Fig.17]{Fig. 7} we show the $f_{\rm gas}$ as a function of redshift. For our DSFGs we apply the gas scaling relation of \cite {liu19b}, inferring the median value of $f_{\rm gas}=0.51\pm 0.12$ for the full sample, and $f_{\rm gas}=0.44\pm 0.09$ and $f_{\rm gas}=0.75\pm 0.09$ for MS and SB DSFGs, respectively. The observed $f_{\rm gas}$ in SB DSFGs is $\sim2-3$ times higher than in MS DSFGs, independently of the compared redshift range. Such a clear distinction of $f_{\rm gas}$ in MS and SB DSFGs is in agreement with other studies from the literature (\citealt{magdis12}, \citealt{santini14}, \citealt{bethermin15},  \citealt{scoville16}, \citealt{saintonge17}, \citealt{liu19b}, \citealt{simpson20}). 

We find that modelled $f_{\rm gas}$ are consistent to our estimates for MS DSFGs, with the difference that SIMBA predicts a slower rise of $f_{\rm gas}$ with redshift, as compared to P19 and B17. The models have different success reproducing $f_{\rm gas}$ of SB DSFGs. The B17 and P19 predict continuous increase of $f_{\rm gas}$, which broadly agrees with our estimates.  Considering SIMBA we see that larger $\Delta_{\rm MS}$ is accompanied by only moderate increase of galaxy molecular gas fraction up to $z\simeq2.5$ and riches values close to ours only at $z\geq2.5$ where the statistics of modelled starbursts is low. The corresponding median $f_{\rm gas}$ in SB DSFGs modelled by SIMBA is of $f_{\rm gas}=0.45$, which is half as much as what our data suggest. %All in all, over the same redshift range, both our data as well as B17, favour the scenario where $f_{\rm gas}$ is in average 1.5 times higher as compared to SIMBA predictions. 

We further investigate how the $\delta_{\rm DGR}$ scales with $f_{\rm gas}$. For the full sample of observed DSFGs, we determine the median value of $\delta_{\rm DGR}=1/148$. For MS and SB DSFGs separately, the medians are $\delta_{\rm DGR}=1/159$ and $\delta_{\rm DGR}=1/139$, respectively. Although derived values are strongly model dependent, they indicate that MS and SB DSFGs exhibit slightly different average $\delta_{\rm DGR}$ (but, see our discussion in \hyperref[sec:4.4]{Section 4.4}). The median $\delta_{\rm DGR}$ in both MS and SB DSFGs agree very well with the calibration by \cite{schreiber18}, but are sightly lower than canonical value obtained for local ULIRGs ($\delta_{\rm DGR}=1/100$, \citealt{leroy11}). All these should be borne in mind when applying $\delta_{\mathrm{DGR}}$ for $M_{\rm gas}$ estimation for high-$z$ DSFGs. 

From the bottom panel of \hyperref[fig:Fig.17]{Fig. 7} we see that observed $\delta_{\rm DGR}$ mildly decreases with increasing $f_{\rm gas}$ and flattens at $f_{\rm gas}\gtrsim0.5$. The range of $\delta_{\rm DGR}$ predicted by B17 and P19 is within $1\sigma$ uncertainty with our data, while predictions from SIMBA are consistent at $f_{\rm gas}<0.5$, but significantly differ at $f_{\rm gas}>0.5$ due to steeper decrease compared to our data. % In the bottom panel of \hyperref[fig:Fig.17]{Fig.8}  we plot the binned average of our data and find that $\delta_{\rm DGR}$ mildly declines as a function of molecular gas fraction, in line with predictions given by B17 and P19. %The median values for MS and SB DSFGs predicted by SIMBA are $\delta_{\rm GDR}=192$ and $\delta_{\rm GDR}=168$, which is again very close to our estimates. Nevertheless, 
Such a sharp reduction in $\delta_{\rm DGR}$ %may be one of the main reasons why cosmological simulations predict relatively low number of DSFGs with $M_{\rm dust}/M_{\star}\gtrsim0.01$.
translates to relatively low number of DSFGs with $M_{\rm dust}/M_{\star}\gtrsim0.01$ produced in cosmological simulations. By investigating the connection between the $\delta_{\rm DGR}$ and $Z_{\rm gas}$ in SIMBA, we find that very low $\delta_{\rm DGR}$ in galaxies from the highest-end of gas fraction is a result of their gas metallicities being lower by a factor of $\sim4$ relative to solar.

% Mean value of $\delta_{\rm GDR}=250$ is very close to the one we estimate for our data $\delta_{\rm GDR}=220$. However, the potential problem is average gas fraction, that is one that we infer by applying $\rm MZR$s over empirically scaled gas fractions, as presented in previous Chapter. Many SAMs (e.g. \citealt{lacey16}) or cosmological simulations (e.g. \citealt{mckinon17}) were unable to catch the brightest dust mass even after including top heavy IMF. As pointed by \cite{lacey16}, top heavy IMF could produce a lot of metals very quickly, without showing a strong increase of a total stellar content due to mass losses. 

The comprehensive treatment of different physical mechanisms that could be responsible for relative shortfall of model predictions, along with numerical limitations, is out of the scope of this paper. In the following we briefly emphasize their importance. %worth discussing in sec 5.3 is numerical limitations.  Owing to ~kpc resolution, I suspect Simba cannot reproduce the very high gas surface densities seen in SBs.  @Desika, if we could resolve this, would this create a net increase in the amount of dust?  Intuitively, it should since there would be more shielding thus less destruction and also the dust production rate I would think would be higher in very dense gas.  It's not obvious to pick out this dependence from all the equations in Simba, but I would think that's how it would work.  This may be another reason why we cannot reproduce the dustiest galaxies.

\begin{itemize}
	\item \textbf{The dust growth timescale is too long}
	
One of the most critical parameters that describes the dust mass growth is the accretion timescale ($\tau_{\rm acc}$). It is often modelled as:
	\begin{equation}
	\label{eqn:5}
	\tau_{\rm acc} = \tau_{\rm acc,0} \times a^{-1} \times n_{\rm H}^{-1}\times T_{\rm gas}^{-1/2} \times Z_{\rm gas}^{-1}
	%f(a, n_{H}, T_{\rm gas}, Z_{\rm gas})\propto \tau_{\rm acc,0} \times Z_{\rm gas}^{-1}
	\end{equation}
	where $a$ is a dust grain size which is usually assumed to be spherical with a typical size of $\sim0.1\mu$m (\citealt{asano13}). The $n_{\rm H}$ and $T_{\rm gas}$ are the number density and temperature of the cold gas phase, and $\tau_{\rm acc,0}$ defines the timing of growth activation. %\footnote{In the literature, there is an increasing number of recent studies attempting to model the dust evolution in early DSFGs (\citealt{triani20}, \citealt{vijayan19}, \citealt{aoyama19}, \citealt{hou19}, \citealt{dwek19}, \citealt{graziani19}, \citealt{hirashita17}, \citealt{mckinon17}, \citealt{popping17}). %Despite the fact that overall recipes are very similar, most of the disagreement comes from the treatment of $\tau_{\rm acc}$ between the models.}. For typical values of $n_{H}=10^{3}\rm cm^{-3}$ and $T_{\rm gas}\approx40-50$K, the different models adopt different $\tau_{\rm acc,0}$ $~1-20$ Myr (see \citealt{li19}, \citealt{graziani19}, \citealt{aoyama19}). %
	 The timescale for dust growth in the ISM changes as a function of gas surface density for different $Z_{\rm gas}$ (see \citealt{popping17}, their Fig.1). If we apply $M_{\rm gas}$ of our DSFGs, along with their compact ALMA continuum sizes (see the next Section), we infer high median molecular gas surface density of $\sim6.7\times 10^{3}M_{\odot}\rm pc^{-2}$. Such a high surface density implies short accretion timescales, on average $\tau_{\rm acc}\sim6\times10^{5}$yr. These will be obtained if $\tau_{\rm acc,0}<10^{6}\rm yr$ which is shorter than what has usually been adopted in cosmological simulations\footnote{For typical values of $n_{H}=10^{3}\rm cm^{-3}$ and $T_{\rm gas}\approx40-50$K, the $\tau_{\rm acc,0}$ adopted by models is usually $~1-20$ Myr (see \citealt{li19}, \citealt{graziani19}, \citealt{aoyama19}).}. %Interestingly, the average value we obtain from our data is very close to the fiducial value adopted in \cite{pantoni19}. 
	The short $\tau_{\rm acc}$ are proposed by \cite{pantoni19}, and are also reproduced in very recent semi-analytical models that claim fairly good overall match to the observations (\citealt{triani20}). These are also in line with \cite{deVis17}, who found that variations in dust growth timescales might help to explain the $M_{\rm dust}$ deficit at high gas-fractions in their large sample of nearby galaxies. Given the SIMBA’s $\sim\rm kpc$ resolution, a multiphase galaxy ISM cannot be resolved, which prevent us from knowing the exact dependence of the modelled dust content to the gas surface density. As a result, parameters such as the reference $\tau_{\rm acc}$ are tuned in order to boost the effective gas density. Thereby, it seems likely that use of shorter accretion timescales (equivalent of increase in dust growth efficiency) in simulations could help partially overcome the shortfall to data.\\

	\item \textbf{Dust destruction is too efficient}
	
It is also possible that the dust destruction in simulations is too efficient. The total rate of dust mass destruction is given by $\dot{M}_{\rm dest}\propto M_{\rm dust}/\tau_{\rm destr}$, where dust destruction timescale is usually approximated as (\citealt{slavin15}):

\begin{equation}
\tau_{\rm destr} = \frac{\Sigma M_{\rm gas}}{f_{\rm ISM}R_{\rm SN}M_{\rm cl}} = \frac{\tau_{\rm SN}M_{\rm gas}}{M_{\rm cl}}
\end{equation}

Here $\Sigma M_{\rm gas}$ is the surface density of molecular gas mass, $f_{\rm ISM}$ is the value that accounts for the effects of correlated SNe, $R_{\rm SN}$ is the SNe rate, $\tau_{\rm SN}$ is the mean interval between supernovae in the Galaxy (the inverse of the rate) and $M_{\rm cl}$ is the total ISM mass swept-up by a SN event. Here is important to note that $M_{\rm cl}$ varies with the ambient gas density and metallicity, and as metals being efficient cooling channel in the ISM, higher $Z_{\rm gas}$ would result in smaller swept mass. (see \citealt{asano13}, \citealt{hou19}). Our galaxies are both gas and metal rich, and by adopting our average values for $M_{\rm gas}$ and $Z_{\rm gas}$ we can roughly approximate distraction timescale. Following \cite{slavin15}, we infer that $\tau_{\rm destr}$ is in the range of $0.17\:\rm Gyr$ to $1.87\:\rm Gyr$, with median of $0.89\:\rm Gyr$ for MS DSFGs, and $0.36\:\rm Gyr$ for SB DSFGs. %These timescales are on average longer than what most of models, including SIMBA, usually suggest for galaxies with similar sSFR and $M_{\star}$ (e.g. $\tau_{\rm destr}$$\simeq10^{7}$ yr). I
In fact, the $\tau_{\rm destr}$ could increase even more if galaxy magnetic field is stronger at high-$z$, causing less dust acceleration thus less destruction (\citealt{slavin15}). Therefore, the excess $M_{\rm dust}/M_{\star}$ could point to those systems where dust survival rate is different at earlier times, as postulated by \cite{dwek14}. We caution, however, that the main sources which dominate the uncertainties (e.g. SNe rate and the ISM model) are very difficult to determine accurately from observations.\\

\item  \textbf{Additional physical mechanisms} 

The modelled strong anti-correlation of $\delta_{\rm DGR}$ with $f_{\rm gas}$ could also be a hint to overefficient feedback mechanism in cosmological simulations (\citealt{hirashita17}, \citealt{aoyama18}). In this work we consider full SIMBA suite which includes both AGN feedback and X-ray heating by black holes. Without these two effects included, we would expect much weaker anti-correlation between the $f_{\rm gas}$ and $M_{\star}$ which would naturally lead to weaker anti-correlation between $M_{\rm dust}/M_{\star}$ and $M_{\star}$. This will strongly disagree with our observations (see \hyperref[fig:Fig.6]{Fig. 4}). On top of this, variations in destruction timescales and efficiencies of ISM dust through SN shocks are also dependent on dust grain size distribution. While most of simulations including SIMBA adopt same grain physical sizes ($a=0.1\mu$m), it is postulated that the large fraction of $M_{\rm dust}$ can survive if grain sizes are larger (\citealt{biscaro16}, \citealt{zhukovska16}, \citealt{aoyama19}). Finally, the excess ratio between the dust-to-stellar mass could be due to significant $M_{\rm dust}$ %in the hot DM halo. The outflowing gas that is confined in the circumgalactic medium would eventually virialise after mixing with both the quiescent and the inflowing gaseous material. Such 
in the large reservoir of metal-enriched circumgalactic gas, as recently observed through ALMA [C II] search on scales of $\rm10-20\:kpc$ (\citealt{fujimoto19}, \citealt{ginolfi20}). 

\end{itemize}

%On a final note, it is also possible that dust produced by SNe in high-redshift galaxies has different properties to that assumed in this work, where the dust emissivity $\kappa$ used to determine the dust mass is calibrated from observations of the Milky Way and other nearby galaxies. If, for example, the dust emissivity in high-redshift DSFGs was systematically higher due to changes in the dust grain size distribution, shape or grain composition, this would serve to decrease the observed dust masses thereby alleviating the tension between observed and predicted properties of DSFGs.}
	
%{\color{red} \item From \cite{pantoni19}:  The gas metallicity shows an increasing behavior as a function of the final stellar mass, related to the more efficient production of metals in galaxies with higher SFRs, that will also yield larger stellar masses; the corresponding redshift evolution is negligible, being the gas metallicity essentially related to in-situ processes.}

%\end{itemize}

\section{The role of compact dusty star-formation in  "Giants"}
\label{sec:6}

%inspect the nature and evolutionary status of our DSFGs by answering to the question: \textit{is the observed evolution of the $M_{\rm dust}/M_{\star}$ with respect to the MS (\hyperref[sec:4.3]{Section 4.3}) primarily caused by a large $f_{\rm gas}$, enhanced SFE, or a combination of both?} One of the direct ways to answering this question is by 
To get an additional insight into the ISM of our DSFGs, we explore the influence of galaxy IR size on $M_{\rm dust}/M_{\star}$. %Because the star formation law (or KS relation) is originally defined using surface densities, here we chose to examine the link between galaxy dust-related parameters and their IR sizes, and the origin of any evolution which may be occurring in these relations. 
%On top of this, cosmic rays and turbulence could, in principle, lead to different gas and dust temperatures, assuming that the cosmic ray energy density scales with the SFR density.
%We compute the star-formation surface density ($\Sigma_{\rm SFR}$) for each galaxy based on its IR size ($\vartheta$). 
For this purpose, we adopt ALMA dust continuum sizes obtained through homogeneous $\mathit{uv}$-visibility size analyses with the exponential disk model ($n = 1$, see \citealt {fujimoto17} for the detailed description of the procedure). %The size of the emitting regionswas obtained through the effective radius of the models . %Dust luminosity surface density ($\Sigma_{L_{\rm IR}}$, defined as $L_{\rm IR}/ 2\pi R^{2}_{\rm kpc}$) and 
In order to probe the surface densities of dusty star-formation, we follow the approach from \cite{elbaz17}. %and through \cite{miettinen17b}, \cite{fujimoto17} and \cite{gomezguijarro18}. We and convert to dust-obscured star-formation rate ($\rm SFR_{IR}$) through \citealt{kennicutt12} relation for Chabrier IMF.
Using the SED derived $L_{\rm IR}$  of our sources and their rest-frame IR continuum sizes, we compute the $\rm IR$ luminosity surface density ($\Sigma_{\rm IR}$) defined as $\Sigma_{\rm IR}\approx\mathrm{L_{\rm IR}}/ 2\pi R_{\rm eff}^{2}$, where $R_{\rm eff}$ is an circularized effective ALMA radius of the source (in $\mathrm{kpc}$). %\footnote{We briefly compare our SFR estimates both from SED fitting and by converting IR luminosity through the standard conversion formula from \cite{kennicutt98} with Chabrier IMF to ensure we have consistency in deriving SFR with both methods.}.

The median IR size of the full sample is $R_{\rm eff}=1.51$ kpc. We find that SB DSFGs are more compact than MS DSFGs ($R_{\rm eff}^{\rm SB}=1.24$ kpc vs. $R_{\rm eff}^{\rm MS}=1.61\rm \: kpc$ respectively). The $\Sigma_{\rm IR}$ ranges from  $3.1\times 10^{10}-9.3\times 10^{12}\:L_{\odot} \rm kpc^{-2}$, with the median of $\Sigma_{\rm IR}=6.9\times 10^{11}\:L_{\odot} \rm kpc^{-2}$ for the full sample\footnote{The inferred range of $\Sigma_{\rm IR}$ corresponds to $13-885\:M_{\odot}\rm yr^{-1} \rm kpc^{-2}$ if we convert $L_{\rm IR}$ to dust-obscured SFR through \citealt{kennicutt12} relation for Chabrier IMF.}. The average sizes and $\Sigma_{\rm IR}$ are typical to those derived for IR-selected DSFGs at $z\sim2.5$ for which the majority of the dusty star formation occurs in a central region (e.g. \citealt{simpson15}, \citealt{ikarashi16}). We find that 5 DSFGs have very high surface densities ($\Sigma_{\rm IR}>5\times 10^{12}\:L_{\odot}\rm kpc^{-2}$). They are suitable candidates for approaching Eddington-limit which is estimated to be $\sim10^{13}\:L_{\odot}\rm kpc^{-2}$ based on the balance between the radiation pressure from the star-formation and the self-gravitation (\citealt{andrews11}). Such non-AGN candidates for Eddington limited starbursts are known in the literature (e.g. \citealt{riechers13}, \citealt{gomezguijarro18}) and it has been proposed that for at least some of them significant dust emission could be excited by an outflow (\citealt{oteo17b}). 

To further inspect how the galaxy $\Sigma_{\rm IR}$ impact their dust-to-stellar mass content withing the MS paradigm, we split our full sample in two groups of objects based on their  $\Sigma_{\rm IR}$. We arbitrarily define "less compact" DSFGs with intermediate surface densities ($\Sigma_{\rm IR}<10^{12}\:L_{\odot}\rm kpc^{-2}$, 236 objects in total), and "more compact" DSFGs, due to their higher surface densities ($\Sigma_{\rm IR}>10^{12}\:L_{\odot}\rm kpc^{-2}$, 64 objects in total).
\begin{figure}[h]
	\vspace{-0.2cm}
	\centering
	%\hspace{-1.0cm}
	%	\includegraphics [width=13.89cm]{dsfg-tracks.pdf}
	%	\includegraphics [width=9.33cm]{Plot_06_6.png}
	%	\includegraphics [width=9.33cm]{Plot_06_4.png}
	\includegraphics [width=7.69cm]{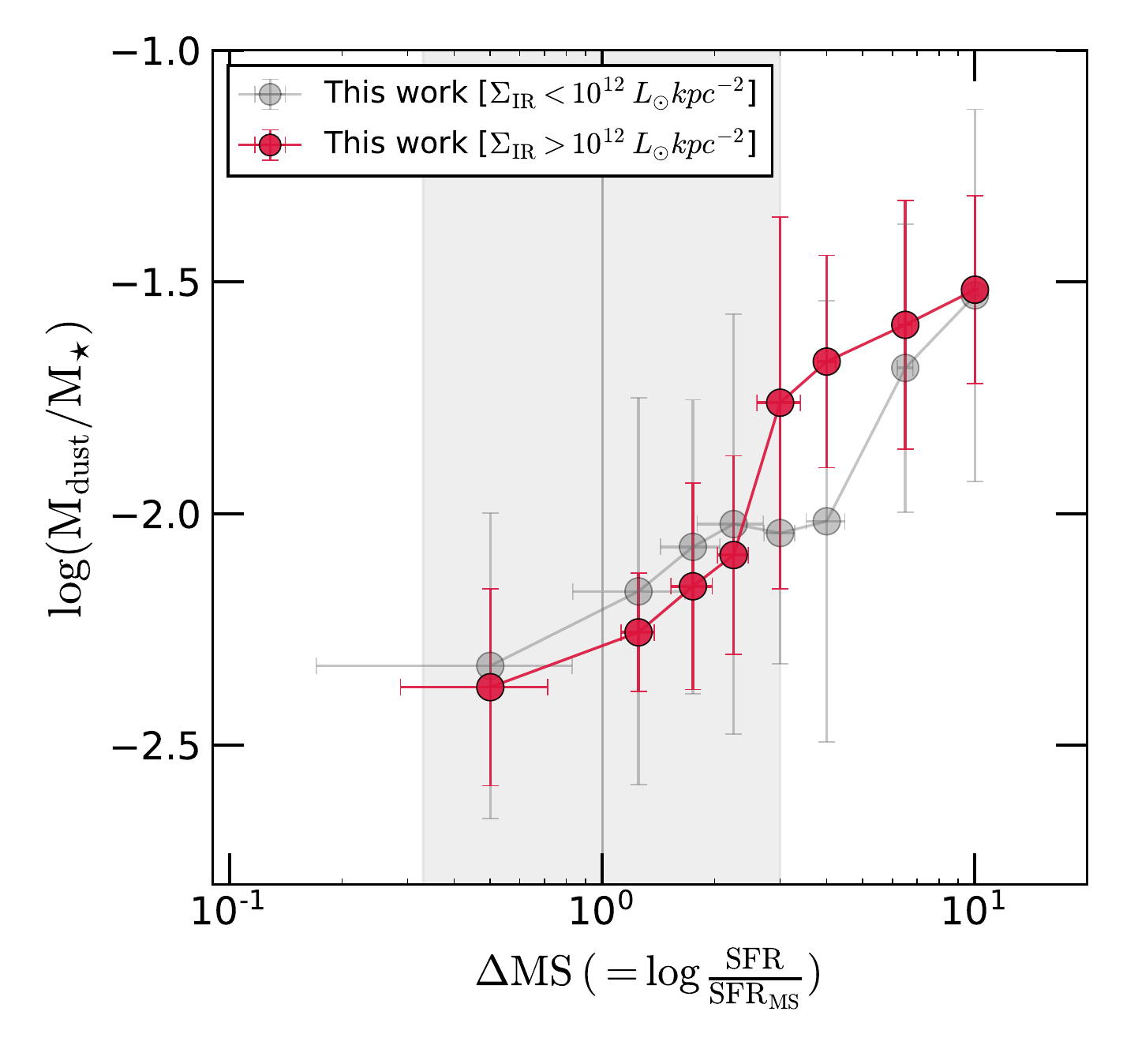}
	\caption{The observed $M_{\rm dust}/M_{\star}$ as a function of a galaxy offset to the MS, defined as $\Delta \rm MS=\log (\rm SFR/SFR_{\rm MS})$. Galaxies with intermediate and high $\Sigma_{\rm IR}$ are denoted with grey and red circles, respectively. The shaded region represents the sequence of MS defined by \cite{speagle14} with a 0.5 dex (3 times) scatter.The points that lie outside the grey region represent the SB DSFGs.}
	%$\Sigma_{\rm SFR}<100\:M_{\odot}\rm yr^{-1} \rm kpc^{-2}$) are marked with grey circles, while red symbols represent DSFGs with higher $\Sigma_{\rm SFR}$ ($\Sigma_{\rm SFR}>100\:M_{\odot}\rm yr^{-1} \rm kpc^{-2}$). }
	\label{fig:Fig.9}
\end{figure}

In \hyperref[fig:Fig.9]{Fig. 8} we show how the $M_{\rm dust}/M_{\star}$ relates to $\Delta \rm MS$ along with galaxy $\Sigma_{\rm IR}$. We note that for easier graphical representation of our results, on $x-$axis we show the galaxy offset to MS in $\log$ scale, labelled as $\Delta \rm MS$. From the \hyperref[fig:Fig.9]{Fig. 8} we see that $M_{\rm dust}/M_{\star}$ tightly relates to $\Delta \rm MS$ regardless of the galaxy $\Sigma_{\rm IR}$. Within the MS, the objects with intermediate and high $\Sigma_{\rm IR}$ have almost identical $M_{\rm dust}/M_{\star}$. Above the MS, the $M_{\rm dust}/M_{\star}$ is slightly higher in more compact sources with higher $\Sigma_{\rm IR}$.
\footnote{This picture is mostly valid if dusty star-formation is not spread in a series of clumps, which would be tested with higher S/N observations at higher spatial resolution.}

The tight relation of $M_{\rm dust}/M_{\star}$ with $\Delta{\rm MS}$ can be interpreted by \textit{"in-situ"} framework which predicts that sources could appear above the MS when caught in an early evolutionary stage. In passing from SB to MS DSFGs one is observing more aged systems, and the decrease in $M_{\rm dust}/M_{\star}$ is due to dust being formed on shorter timescales with respect to $M_{\star}$. %That said, younger objects have already accumulated their $M_{\rm dust}$ while still assembling their $M_{\star}$. With galaxy ageing, the $M_{\star}$ progressively increases, causing the subsequent decline of $M_{\rm dust}/M_{\star}$. 
Such interpretation is strengthen by SED derived young, mass-weighted ages of our SB DSFGs, with the median of $409\pm60$ Myr, half as long as in MS DSFGs. We caution that SED derived mass-weighted ages are strongly model dependent (due to age-metallicity degeneracy), even though our estimates agree with those reported in the literature (\citealt{as2uds19}, \citealt{martis19}). %Nevertheless, if confirmed, these values can strengthen the prediction that the bulk of high-$z$ massive, gas-rich galaxies may have built up significant stellar masses at the early cosmological times when these galaxies are observed. 

The link between $\Sigma_{\rm IR}$ and $M_{\rm dust}/M_{\star}$ less obvious and is challenging to interpret. %For MS DSFGs, we see that  $M_{\rm dust}/M_{\star}$ is similar for both groups of sources, while for SB DSFGs there is an tentative evidence that more compact dust emission is linked to higher $M_{\rm dust}/M_{\star}$.
%This is in overall agreement with \cite{kirkpatrick17} who reach the similar conclusion studying much smaller sample of DSFGs at $z=2$. The higher SFE in more compact sources can be attributed to gas compression (see e.g. \citealt{hayward12}, \citealt{scoville16}. 
In the local Universe, decrease in size can enhance the efficiency of transforming atomic gas into molecular gas, boosting the $M_{\rm gas}$ (\citealt{larson16}, \citealt{kirkpatrick17}). %Alternately, a smaller size translates to a higher SFR surface density, which would boost $L_{\rm IR}$ without requiring a boost in gas mass. 
Recently, \cite{cochrane19} performed detailed study of the spatially-resolved dust continuum emission of simulated DSFGs at $z>1$ and found that the most compact dust emission is driven by particularly compact recent star-formation. %On top of that, \cite{hayward12} proposed that for a highly turbulent ISM, a smaller size translates to a higher $\Sigma_{\rm IR}$, which would boost $L_{\rm IR}$ (thus $M_{\rm dust}$) without requiring enhancement in $M_{\rm gas}$. This seems valid for a mixed geometry, for which is expected that $T_{\rm dust}$ is almost insensitive to the size (\citealt{safarzadeh15}, \citealt{kirkpatrick17}. 
Distant DSFGs are also expected to have highly turbulent ISM (\citealt{scoville16}). Turbulence can rapidly accelerate the grain growth (\citealt{mattsson20}), which would increase the amount of large dust grains relative to small ones, and produce colder $T_{\rm dust}$ for a given radiation field. If dust emission is optically thin, this would result in higher $M_{\rm dust}$ at a given $L_{\rm IR}$. In our companion paper (Paper II, Donevski et al., in prep.), we will present detailed analysis of various mechanisms that produce the high dust/gas densities in distant DSFGs. %As mentioned, some of our DSFGs have extreme gas surface densities. Such high densities would make the gaseous ISM highly optically thick even in the re-radiated IR, and the radiation pressure on dust grains makes the system become Eddington-limited.

Future James Webb Space Telescope (JWST) data combined with larger ALMA samples will be of crucial importance to discriminate between different scenarios. The JWST will be able to derive accurate estimates of the AGN contribution to most massive DSFGs, and place important constraints on the gas reservoirs of these sources from various near-IR and mid-IR lines, resulting from a PAH cooling process.

%UBACI KASNIJE: On top of all these scenarios, we also have to note the possibility that high-$z$ galaxies have more turbulent ISMs, leading to compression in the ISM and enhancing the star formation efficiency per unit mass. While this efficient mode of star formation can occur throughout the galaxy, and no obvious correlation between galaxy size and $T_{\rm dust}$ may be expected, we should expect the correlation between galaxy size and SFE represented as $\rm L_{\rm IR}/M_{\rm dust}$. This is, indeed, what we see in our data, and could point to one of the potential cause of higher SFEs in our SB DSFGs.

%The Popping et al. (2016) models shown in our comparison are based on a semi-analytic model of galaxy formation in a cosmological context, which includes a standard suite of physi- cal processes (gas accretion and cooling, star formation, stellar feedback, chemical enrichment, etc). In addition, the Popping et al. (2016) model includes self-consistent tracking of the main processes thought to produce and destroy dust in galaxies, including dust condensation in stellar ejecta, dust growth through accretion in the ISM, dust destruction by supernovae, and ejection of dust by stellar driven winds. The much milder evolution of Md-Ms with redshift predicted by these models relative to our findings is interesting, as it indicates that one or more of the model ingredients needs to be revised.

\section{Conclusions}

We perform a systematic study of the dust-to-stellar mass ratio in 300 massive ($M_{\star}>10^{10}\:M_{\odot}$) DSFGs in the COSMOS field, observed with ALMA over a wide redshift range ($0.5<z<5.25$). We apply self-consistent, multi-band SED fitting method and explore trends of $M_{\rm dust}/M_{\star}$ with different physical parameters in galaxies within and above the main-sequence. We fully evaluate our findings with the models of dusty galaxy formation and evolution. Our main results are summarised as follows: 

\begin{itemize}
	\item We find that $M_{\rm dust}/M_{\star}$ evolves with the redshift, stellar mass and specific star formation rate. For both galaxy populations the $M_{\rm dust}$/$M_{\rm \star}$ rises up to $z\sim2$, steeper in SB than in MS DSFGs, followed by mild decline/flattening at $z\gtrsim2$. We infer the median of $M_{\rm dust}/M_{\star}=0.006^{+0.004}_{-0.003}$ and $M_{\rm dust}/M_{\star}=0.017^{+0.010}_{-0.006}$ for MS and SB DSFGs, respectively. %Starburst galaxies have systematically higher $M_{\rm dust}$/$M_{\rm \star}$ irrespective of the observed redshift, galaxy integrated dust size and stellar mass. %
	%We find that for both populations $M_{\rm dust}/M_{\star}$ rises up to $z\sim2-2.25$ and flattens/bend above this redshift. 
	Regardless of the observed redshift, the SB DSFGs typically have $\sim3$ times higher $M_{\rm dust}/M_{\star}$ as compared to MS DSFGs. %Therefore, $M_{\rm dust}/M_{\star}$ can be applied as a useful, complementary tool for disentangling the general population of DSFGs into main-sequence objects and starbursts. %This finding complement those of \cite{tan14}%and testing the dichotomy of the general population of DSFGs at the highest redshifts where normalisation of the MS is still uncertain. %the trend can be traced back to their denser environment, in turn yielding shorter star formation timescales. 
	\item Differently than local ULIRGs, the $M_{\rm dust}$ and $\rm SFR$ in our high-$z$ DSFGs obey sub-linear trend that exhibits a plateau above the characteristic dust mass ($M_{\rm dust}\approx10^{9}\:M_{\odot}$). This implies possible evolution in dust properties (e.g. dust opacities). %This is different than % The trend is shallower to what is observed in the local Universe.%suggesting that  dust production at earlier cosmic times is likely dominated by dust growth through accretion in the ISM.
	
	\item We confirmed, for the first time, that the inverse relation between $M_{\rm dust}$/$M_{\rm \star}$ and the $M_{\star}$ holds until $z\approx5$. The normalisation of this inverse relation gradually increases by $\sim0.5\:\rm dex$ from $z=1$ to $z=5$. We interpret the observed trend as an evolutionary transition from earlier to later starburst phases of DSFGs.
	
	\item We model the observed $M_{\rm dust}/M_{\star}$ by applying empirical relations for $f_{\rm gas}$ and $\rm MZR$. Both MS and SB DSFGs require high, solar-like $Z_{\rm gas}$ in order to match the estimated $M_{\rm dust}/M_{\star}$. The modelled $M_{\rm dust}/M_{\star}$ % is very sensitive to the functional form of $f_{\rm gas}$. It 
	faithfully represents observed trend in MS DSFGs over the full redshift range. While  adopted gas scaling relations anticipate somewhat larger average gas supply in SB than in MS DSFGs  ($M_{\rm gas}=1.03\times10^{11}M_{\odot}$ vs. $M_{\rm gas}=8.92\times10^{10}M_{\odot}$, respectively), at the same time they slightly overpredict our data for SB DSFGs at $z>2.5$. The latter indicates the possibility of super-solar $Z_{\rm gas}$ in some high-$z$ starbursts, pointing towards the need of a rapid metal enrichment. %\textcolor{red}{Even considering substantial uncertainties in scaling relations and unknown $Z_{\rm gas}$, the $M_{\rm dust}/M_{\star}$ for SBs are high toward higher redshifts, meaning that even in the early Universe SBs were substantially metal rich, pointing toward the need of rapid metal enrichment in the starburst phase.}	 %From $M_{\rm dust}$and gas scaling relation by \citealt{liu19b}, we find that SB DSFGs have slightly higher $M_{\rm gas}$ than MS DSFGs ($M_{\rm gas}=1.03\times10^{11}M_{\odot}$ vs. $M_{\rm gas}=8.92\times10^{10}M_{\odot}$, respectively).
	\item We %infer the corresponding median dust-to-gas ratios of $\delta_{\rm DGR}=1/159$ and $\delta_{\rm DGR}=1/139$  for MS and SB DSFGs, respectively. 
    show that $M_{\rm dust}/M_{\star}$ mirrors the increase in molecular gas fraction with the redshift. By linking the gas scaling relation from \cite{liu19b} and MZR from \cite{genzel15} we infer a median of $f_{\rm gas}=0.44\pm0.09$ for MS DSFGs and $f_{\rm gas}=0.75\pm0.09$ for SB DSFGs. %\textcolor{red}{The increased dust mass might then be a natural consequence of the increase in gas fractions with lookback time.} %We infer the corresponding median dust-to-gas ratios of $\delta_{\rm DGR}=1/159$ and $\delta_{\rm DGR}=1/139$  for MS and SB DSFGs, respectively. 
	
	\item We fully evaluate our findings with different models of dusty galaxy formation. The cosmological simulation SIMBA (\citealt{simba}) predicts the cosmic evolution of $M_{\rm dust}/M_{\star}$ in MS DSFGs consistent within $2\sigma$ with our data. SIMBA underpredicts the $M_{\rm dust}/M_{\star}$ in SB DSFGs.
	% and we demonstrate this is partly due to lower $Z_{\rm gas}$ relative to observations, and due to 
	This point to the necessity of refining the dust treatment in simulations, for instance by adding the recipes for dust size distribution or accounting for more rapid metal enrichment in early starburst phase.
	\item The observed $M_{\rm dust}/M_{\star}$ in both MS and SB DSFGs is well reproduced by the phenomenological model of \cite{bethermin17} and the analytic model of \cite{pantoni19}. The overall agreement with these models has two important implications: (1) the existent knowledge about galaxy star-formation MS and the $M_{\rm dust}/M_{\star}$ converges towards the consistent quantitative picture; (2) the fast dust growth through accretion in the metal-rich ISM  is needed to capture the observed $M_{\rm dust}/M_{\star}$ in high-$z$ DSFGs. 
 %One step towards right direction could be refined treatment of dust grain distribution in simulations. %This fast drop is predicted in the newest generation of analytical models (\cite{imara19}, \citealt{pantoni19}), chemical models (\citealt{calura17}), and cosmological simulations that include dust growth based on the grain ISM growth (\citealt{simba}). However, the model predictions significantly differ in respect to the condensation time and gas-phase metallicity, resulting in different slopes following this turn.

	\item We examine the link between $M_{\rm dust}/M_{\star}$ and compact dusty star-formation along the MS paradigm. The observed $M_{\rm dust}/M_{\star}$ in MS DSFGs relates to $\Delta \rm MS$ regardless of the galaxy $\Sigma_{\rm IR}$, while for SB DSFGs we find an evidence that $M_{\rm dust}/M_{\star}$ is enhanced in systems with higher $\Sigma_{\rm IR}$. Further investigation of these objects is crucial for understanding the role of compact dusty star-formation in galaxy evolution.

	%\item We identify the sub-population of compact, $z\gtrsim2$ DSFGs located within the MS that exhibit high $\rm SFE$ (short $\tau_{\rm dep}$), but have low $M_{\rm dust}/M_{\star}$. These objects are similar to those initially discovered by \cite{elbaz17}, but show no signatures of AGNs. We consider them suitable candidates for high-$z$ quiescent galaxies, and their further investigation is important for understanding the role of compact star-formation in galaxy evolution.

\end{itemize}

This work highlights that analysing the different trends with $M_{\rm dust}$/$M_{\rm \star}$ is a useful diagnostic tool for the present and future studies of DSFGs. Firstly, it can be applied for separating the main-sequence galaxies and starbursts over wide redshift range. This confirms and complements the conclusion from seminal works of \cite{tan14} and \cite{bethermin15}. Secondly, having combined with the independent molecular gas estimations, the $M_{\rm dust}$/$M_{\rm \star}$ can be a powerful probe of the evolutionary phase of massive objects. %This work also shows that the close independence of $M_{\rm dust}$/$M_{\rm \star}$ with respect to redshift for both MS and SB DSFGs strongly supports the picture where star formation and dust production are mainly \textit{"in-situ"} processes. 
Finally, in a future paper, we will present the direct predictions related to synergy between the next JWST and present (sub)millimeter surveys.
%%%%%%%%%%%%%%%%%%%%%%%%%%%%%%%%%%%%%%%%

%%%%%%%%%%%%%%%%%%%%%
\begin{acknowledgements}
We would like to acknowledge Veronique Buat, Pauline Vielzeuf, Robert Feldmann, Ciro Pappalardo and Federico Bianchini for useful discussions, comments and/or support. This work has been partially supported by PRIN MIUR 2017 prot. 20173ML3WW002, "Opening the ALMA window on the cosmic evolution of gas, stars and supermassive black holes". D.D. acknowledge the Dunlap visitor program, at the Dunlap Institute for Astronomy \& Astrophysics at the University of Toronto. A.L. acknowledges the MIUR grant ‘Finanziamento annuale individuale attività base di ricerca’ and the EU H2020-MSCA-ITN-2019 Project 860744 ‘BiD4BEST: Big Data applications for Black hole Evolution STudies". K.M. has been supported by the National Science Centre (UMO-2018/30/E/ST9/00082). D. L. acknowledges funding from the European Research Council (ERC) under the European Union's Horizon 2020 research and innovation programme (grant agreement No. 694343). A.M. is supported by a Dunlap Fellowship at the Dunlap Institute for Astronomy \& Astrophysics, funded through an endowment established by the David Dunlap family and the University of Toronto. The University of Toronto operates on the traditional land of the Huron-Wendat, the Seneca, and most recently, the Mississaugas of the Credit River; A.M. and D.D are grateful to have the opportunity to work on this land. S.F. acknowledge support from the European Research Council (ERC) Consolidator Grant funding scheme (project ConTExt, grant No. 648179). The Cosmic Dawn Center is funded by the Danish National Research Foundation under grant No. 140. A.F. acknowledges the support from grant PRIN MIUR 2017 20173ML3WW. 

\\
This paper make use of following ALMA data: 

ADS/JAO.ALMA: \#2011.0.00064.S, \#2011. 0.00097.S, \#2011.0.00539.S, \#2011.0.00742.S, \#2012.1.00076.S, \#2012.1.00323.S, \# 2012.1.00523.S, \#2012.1.00536.S, \#2012.1.00919.S, \#2012.1.00952.S, \#2012.1.00978.S, \#2013.1.00884. S, \#2013.1.00914.S, 
\#2015.1.00137.S, \#2015.1.00568.S, \#2015.1.00664.S, \#2015.1.00704.S, \#2015.1.00853. S, \#2015.1.00861.S, \# 2015.1.00862.S, \#2015.1.00928.S, \#2015.1.01074.S,  \# 2015.1.01590.S, \#2015.A.00026.S, \#2016.1.00478.S, \#2016.1.00624.S, \#2016.1.00735.S.
ALMA is a partnership of ESO (representing its member states), NSF (USA) and NINS (Japan), together with NRC (Canada), MOST and ASIAA (Taiwan), and KASI (Republic of Korea), in cooperation with the Republic of Chile. The Joint ALMA Observatory is operated by ESO, AUI/NRAO and NAOJ.

\\
%	$\mathit{Herschel}$ is an ESA space observatory with science instruments provided by European-led Principal Investigator consortia and with important participation from NASA. SPIRE  has  been  developed  by  a  consortium  of  institutes led by Cardiff University (UK) and including Univ. Lethbridge (Canada);  NAOC  (China);  CEA,  LAM  (France);  IFSI,  Univ. Padua (Italy); IAC (Spain); Stockholm Observatory (Sweden); Imperial  College  London,  RAL,  UCL-MSSL,  UKATC,  Univ.Sussex (UK); and Caltech, JPL, NHSC, Univ. Colorado (USA). This development has been supported by national funding agencies:  CSA  (Canada);  NAOC  (China);  CEA,  CNES,  CNRS (France); ASI (Italy); MCINN (Spain); SNSB (Sweden); STFC (UK); and NASA (USA). This publication  makes  use  of  public data  products  from  the Two Micron All Sky Survey (2MASS),  which is a joint project of the  University  of  Massachusetts  and the Infrared Processing and Analysis Center California Institute of Technology, funded by the National Aeronautics and Space Administration and the National Science Foundation. We express are gratitude to \texttt{Daft} \url{https://github.com/dfm/daft}, a probabilistic graphical model available under the MIT License.
\end{acknowledgements}

%%%%%%%%%%%%%%%%%

%-------------------------------------------------------------------
\bibliographystyle{aa}
\bibliography{ddrisers}
%\begin{thebibliography}{12}

%%%%%%%%%%%%%%%%%%%%%%%%%%%%%%%%%%%%%%
%\begin{appendix} 
%	\section{The properties of selected sources}
	\clearpage
 \newpage

	\onecolumn
\setlength\LTleft{-30pt}            % default: \fill
\setlength\LTright{-30pt}           % default: \fill
	\begin{small}
	\begin{longtable}
	{rrrrrrrrrrrrr}
		\caption{List of selected DSFGs and their physical properties.} 
		\label{tab:4}
		%	\hline\hline
	\\
		Source ID & RA &  Dec & \textit{z} & $\log M_{\star}$  & $\log M_{\star}^{err}$ &$\log L_{\rm IR}$ & $\log L_{\rm IR}^{err}$ & $\log M_{\rm dust}$ & $\log M_{\rm dust}^{err}$ & $S_{\rm ALMA}$ & $S_{\rm ALMA}^{err}$ & $\nu_{\rm obs}$ \\
	&[deg]&[deg]&&[$\log M_{\odot}$]&[$\log M_{\odot}$]&[$\log L_{\odot}$]&[$\log L_{\odot}$]&[$\log M_{\odot}$]&[$\log M_{\odot}$]&[mJy]&[mJy]&[GHz]\\
		\hline
		\hline
		\\
		\endfirsthead
		%caption{continued.}
		\\
		%	\hline\hline
		\\
			Source ID & RA &  Dec & \textit{z} & $\log M_{\star}$  & $\log M_{\star}^{err}$ &$\log L_{\rm IR}$ & $\log L_{\rm IR}^{err}$ & $\log M_{\rm dust}$ & $\log M_{\rm dust}^{err}$ & $S_{\rm ALMA}$ & $S_{\rm ALMA}^{err}$ & $\nu_{\rm obs}$ \\
	&[deg]&[deg]&&[$\log M_{\odot}$]&[$\log M_{\odot}$]&[$\log L_{\odot}$]&[$\log L_{\odot}$]&[$\log M_{\odot}$]&[$\log M_{\odot}$]&[mJy]&[mJy]&[GHz]\\
		\hline
		\hline
		\\
		\endhead
			\hline
				\endfoot
	
HELP\_J100041.969 & 150.175 & 2.353 & 2.892 & 10.630 & 0.113 & 12.596 & 0.428 & 9.188 & 0.152 & 0.148 & 0.056 & 148.281\\
HELP\_J095859.136& 149.746 & 2.084 & 2.192 & 10.695 & 0.106 & 11.856 & 0.140 & 7.735 & 0.317 & 0.249 & 0.047 & 260.370\\
HELP\_J100033.409& 150.139 & 2.432 & 2.930 & 11.355 & 0.401 & 12.497 & 0.333 & 8.725 & 0.022 & 0.282 & 0.087 & 148.281\\
HELP\_J095957.847& 149.991 & 1.797 & 3.528 & 11.056 & 0.251 & 12.198 & 0.271 & 8.030 & 0.245 & 0.533 & 0.124 & 239.989\\
HELP\_J100126.753 & 150.361 & 2.062 & 2.999 & 10.685 & 0.279 & 12.492 & 0.316 & 8.431 & 0.303 & 0.611 & 0.108 & 239.989\\
HELP\_J100206.822 & 150.528 & 2.574 & 3.251 & 11.191 & 0.270 & 12.333 & 0.259 & 8.339 & 0.255 & 0.613 & 0.139 & 239.989\\
HELP\_J095847.056& 149.696 & 2.122 & 1.351 & 11.264 & 0.291 & 12.195 & 0.239 & 8.545 & 0.154 & 0.669 & 0.159 & 252.051\\
HELP\_J100231.047& 150.629 & 2.550 & 2.820 & 11.399 & 0.238 & 12.542 & 0.312 & 8.216 & 0.251 & 0.682 & 0.124 & 239.989\\
HELP\_J100006.057& 150.025 & 2.312 & 3.949 & 10.928 & 0.137 & 12.454 & 0.307 & 8.326 & 0.273 & 0.689 & 0.108 & 239.989\\
HELP\_J100036.344& 150.151 & 1.936 & 0.330 & 10.767 & 0.243 & 11.405 & 0.035 & 8.401 & 0.180 & 0.703 & 0.157 & 255.123\\
HELP\_J100213.787& 150.557 & 2.691 & 3.314 & 10.496 & 0.334 & 12.436 & 0.282 & 8.477 & 0.223 & 0.714 & 0.105 & 239.989\\
HELP\_J100123.355& 150.347 & 2.747 & 3.033 & 10.651 & 0.303 & 12.589 & 0.448 & 8.572 & 0.196 & 0.739 & 0.118 & 239.989\\
HELP\_J095756.196& 149.484 & 1.630 & 4.561 & 11.161 & 0.181 & 12.583 & 0.322 & 8.217 & 0.240 & 0.753 & 0.107 & 239.991\\
HELP\_J095955.543& 149.981 & 2.253 & 1.404 & 11.006 & 0.291 & 12.309 & 0.260 & 8.763 & 0.319 & 0.755 & 0.037 & 226.768\\
HELP\_J095931.526& 149.881 & 2.450 & 1.370 & 10.799 & 0.170 & 12.181 & 0.122 & 8.656 & 0.336 & 0.781 & 0.255 & 343.485\\
HELP\_J095957.524& 149.990 & 1.798 & 3.169 & 10.711 & 0.223 & 12.014 & 0.503 & 8.478 & 0.145 & 0.788 & 0.108 & 239.989\\
HELP\_J095817.059& 149.571 & 1.674 & 3.057 & 10.909 & 0.303 & 12.212 & 0.405 & 8.670 & 0.194 & 0.795 & 0.108 & 239.991\\
HELP\_J095755.941& 149.483 & 2.505 & 3.195 & 11.279 & 0.355 & 12.628 & 0.361 & 8.624 & 0.293 & 0.822 & 0.117 & 239.989\\
HELP\_J100158.471& 150.494 & 1.819 & 3.242 & 10.627 & 0.191 & 12.434 & 0.350 & 8.472 & 0.252 & 0.832 & 0.140 & 239.989\\
HELP\_J095821.776& 149.591 & 2.806 & 0.349 & 10.910 & 0.341 & 11.602 & 0.074 & 8.179 & 0.428 & 0.860 & 0.122 & 264.193\\
HELP\_J100027.446& 150.114 & 2.370 & 4.211 & 10.247 & 0.158 & 11.407 & 0.385 & 8.069 & 0.086 & 0.864 & 0.275 & 344.772\\
HELP\_J100046.047& 150.192 & 1.722 & 1.990 & 10.822 & 0.275 & 12.030 & 0.237 & 9.229 & 0.238 & 0.866 & 0.257 & 343.523\\
HELP\_J100040.991& 150.171 & 2.369 & 3.331 & 10.675 & 0.155 & 12.482 & 0.432 & 9.032 & 0.216 & 0.869 & 0.297 & 344.772\\
HELP\_J095852.010& 149.717 & 1.861 & 2.011 & 10.891 & 0.154 & 12.418 & 0.326 & 8.650 & 0.270 & 0.886 & 0.120 & 245.031\\
HELP\_J100017.357& 150.072 & 1.974 & 3.018 & 10.694 & 0.219 & 12.152 & 0.348 & 8.758 & 0.149 & 0.886 & 0.110 & 239.989\\
HELP\_J100003.851& 150.016 & 2.042 & 2.744 & 10.344 & 0.250 & 12.540 & 0.368 & 8.724 & 0.325 & 0.888 & 0.117 & 239.989\\
HELP\_J100042.499& 150.177 & 2.221 & 3.466 & 10.658 & 0.348 & 11.831 & 0.215 & 8.395 & 0.127 & 0.888 & 0.276 & 344.772\\
HELP\_J100212.172& 150.551 & 2.190 & 3.238 & 10.895 & 0.383 & 12.645 & 0.402 & 8.571 & 0.298 & 0.898 & 0.112 & 239.989\\
HELP\_J095904.718& 149.770 & 1.792 & 2.946 & 10.856 & 0.172 & 12.455 & 0.346 & 8.723 & 0.242 & 0.911 & 0.124 & 239.989\\
HELP\_J095927.290& 149.864 & 1.950 & 0.991 & 10.925 & 0.245 & 11.634 & 0.106 & 8.521 & 0.126 & 0.930 & 0.275 & 343.524\\
HELP\_J100208.456& 150.535 & 2.011 & 4.224 & 11.155 & 0.239 & 12.675 & 0.347 & 8.554 & 0.306 & 0.944 & 0.147 & 245.029\\
HELP\_J100203.598& 150.515 & 2.618 & 3.796 & 11.461 & 0.307 & 12.899 & 0.422 & 8.804 & 0.301 & 0.946 & 0.132 & 239.989\\
HELP\_J100124.813& 150.353 & 1.654 & 1.451 & 11.024 & 0.303 & 12.119 & 0.204 & 8.617 & 0.164 & 0.955 & 0.286 & 343.523\\
HELP\_J095953.308& 149.972 & 1.744 & 1.591 & 10.647 & 0.296 & 12.173 & 0.484 & 8.757 & 0.086 & 0.951 & 0.284 & 343.524\\
HELP\_J100139.714& 150.415 & 2.105 & 1.994 & 11.414 & 0.263 & 12.490 & 0.306 & 8.792 & 0.261 & 0.954 & 0.184 & 239.989\\
HELP\_J100109.857& 150.291 & 2.063 & 3.031 & 11.316 & 0.275 & 12.651 & 0.347 & 8.519 & 0.309 & 0.964 & 0.202 & 239.984\\
HELP\_J100110.238& 150.293 & 2.547 & 2.582 & 10.769 & 0.236 & 11.977 & 0.486 & 8.944 & 0.132 & 0.965 & 0.266 & 343.528\\
HELP\_J100227.936& 150.616 & 2.168 & 1.505 & 10.989 & 0.295 & 12.928 & 0.454 & 9.269 & 0.356 & 0.967 & 0.048 & 235.006\\
HELP\_J100008.787& 150.037 & 2.271 & 1.798 & 10.433 & 0.205 & 11.968 & 0.196 & 8.210 & 0.197 & 0.982 & 0.241 & 343.524\\
HELP\_J100145.957& 150.441 & 2.557 & 3.134 & 10.414 & 0.266 & 11.302 & 0.217 & 8.396 & 0.251 & 0.983 & 0.105 & 239.989\\
HELP\_J100105.480& 150.273 & 2.782 & 1.318 & 11.277 & 0.305 & 12.040 & 0.187 & 8.697 & 0.206 & 1.025 & 0.211 & 343.524\\
HELP\_J095858.998& 149.746 & 2.126 & 1.579 & 11.122 & 0.151 & 11.937 & 0.367 & 9.110 & 0.233 & 1.027 & 0.298 & 343.524\\
HELP\_J095935.731& 149.899 & 1.968 & 3.253 & 10.935 & 0.168 & 12.151 & 0.305 & 8.771 & 0.172 & 1.027 & 0.121 & 239.989\\
HELP\_J100024.684& 150.103 & 2.385 & 4.971 & 11.348 & 0.334 & 13.211 & 0.601 & 8.790 & 0.320 & 1.028 & 0.287 & 344.772\\
HELP\_J095904.332& 149.768 & 1.617 & 2.322 & 10.315 & 0.325 & 12.281 & 0.324 & 9.232 & 0.131 & 1.033 & 0.241 & 343.533\\
HELP\_J095933.781& 149.891 & 2.649 & 1.567 & 10.774 & 0.243 & 11.589 & 0.079 & 8.684 & 0.298 & 1.033 & 0.232 & 343.485\\
HELP\_J100026.925& 150.112 & 2.314 & 2.276 & 11.171 & 0.206 & 12.235 & 0.232 & 8.936 & 0.155 & 1.036 & 0.209 & 265.022\\
HELP\_J100211.616& 150.548 & 2.745 & 3.232 & 10.787 & 0.301 & 12.181 & 0.268 & 8.844 & 0.184 & 1.036 & 0.142 & 239.989\\
HELP\_J100200.662& 150.503 & 2.219 & 1.259 & 11.070 & 0.122 & 12.299 & 0.258 & 8.412 & 0.246 & 1.044 & 0.227 & 343.484\\
HELP\_J100201.903& 150.508 & 2.202 & 3.408 & 10.343 & 0.304 & 12.533 & 0.341 & 8.838 & 0.278 & 1.058 & 0.109 & 239.989\\
HELP\_J100043.031& 150.179 & 2.088 & 1.782 & 10.817 & 0.318 & 11.763 & 0.155 & 8.292 & 0.020 & 1.065 & 0.302 & 343.52\\
HELP\_J100258.303& 150.743 & 1.885 & 1.790 & 10.827 & 0.193 & 12.353 & 0.322 & 9.089 & 0.060& 1.077 & 0.228 & 343.523\\
HELP\_J100019.048& 150.079 & 2.341 & 2.590 & 10.642 & 0.245 & 11.510 & 0.049 & 8.001 & 0.052 & 1.078 & 0.289 & 343.532\\
HELP\_J100026.781& 150.112 & 1.738 & 1.359 & 11.126 & 0.373 & 11.941 & 0.164 & 8.292 & 0.243 & 1.094 & 0.231 & 343.524\\
HELP\_J095924.950& 149.854 & 1.754 & 3.140 & 10.430 & 0.373 & 11.889 & 0.224 & 8.961 & 0.315 & 1.095 & 0.111 & 239.989\\
HELP\_J100026.973& 150.112 & 2.375 & 2.208 & 10.730 & 0.148 & 12.328 & 0.269 & 8.468 & 0.032 & 1.097 & 0.303 & 343.531\\
HELP\_J100008.991& 150.037 & 2.272 & 1.756 & 10.759 & 0.230 & 12.141 & 0.212 & 8.746 & 0.141 & 1.100& 0.253 & 343.524\\
HELP\_J100116.278& 150.318 & 2.716 & 2.528 & 11.249 & 0.274 & 12.599 & 0.382 & 9.258 & 0.204 & 1.100 & 0.410 & 239.988\\
HELP\_J100219.083& 150.579 & 2.708 & 2.289 & 10.680 & 0.164 & 12.431 & 0.355 & 8.772 & 0.255 & 1.101 & 0.253 & 343.523\\
HELP\_J100208.330& 150.535 & 2.678 & 2.319 & 11.312 & 0.112 & 12.647 & 0.336 & 8.795 & 0.243 & 1.123 & 0.126 & 239.989\\
HELP\_J100114.698& 150.311 & 2.588 & 3.007 & 10.890 & 0.193 & 12.744 & 0.428 & 9.086 & 0.269 & 1.165 & 0.228 & 239.984\\
HELP\_J095915.963 & 149.816 & 1.780 & 2.093 & 10.461 & 0.235 & 12.393 & 0.272 & 8.307 & 0.280 & 1.171 & 0.311 & 343.533\\
HELP\_J100131.384& 150.381 & 2.057 & 3.152 & 11.291 & 0.225 & 12.626 & 0.332 & 8.569 & 0.208 & 1.171 & 0.111 & 239.989\\
HELP\_J100124.976& 150.354 & 1.667 & 3.123 & 11.162 & 0.254 & 12.579 & 0.351 & 8.647 & 0.266 & 1.190 & 0.118 & 239.989\\
HELP\_J100052.592& 150.219 & 2.523 & 1.439 & 11.103 & 0.348 & 12.269 & 0.250 & 8.279 & 0.076 & 1.195 & 0.324 & 343.496\\
HELP\_J100038.559& 150.161 & 2.156 & 2.166 & 10.643 & 0.381 & 11.871 & 0.143 & 7.865 & 0.121 & 1.200 & 0.368 & 343.519\\
HELP\_J100140.278& 150.418 & 2.559 & 1.210 & 11.027 & 0.297 & 11.973 & 0.349 & 8.491 & 0.148 & 1.203 & 0.274 & 343.484\\
HELP\_J100058.684& 150.244 & 2.160 & 3.807 & 11.608 & 0.382 & 12.775 & 0.368 & 8.671 & 0.325 & 1.227 & 0.142 & 245.029\\
HELP\_J095909.620& 149.790 & 1.712 & 1.784 & 10.544 & 0.253 & 12.482 & 0.410 & 8.370 & 0.222 & 1.228 & 0.237 & 343.524\\
HELP\_J095800.755& 149.503 & 2.506 & 2.310 & 10.943 & 0.309 & 12.375 & 0.350 & 8.260 & 0.253 & 1.240 & 0.291 & 343.524\\
HELP\_J100144.134& 150.434 & 2.765 & 1.011 & 11.413 & 0.342 & 12.114 & 0.203 & 8.769 & 0.377 & 1.244 & 0.316 & 343.523\\
HELP\_J095904.348& 149.768 & 2.220 & 1.806 & 10.869 & 0.285 & 12.185 & 0.261 & 8.746 & 0.120 & 1.252 & 0.263 & 343.485\\
HELP\_J100001.660& 150.007 & 2.408 & 2.891 & 11.347 & 0.230 & 12.682 & 0.345 & 8.476 & 0.291 & 1.254 & 0.120 & 239.989\\
HELP\_J100140.884& 150.420 & 2.118 & 3.883 & 11.087 & 0.284 & 12.613 & 0.369 & 8.602 & 0.238 & 1.274 & 0.163 & 245.029\\
HELP\_J095758.146& 149.492 & 2.803 & 4.515 & 11.365 & 0.306 & 13.126 & 0.605 & 8.533 & 0.396 & 1.299 & 0.400 & 341.950\\
HELP\_J095838.735& 149.661 & 1.949 & 5.234 & 10.840 & 0.243 & 12.175 & 0.396 & 9.053 & 0.237 & 1.304 & 0.136 & 239.989\\
HELP\_J100124.321& 150.351 & 2.689 & 3.152 & 10.597 & 0.334 & 12.538 & 0.309 & 8.896 & 0.240 & 1.304 & 0.127 & 239.988\\
HELP\_J100034.254& 150.143 & 1.816 & 2.341 & 11.387 & 0.333 & 12.529 & 0.308 & 8.444 & 0.324 & 1.316 & 0.247 & 239.984\\
HELP\_J100128.498& 150.369 & 2.396 & 5.021 & 11.006 & 0.143 & 12.532 & 0.320 & 8.472 & 0.264 & 1.329 & 0.203 & 239.984\\
HELP\_J100211.566& 150.548 & 2.533 & 1.341 & 11.414 & 0.293 & 12.375 & 0.267 & 8.613 & 0.414 & 1.333 & 0.265 & 343.524\\
HELP\_J100027.014& 150.113 & 2.377 & 4.904 & 11.020 & 0.231 & 12.453 & 0.294 & 8.944 & 0.228 & 1.341 & 0.459 & 343.531\\
HELP\_J100018.587& 150.077 & 2.185 & 2.299 & 11.164 & 0.335 & 12.500 & 0.408 & 9.041 & 0.020 & 1.350& 0.344 & 343.532\\
HELP\_J095943.468& 149.931 & 2.646 & 1.190 & 11.207 & 0.349 & 12.062 & 0.264 & 8.655 & 0.173 & 1.351 & 0.274 & 343.524\\
HELP\_J100015.799& 150.066 & 2.594 & 2.766 & 11.252 & 0.251 & 12.602 & 0.358 & 8.907 & 0.270 & 1.367 & 0.119 & 239.989\\
HELP\_J100258.541& 150.744 & 1.884 & 1.922 & 10.905 & 0.275 & 12.419 & 0.286 & 8.515 & 0.223 & 1.376 & 0.352 & 343.523\\
HELP\_J095756.967& 149.487 & 2.549 & 2.109 & 11.448 & 0.339 & 12.783 & 0.398 & 8.695 & 0.307 & 1.415 & 0.125 & 239.989\\
HELP\_J100030.383& 150.127 & 2.696 & 2.323 & 11.221 & 0.266 & 12.557 & 0.422 & 8.890 & 0.142 & 1.415 & 0.125 & 239.989\\
HELP\_J100200.630 & 150.503 & 2.357 & 1.242 & 11.347 & 0.379 & 12.049 & 0.189 & 8.043 & 0.330 & 1.418 & 0.336 & 343.524\\
HELP\_J095745.510 & 149.440 & 2.550 & 3.118 & 11.469 & 0.306 & 12.819 & 0.381 & 8.729 & 0.328 & 1.426 & 0.126 & 239.989\\
HELP\_J095817.712 & 149.574 & 2.086 & 1.224 & 10.993 & 0.242 & 11.702 & 0.099 & 8.810 & 0.327 & 1.433 & 0.303 & 343.525\\
HELP\_J095853.694& 149.724 & 2.281 & 3.308 & 11.155 & 0.256 & 12.577 & 0.321 & 8.818 & 0.259 & 1.437 & 0.177 & 239.989\\
HELP\_J095914.020 & 149.808 & 2.720& 2.955 & 11.513 & 0.214 & 12.848 & 0.387 & 8.777 & 0.335 & 1.438 & 0.115 & 239.989\\
HELP\_J100135.882& 150.400 & 2.465 & 1.205 & 11.120 & 0.253 & 11.829 & 0.156 & 8.913 & 0.247 & 1.438 & 0.335 & 343.524\\
HELP\_J095806.241 & 149.526 & 2.777 & 2.076 & 11.245 & 0.359 & 12.247 & 0.235 & 8.817 & 0.136 & 1.443 & 0.342 & 343.528\\
HELP\_J095855.808 & 149.733 & 2.483 & 1.354 & 11.284 & 0.244 & 11.923 & 0.157 & 8.508 & 0.200 & 1.449 & 0.417 & 338.855\\
HELP\_J100157.447 & 150.489 & 1.822 & 1.969 & 11.179 & 0.216 & 12.417 & 0.365 & 8.707 & 0.038 & 1.456 & 0.261 & 343.523\\
HELP\_J100133.566 & 150.390 & 2.025 & 2.445 & 11.131 & 0.209 & 12.664 & 0.336 & 8.920 & 0.284 & 1.472 & 0.131 & 245.029\\
HELP\_J095906.418 & 149.777 & 1.761 & 2.282 & 10.642 & 0.301 & 12.177 & 0.311 & 8.728 & 0.143 & 1.489 & 0.311 & 343.471\\
HELP\_J095918.368 & 149.827 & 2.019 & 2.307 & 10.541 & 0.175 & 12.507 & 0.349 & 8.997 & 0.241 & 1.489 & 0.212 & 239.984\\
HELP\_J100027.138& 150.113 & 2.528 & 2.017 & 10.944 & 0.307 & 12.479 & 0.355 & 9.162 & 0.282 & 1.490 & 0.270 & 239.984\\
HELP\_J100207.354 & 150.531 & 2.776 & 1.352 & 11.161 & 0.115 & 12.310 & 0.251 & 9.068 & 0.132 & 1.509 & 0.286 & 343.484\\
HELP\_J100025.784 & 150.107 & 2.646 & 1.544 & 10.922 & 0.191 & 12.071 & 0.194 & 8.878 & 0.391 & 1.532 & 0.253 & 343.524\\
HELP\_J095918.914 & 149.829 & 1.928 & 1.424 & 11.022 & 0.300 & 12.171 & 0.240 & 8.415 & 0.271 & 1.536 & 0.264 & 343.485\\
HELP\_J100014.754 & 150.061 & 2.379 & 3.333 & 10.452 & 0.305 & 11.835 & 0.134 & 7.632 & 0.277 & 1.558 & 0.311 & 344.772\\
HELP\_J100149.675 & 150.457 & 1.934 & 2.133 & 10.574 & 0.269 & 12.238 & 0.236 & 8.173 & 0.186 & 1.560 & 0.302 & 343.532\\
HELP\_J100029.499 & 150.123 & 2.361 & 2.055 & 10.995 & 0.305 & 12.161 & 0.215 & 8.087 & 0.333 & 1.568 & 0.294 & 343.523\\
HELP\_J100125.263 & 150.355 & 1.959 & 2.057 & 10.778 & 0.132 & 12.081 & 0.195 & 8.791 & 0.198 & 1.568 & 0.281 & 343.532\\
HELP\_J095825.001 & 149.604 & 2.275 & 2.007 & 10.568 & 0.268 & 12.103 & 0.201 & 8.821 & 0.278 & 1.576 & 0.297 & 343.524\\
HELP\_J100123.782 & 150.349 & 1.705 & 2.345 & 10.579 & 0.324 & 12.371 & 0.286 & 8.607 & 0.304 & 1.580 & 0.261 & 343.532\\
HELP\_J100038.748 & 150.161 & 2.691 & 1.989 & 10.916 & 0.364 & 11.946 & 0.213 & 8.827 & 0.202 & 1.598 & 0.284 & 343.485\\
HELP\_J100101.205 & 150.255 & 1.858 & 1.200 & 11.253 & 0.321 & 12.215 & 0.228 & 8.843 & 0.333 & 1.601 & 0.255 & 343.524\\
HELP\_J095951.961& 149.966 & 1.779 & 1.438 & 10.700 & 0.347 & 11.978 & 0.253 & 9.021 & 0.253 & 1.606 & 0.268 & 343.485\\
HELP\_J095848.358& 149.701 & 2.087 & 1.483 & 10.922 & 0.310 & 12.064 & 0.193 & 9.112 & 0.234 & 1.618 & 0.309 & 343.533\\
HELP\_J100149.235 & 150.455 & 2.083 & 2.828 & 10.978 & 0.358 & 12.246 & 0.286 & 8.194 & 0.259 & 1.623 & 0.353 & 343.520\\
HELP\_J095845.122 & 149.688 & 2.242 & 2.207 & 11.468 & 0.353 & 12.615 & 0.328 & 9.182 & 0.328 & 1.624 & 0.338 & 239.984\\
HELP\_J100114.838& 150.312 & 2.196 & 1.196 & 10.437 & 0.108 & 12.421 & 0.349 & 8.820 & 0.398 & 1.626 & 0.274 & 343.484\\
HELP\_J100005.112 & 150.021 & 1.922 & 2.626 & 11.415 & 0.375 & 12.558 & 0.314 & 9.024 & 0.139 & 1.635 & 0.177 & 239.984\\
HELP\_J100037.573 & 150.157 & 1.825 & 3.129 & 11.401 & 0.402 & 12.751 & 0.363 & 8.732 & 0.272 & 1.632 & 0.135 & 239.989\\
HELP\_J100015.701 & 150.065 & 1.746 & 1.217 & 11.624 & 0.380 & 12.326 & 0.257 & 8.863 & 0.361 & 1.651 & 0.279 & 343.524\\
HELP\_J100011.406& 150.048 & 2.621 & 1.510 & 11.256 & 0.354 & 12.019 & 0.180 & 8.847 & 0.183 & 1.654 & 0.268 & 343.524\\
HELP\_J100054.307 & 150.226 & 2.113 & 1.222 & 11.330 & 0.336 & 11.969 & 0.168 & 8.641 & 0.289 & 1.665 & 0.281 & 343.523\\
HELP\_J100132.959& 150.387 & 1.936 & 2.641 & 10.784 & 0.328 & 12.724 & 0.353 & 8.996 & 0.290 & 1.669 & 0.118 & 239.989\\
HELP\_J095958.439& 149.993 & 2.239 & 2.256 & 11.129 & 0.340 & 12.091 & 0.233 & 8.831 & 0.155 & 1.675 & 0.357 & 343.476\\
HELP\_J095743.912& 149.433 & 1.693 & 3.861 & 10.500 & 0.237 & 12.165 & 0.384 & 9.082 & 0.156 & 1.677 & 0.406 & 343.52\\
HELP\_J095930.531& 149.877 & 2.284 & 2.756 & 10.721 & 0.221 & 12.637 & 0.371 & 9.002 & 0.276 & 1.686 & 0.136 & 239.988\\
HELP\_J100139.852 & 150.416 & 2.558 & 1.091 & 11.198 & 0.327 & 12.066 & 0.188 & 8.808 & 0.277 & 1.687 & 0.337 & 343.484\\
HELP\_J095958.003 & 149.992 & 2.694 & 2.013 & 10.946 & 0.358 & 12.089 & 0.244 & 8.787 & 0.137 & 1.688 & 0.348 & 343.532\\
HELP\_J100036.609& 150.153 & 1.769 & 1.556 & 11.053 & 0.266 & 12.202 & 0.225 & 8.488 & 0.350 & 1.693 & 0.311 & 343.523\\
HELP\_J095849.300& 149.705 & 2.217 & 2.316 & 10.907 & 0.343 & 12.442 & 0.300 & 9.215 & 0.164 & 1.712 & 0.221 & 232.986\\
HELP\_J100145.981& 150.442 & 2.129 & 1.507 & 11.468 & 0.366 & 12.230 & 0.236 & 9.004 & 0.118 & 1.717 & 0.364 & 343.523\\
HELP\_J100151.278 & 150.464 & 2.786 & 3.545 & 10.897 & 0.193 & 12.426 & 0.334 & 8.625 & 0.264 & 1.717 & 0.149 & 245.025\\
HELP\_J095958.117& 149.992 & 2.693 & 2.128 & 10.783 & 0.278 & 11.99 & 0.329 & 8.818 & 0.155 & 1.721 & 0.298 & 343.532\\
HELP\_J095741.106 & 149.421 & 2.041 & 1.476 & 10.853 & 0.273 & 12.517 & 0.307 & 8.813 & 0.424 & 1.728 & 0.237 & 343.485\\
HELP\_J100106.800& 150.278 & 2.259 & 3.171 & 10.635 & 0.345 & 12.577 & 0.342 & 8.566 & 0.178 & 1.732 & 0.112 & 239.989\\
HELP\_J100013.477& 150.056 & 1.618 & 3.304 & 10.974 & 0.298 & 12.781 & 0.47 & 9.243 & 0.154 & 1.735 & 0.113 & 239.989\\
HELP\_J100025.483 & 150.106 & 2.053 & 3.333 & 10.581 & 0.196 & 12.075 & 0.365 & 9.151 & 0.281 & 1.747 & 0.124 & 239.989\\
HELP\_J100214.729 & 150.561 & 2.346 & 1.600 & 10.815 & 0.224 & 12.023 & 0.181 & 8.408 & 0.083 & 1.749 & 0.208 & 343.484\\
HELP\_J095958.275 & 149.993 & 2.601 & 3.962 & 11.428 & 0.257 & 12.763 & 0.365 & 8.895 & 0.264 & 1.761 & 0.104 & 239.989\\
HELP\_J100159.766& 150.499 & 1.724 & 1.164 & 11.479 & 0.319 & 12.137 & 0.209 & 8.843 & 0.270 & 1.762 & 0.387 & 343.524\\
HELP\_J100104.393& 150.268 & 2.749 & 2.019 & 10.708 & 0.327 & 12.648 & 0.335 & 8.392 & 0.276 & 1.764 & 0.283 & 343.532\\
HELP\_J100114.528 & 150.311 & 2.452 & 2.761 & 10.556 & 0.327 & 12.126 & 0.308 & 9.187 & 0.303 & 1.771 & 0.211 & 239.984\\
HELP\_J095911.562 & 149.798 & 2.390 & 1.461 & 11.127 & 0.224 & 12.335 & 0.267 & 8.738 & 0.333 & 1.775 & 0.244 & 343.485\\
HELP\_J100011.577& 150.048 & 2.251 & 3.314 & 11.335 & 0.282 & 12.862 & 0.431 & 8.996 & 0.308 & 1.795 & 0.106 & 239.989\\
HELP\_J095929.234 & 149.872 & 2.212 & 4.866 & 11.392 & 0.185 & 12.814 & 0.378 & 8.512 & 0.261 & 1.798 & 0.118 & 239.989\\
HELP\_J095904.442& 149.768 & 2.110 & 2.106 & 10.403 & 0.326 & 12.337 & 0.391 & 8.703 & 0.283 & 1.806 & 0.281 & 343.533\\
HELP\_J100226.177 & 150.609 & 2.208 & 1.593 & 11.074 & 0.344 & 12.036 & 0.186 & 8.947 & 0.252 & 1.814 & 0.226 & 343.523\\
HELP\_J100036.986 & 150.154 & 1.804 & 1.104 & 11.251 & 0.277 & 12.393 & 0.271 & 8.798 & 0.388 & 1.815 & 0.221 & 343.524\\
HELP\_J100136.641 & 150.403 & 2.611 & 2.611 & 11.047 & 0.162 & 12.383 & 0.366 & 8.981 & 0.126 & 1.820 & 0.138 & 239.989\\
HELP\_J100053.869 & 150.224 & 2.433 & 1.635 & 10.767 & 0.330 & 12.517 & 0.310 & 8.743 & 0.289 & 1.826 & 0.263 & 343.532\\
HELP\_J095816.952 & 149.571 & 2.125 & 1.312 & 10.980 & 0.350 & 12.129 & 0.224 & 8.979 & 0.259 & 1.838 & 0.260 & 343.485\\
HELP\_J100106.841& 150.278 & 1.877 & 1.072 & 11.416 & 0.344 & 12.118 & 0.204 & 8.951 & 0.326 & 1.839 & 0.266 & 343.524\\
HELP\_J100015.443 & 150.064 & 1.746 & 4.006 & 11.478 & 0.291 & 13.239 & 0.621 & 9.372 & 0.169 & 1.856 & 0.419 & 343.524\\
HELP\_J095933.288 & 149.889 & 2.142 & 3.813 & 11.015 & 0.274 & 12.437 & 0.287 & 9.025 & 0.159 & 1.860 & 0.202 & 239.984\\
HELP\_J100053.546 & 150.223 & 2.695 & 1.290 & 11.230 & 0.238 & 12.438 & 0.285 & 8.555 & 0.371 & 1.863 & 0.343 & 343.496\\
HELP\_J100007.542 & 150.031 & 2.196 & 3.501 & 10.629 & 0.266 & 12.570 & 0.316 & 8.951 & 0.199 & 1.871 & 0.121 & 240.02\\
HELP\_J100033.356 & 150.139 & 2.434 & 3.006 & 11.122 & 0.308 & 12.721 & 0.381 & 8.885 & 0.221 & 1.890 & 0.286 & 239.984\\
HELP\_J100134.456 & 150.394 & 2.361 & 2.174 & 11.273 & 0.165 & 12.268 & 0.236 & 8.842 & 0.208 & 1.891 & 0.361 & 343.528\\
HELP\_J095902.080 & 149.759 & 1.994 & 4.292 & 10.847 & 0.181 & 12.230 & 0.295 & 9.289 & 0.213 & 1.897 & 0.306 & 239.984\\
HELP\_J095943.649 & 149.932 & 2.228 & 2.603 & 11.428 & 0.308 & 12.570 & 0.318 & 9.161 & 0.252 & 1.908 & 0.270 & 239.984\\
HELP\_J095843.048& 149.679 & 2.208 & 0.956 & 10.819 & 0.146 & 12.154 & 0.213 & 8.841 & 0.229 & 1.917 & 0.251 & 343.485\\
HELP\_J100200.846& 150.503 & 1.812 & 1.162 & 10.825 & 0.289 & 12.486 & 0.309 & 8.778 & 0.364 & 1.934 & 0.229 & 343.523\\
HELP\_J095907.955& 149.783 & 2.372 & 2.203 & 10.832 & 0.207 & 12.040 & 0.315 & 8.584 & 0.117 & 1.940 & 0.335 & 341.950\\
HELP\_J100139.450 & 150.414 & 2.801 & 1.473 & 11.383 & 0.356 & 12.199 & 0.229 & 8.793 & 0.232 & 1.940 & 0.233 & 343.524\\
HELP\_J095940.576 & 149.919 & 2.761 & 1.130 & 10.664 & 0.173 & 12.047 & 0.188 & 8.816 & 0.206 & 1.952 & 0.327 & 343.496\\
HELP\_J095905.248 & 149.772 & 2.050 & 3.537 & 10.900 & 0.320 & 12.840 & 0.389 & 8.941 & 0.333 & 1.953 & 0.132 & 239.989\\
HELP\_J100122.061 & 150.342 & 1.880 & 2.249 & 10.570 & 0.229 & 12.583 & 0.392 & 8.693 & 0.345 & 1.958 & 0.285 & 343.532\\
HELP\_J095802.380& 149.510 & 2.102 & 1.943 & 10.940 & 0.303 & 12.362 & 0.264 & 8.419 & 0.259 & 1.967 & 0.407 & 341.950\\
HELP\_J100209.789& 150.541 & 2.559 & 3.248 & 11.417 & 0.417 & 12.579 & 0.415 & 8.786 & 0.266 & 1.973 & 0.298 & 343.527\\
HELP\_J095920.628& 149.836 & 2.114 & 3.161 & 11.365 & 0.337 & 12.700 & 0.349 & 8.921 & 0.237 & 1.979 & 0.124 & 239.989\\
HELP\_J100126.537 & 150.361 & 2.002 & 2.157 & 11.305 & 0.247 & 12.896 & 0.431 & 8.821 & 0.263 & 1.991 & 0.235 & 239.984\\
HELP\_J100012.924 & 150.054 & 2.576 & 2.902 & 11.122 & 0.263 & 12.264 & 0.243 & 9.230 & 0.227 & 1.998 & 0.117 & 239.989\\
HELP\_J095933.505& 149.890 & 2.648 & 2.446 & 11.291 & 0.299 & 12.434 & 0.283 & 8.488 & 0.352 & 2.008 & 0.348 & 343.485\\
HELP\_J095815.334 & 149.564 & 2.546 & 4.854 & 11.012 & 0.247 & 12.539 & 0.420 & 8.966 & 0.138 & 2.045 & 0.398 & 341.950\\
HELP\_J100131.884& 150.383 & 2.194 & 2.459 & 11.293 & 0.338 & 12.368 & 0.446 & 9.007 & 0.087 & 2.050 & 0.212 & 239.984\\
HELP\_J100128.115 & 150.367 & 2.828 & 0.602 & 10.371 & 0.305 & 11.549 & 0.063 & 8.312 & 0.326 & 2.064 & 0.202 & 223.324\\
HELP\_J095824.312& 149.601 & 2.254 & 1.640 & 10.964 & 0.284 & 12.462 & 0.315 & 8.778 & 0.319 & 2.074 & 0.387 & 343.485\\
HELP\_J100209.548& 150.540 & 2.559 & 1.825 & 11.468 & 0.362 & 12.430 & 0.282 & 8.701 & 0.328 & 2.099 & 0.303 & 343.527\\
HELP\_J095935.965& 149.900 & 2.657 & 1.310 & 11.430 & 0.311 & 12.392 & 0.274 & 8.895 & 0.345 & 2.175 & 0.352 & 343.524\\
HELP\_J095853.215 & 149.722 & 1.965 & 1.554 & 11.329 & 0.312 & 12.471 & 0.292 & 8.769 & 0.333 & 2.208 & 0.229 & 343.524\\
HELP\_J095852.879 & 149.720 & 1.967 & 2.286 & 11.481 & 0.349 & 12.443 & 0.286 & 8.905 & 0.315 & 2.235 & 0.448 & 343.524\\
HELP\_J100150.374 & 150.460 & 1.673 & 1.811 & 10.642 & 0.256 & 12.554 & 0.357 & 8.797 & 0.259 & 2.235 & 0.475 & 343.532\\
HELP\_J100045.470 & 150.189 & 2.037 & 1.188 & 11.364 & 0.312 & 12.128 & 0.237 & 8.365 & 0.260 & 2.254 & 0.412 & 343.520\\
HELP\_J095800.833& 149.503 & 2.507 & 1.542 & 10.954 & 0.302 & 12.386 & 0.272 & 8.329 & 0.259 & 2.257 & 0.331 & 343.524\\
HELP\_J095846.439 & 149.694 & 1.723 & 3.513 & 10.767 & 0.249 & 12.736 & 0.359 & 9.047 & 0.279 & 2.272 & 0.130 & 239.988\\
HELP\_J095924.433 & 149.852 & 2.718 & 1.740 & 11.246 & 0.302 & 12.104 & 0.201 & 8.551 & 0.067 & 2.284 & 0.297 & 343.485\\
HELP\_J100109.148 & 150.288 & 2.381 & 1.595 & 11.530 & 0.377 & 12.351 & 0.263 & 8.621 & 0.231 & 2.301 & 0.485 & 343.484\\
HELP\_J100124.605 & 150.353 & 2.005 & 1.884 & 10.793 & 0.156 & 12.777 & 0.368 & 8.641 & 0.315 & 2.306 & 0.275 & 343.471\\
HELP\_J095941.639 & 149.924 & 1.904 & 2.150 & 10.691 & 0.204 & 11.994 & 0.503 & 9.069 & 0.217 & 2.365 & 0.361 & 343.471\\
HELP\_J100151.559 & 150.465 & 2.653 & 2.024 & 11.227 & 0.319 & 12.173 & 0.213 & 8.522 & 0.268 & 2.375 & 0.478 & 343.528\\
HELP\_J100114.795 & 150.312 & 2.451 & 3.005 & 10.576 & 0.244 & 11.879 & 0.339 & 8.689 & 0.169 & 2.382 & 0.356 & 343.473\\
HELP\_J100003.925& 150.016 & 2.321 & 2.080 & 10.904 & 0.288 & 12.430 & 0.342 & 8.930 & 0.141 & 2.390 & 0.338 & 343.519\\
HELP\_J100008.944& 150.037 & 2.670 & 1.847 & 10.945 & 0.307 & 12.877 & 0.394 & 9.223 & 0.306 & 2.405 & 0.157 & 228.399\\
HELP\_J100207.702& 150.532 & 2.184 & 2.049 & 11.164 & 0.316 & 12.341 & 0.318 & 8.938 & 0.175 & 2.442 & 0.362 & 343.523\\
HELP\_J095836.817& 149.653 & 1.723 & 2.639 & 10.625 & 0.230 & 12.591 & 0.426 & 8.804 & 0.261 & 2.447 & 0.326 & 343.471\\
HELP\_J100111.569& 150.298 & 2.478 & 2.784 & 11.610 & 0.355 & 12.787 & 0.396 & 8.929 & 0.298 & 2.477 & 0.204 & 239.984\\
HELP\_J100019.764& 150.082 & 2.535 & 3.206 & 10.870 & 0.205 & 12.808 & 0.566 & 9.506 & 0.134 & 2.496 & 0.203 & 239.984\\
HELP\_J095927.133& 149.863 & 1.879 & 1.908 & 10.666 & 0.108 & 12.605 & 0.369 & 8.912 & 0.281 & 2.524 & 0.382 & 343.533\\
HELP\_J095904.404& 149.768 & 2.220 & 1.787 & 11.002 & 0.291 & 12.318 & 0.320 & 9.061 & 0.154 & 2.530 & 0.385 & 343.485\\
HELP\_J100058.111 & 150.242 & 2.237 & 1.352 & 11.020 & 0.155 & 12.442 & 0.285 & 8.804 & 0.317 & 2.549 & 0.265 & 343.523\\
HELP\_J095808.900& 149.537 & 1.864 & 1.972 & 11.175 & 0.167 & 12.446 & 0.370 & 8.835 & 0.238 & 2.563 & 0.314 & 343.533\\
HELP\_J100033.900& 150.141 & 2.676 & 2.128 & 11.325 & 0.322 & 12.321 & 0.250 & 8.924 & 0.194 & 2.565 & 0.351 & 343.528\\
HELP\_J100122.284& 150.343 & 1.945 & 3.184 & 10.090 & 0.251 & 12.427 & 0.280 & 8.802 & 0.236 & 2.578 & 0.358 & 343.474\\
HELP\_J100035.405& 150.148 & 2.592 & 1.422 & 11.126 & 0.308 & 12.232 & 0.230 & 9.017 & 0.319 & 2.582 & 0.281 & 343.524\\
HELP\_J100120.656 & 150.336 & 2.440 & 2.843 & 10.577 & 0.277 & 11.910 & 0.272 & 8.859 & 0.200 & 2.588 & 0.283 & 343.532\\
HELP\_J100117.348& 150.322 & 2.176 & 2.707 & 11.249 & 0.232 & 12.584 & 0.425 & 8.606 & 0.178 & 2.591 & 0.348 & 343.528\\
HELP\_J100123.867& 150.349 & 1.937 & 1.120 & 10.993 & 0.299 & 12.129 & 0.205 & 8.837 & 0.139 & 2.594 & 0.270 & 343.485\\
HELP\_J100231.530& 150.631 & 2.477 & 1.460 & 11.338 & 0.341 & 12.480 & 0.296 & 8.749 & 0.266 & 2.636 & 0.270 & 343.524\\
HELP\_J100108.962& 150.287 & 2.382 & 1.928 & 11.527 & 0.325 & 12.528 & 0.306 & 8.901 & 0.212 & 2.646 & 0.305 & 343.484\\
HELP\_J100146.978& 150.446 & 2.413 & 2.577 & 10.972 & 0.183 & 12.275 & 0.355 & 8.915 & 0.133 & 2.650 & 0.318 & 343.528\\
HELP\_J095919.798 & 149.832 & 2.066 & 1.139 & 11.233 & 0.292 & 12.471 & 0.293 & 9.135 & 0.502 & 2.659 & 0.414 & 343.520\\
HELP\_J100137.356 & 150.406 & 2.151 & 2.847 & 10.585 & 0.295 & 12.547 & 0.349 & 8.866 & 0.280 & 2.686 & 0.283 & 343.532\\
HELP\_J100136.147& 150.401 & 1.862 & 3.186 & 11.886 & 0.418 & 13.048 & 0.436 & 9.050 & 0.324 & 2.694 & 0.114 & 239.989\\
HELP\_J100150.370 & 150.460 & 1.672 & 2.754 & 11.095 & 0.211 & 12.658 & 0.338 & 8.582 & 0.292 & 2.716 & 0.276 & 343.523\\
HELP\_J100238.754 & 150.661 & 2.797 & 1.026 & 11.267 & 0.321 & 11.969 & 0.201 & 9.026 & 0.214 & 2.740 & 0.404 & 343.496\\
HELP\_J100235.885 & 150.650 & 2.057 & 2.321 & 11.108 & 0.171 & 12.530 & 0.307 & 8.887 & 0.276 & 2.764 & 0.459 & 343.527\\
HELP\_J100143.858 & 150.433 & 2.674 & 2.200 & 11.338 & 0.347 & 12.356 & 0.265 & 9.088 & 0.296 & 2.841 & 0.365 & 343.528\\
HELP\_J095943.279 & 149.930 & 1.769 & 4.641 & 10.790 & 0.233 & 12.732 & 0.355 & 9.104 & 0.235 & 2.875 & 0.140 & 245.031\\
HELP\_J095839.784& 149.666 & 1.914 & 1.851 & 11.295 & 0.218 & 12.631 & 0.334 & 8.998 & 0.315 & 2.901 & 0.299 & 343.471\\
HELP\_J100228.856 & 150.620 & 2.690 & 3.317 & 11.014 & 0.328 & 12.868 & 0.429 & 9.196 & 0.278 & 2.904 & 0.113 & 239.989\\
HELP\_J100254.909& 150.729 & 2.405 & 2.215 & 11.382 & 0.339 & 12.717 & 0.353 & 8.691 & 0.301 & 2.918 & 0.351 & 343.527\\
HELP\_J095740.909 & 149.420 & 2.042 & 1.471 & 10.811 & 0.276 & 11.960 & 0.183 & 8.740 & 0.256 & 2.925 & 0.324 & 343.485\\
HELP\_J095906.289 & 149.776 & 2.677 & 2.755 & 10.415 & 0.348 & 12.752 & 0.364 & 8.634 & 0.317 & 2.943 & 0.313 & 343.533\\
HELP\_J100023.546 & 150.098 & 2.166 & 1.185 & 11.247 & 0.248 & 12.011 & 0.180 & 8.173 & 0.194 & 2.955 & 0.621 & 343.520\\
HELP\_J100135.669 & 150.399 & 2.188 & 1.015 & 11.114 & 0.312 & 12.256 & 0.240 & 9.145 & 0.429 & 3.017 & 0.413 & 343.520\\
HELP\_J095922.232 & 149.843 & 2.522 & 1.789 & 11.076 & 0.307 & 12.225 & 0.418 & 9.069 & 0.225 & 3.022 & 0.287 & 343.524\\
HELP\_J095942.584 & 149.927 & 1.917 & 1.948 & 11.176 & 0.212 & 12.425 & 0.399 & 9.388 & 0.223 & 3.030 & 0.444 & 239.984\\
HELP\_J100121.413 & 150.339 & 2.173 & 2.882 & 11.393 & 0.205 & 12.728 & 0.356 & 8.649 & 0.333 & 3.092 & 0.307 & 343.528\\
HELP\_J100123.949 & 150.350 & 1.875 & 1.448 & 11.119 & 0.255 & 12.269 & 0.242 & 9.021 & 0.262 & 3.094 & 0.261 & 343.484\\
HELP\_J100122.958& 150.346 & 2.335 & 2.616 & 11.218 & 0.192 & 12.640 & 0.336 & 9.163 & 0.289 & 3.126 & 0.223 & 239.984\\
HELP\_J095940.867 & 149.920 & 2.020 & 1.870 & 11.271 & 0.282 & 12.478 & 0.293 & 8.966 & 0.302 & 3.174 & 0.336 & 341.950\\
HELP\_J100129.520& 150.373 & 2.156 & 1.970 & 10.556 & 0.190 & 12.488 & 0.298 & 8.949 & 0.128 & 3.208 & 0.351 & 343.532\\
HELP\_J100045.399 & 150.189 & 2.572 & 2.683 & 10.729 & 0.321 & 12.669 & 0.340 & 8.855 & 0.338 & 3.262 & 0.330 & 343.532\\
HELP\_J095859.660 & 149.749 & 2.235 & 2.441 & 11.274 & 0.336 & 12.609 & 0.325 & 8.960 & 0.271 & 3.270 & 0.293 & 343.528\\
HELP\_J100028.715 & 150.120 & 2.534 & 3.175 & 10.966 & 0.294 & 12.932 & 0.408 & 9.242 & 0.273 & 3.285 & 0.114 & 239.989\\
HELP\_J095931.748 & 149.882 & 2.507 & 2.371 & 11.230 & 0.340 & 12.231 & 0.234 & 9.125 & 0.167 & 3.290& 0.537 & 343.478\\
HELP\_J095927.208& 149.863 & 2.618 & 1.308 & 10.910 & 0.319 & 12.245 & 0.350 & 9.059 & 0.229 & 3.294 & 0.323 & 343.496\\
HELP\_J100121.978 & 150.341 & 1.946 & 2.709 & 11.228 & 0.345 & 12.669 & 0.416 & 8.839 & 0.261 & 3.389 & 0.451 & 343.474\\
HELP\_J100035.300 & 150.147 & 2.731 & 2.384 & 11.587 & 0.193 & 13.107 & 0.455 & 9.400 & 0.318 & 3.448 & 0.216 & 239.984\\
HELP\_J100056.324& 150.235 & 2.144 & 1.575 & 11.110 & 0.383 & 12.072 & 0.197 & 9.170 & 0.440 & 3.452 & 0.381 & 343.524\\
HELP\_J095849.961& 149.708 & 1.768 & 2.562 & 11.020 & 0.351 & 12.323 & 0.344 & 9.019 & 0.280 & 3.471 & 0.390 & 343.533\\
HELP\_J100115.213 & 150.313 & 2.716 & 3.508 & 10.416 & 0.364 & 12.351 & 0.262 & 9.436 & 0.306 & 3.516 & 0.119 & 239.989\\
HELP\_J100010.187& 150.042 & 2.527 & 1.230 & 11.012 & 0.261 & 11.722 & 0.130 & 8.798 & 0.359 & 3.534 & 0.632 & 343.496\\
HELP\_J100238.844& 150.662 & 1.715 & 2.702 & 11.384 & 0.341 & 12.547 & 0.310 & 8.543 & 0.275 & 3.588 & 0.339 & 343.527\\
HELP\_J100039.644& 150.165 & 1.679 & 2.625 & 11.223 & 0.195 & 12.366 & 0.267 & 9.118 & 0.200& 3.625 & 0.340 & 343.532\\
HELP\_J095834.990 & 149.646 & 2.389 & 1.405 & 11.483 & 0.275 & 12.661 & 0.514 & 9.021 & 0.182 & 3.636 & 0.284 & 343.485\\
HELP\_J100103.772& 150.266 & 2.703 & 3.808 & 11.215 & 0.203 & 12.909 & 0.478 & 9.222 & 0.271 & 3.655 & 0.266 & 233.026\\
HELP\_J095825.009& 149.604 & 1.716 & 2.014 & 11.243 & 0.253 & 12.481 & 0.382 & 9.230 & 0.307 & 3.696 & 0.341 & 343.471\\
HELP\_J095930.653 & 149.878 & 2.576 & 1.397 & 11.575 & 0.275 & 12.717 & 0.353 & 9.001 & 0.338 & 3.769 & 0.327 & 343.485\\
HELP\_J100151.520& 150.465 & 1.710 & 2.664 & 10.700 & 0.356 & 12.645 & 0.334 & 9.203 & 0.345 & 3.785 & 0.359 & 343.532\\
HELP\_J095845.947& 149.691 & 2.725 & 3.172 & 10.888 & 0.273 & 12.871 & 0.397 & 9.368 & 0.108 & 3.836 & 0.124 & 239.989\\
HELP\_J095959.334& 149.997 & 2.578 & 3.031 & 11.298 & 0.422 & 12.476 & 0.310 & 8.832 & 0.073 & 3.86 & 0.158 & 239.988\\
HELP\_J100235.717& 150.649 & 2.055 & 2.231 & 11.201 & 0.234 & 12.365 & 0.384 & 9.102 & 0.142 & 3.879 & 0.328 & 343.527\\
HELP\_J100151.719& 150.465 & 2.430 & 1.633 & 11.179 & 0.203 & 12.417 & 0.277 & 8.822 & 0.167 & 3.953 & 0.400 & 341.950\\
HELP\_J095902.171 & 149.759 & 2.471 & 3.139 & 10.470 & 0.220 & 12.660& 0.432 & 9.005 & 0.277 & 3.961 & 0.339 & 343.533\\
HELP\_J095939.126& 149.913 & 2.540 & 2.718 & 10.960 & 0.259 & 12.446 & 0.356 & 9.078 & 0.186 & 3.983 & 0.653 & 338.854\\
HELP\_J095759.256& 149.497 & 2.456 & 3.092 & 10.966 & 0.115 & 12.950 & 0.412 & 8.608 & 0.248 & 4.032 & 0.296 & 343.533\\
HELP\_J100004.344& 150.018 & 2.350 & 3.792 & 10.856 & 0.303 & 12.822 & 0.377 & 8.754 & 0.268 & 4.032 & 0.252 & 343.478\\
HELP\_J100101.263& 150.255 & 2.467 & 3.142 & 10.817 & 0.284 & 12.755 & 0.375 & 8.640 & 0.28 & 4.032 & 0.331 & 343.532\\
HELP\_J095819.782 & 149.582 & 2.603 & 2.145 & 10.035 & 0.255 & 11.105 & 0.005 & 8.273 & 0.405 & 4.099 & 0.111 & 239.989\\
HELP\_J095814.445 & 149.560 & 2.335 & 4.044 & 11.520 & 0.332 & 12.683 & 0.421 & 8.830 & 0.204 & 4.105 & 0.452 & 343.525\\
HELP\_J100015.634& 150.065 & 2.264 & 3.306 & 10.547 & 0.239 & 12.513 & 0.401 & 9.546 & 0.262 & 4.115 & 0.220 & 239.984\\
HELP\_J100209.648& 150.54 & 2.609 & 4.041 & 11.448 & 0.256 & 12.974 & 0.618 & 9.097 & 0.191 & 4.125 & 0.147 & 239.989\\
HELP\_J095837.963& 149.658 & 2.236 & 2.196 & 10.495 & 0.219 & 12.428 & 0.428 & 9.547 & 0.225 & 4.200 & 0.239 & 239.984\\
HELP\_J100226.247& 150.609 & 2.208 & 1.962 & 11.330 & 0.361 & 12.472 & 0.294 & 9.290 & 0.139 & 4.230 & 0.299 & 343.523\\
HELP\_J100206.487 & 150.527 & 2.154 & 2.564 & 10.919 & 0.228 & 12.612 & 0.422 & 9.346 & 0.331 & 4.324 & 0.530 & 343.520\\
HELP\_J100024.366& 150.102 & 1.728 & 1.618 & 10.871 & 0.159 & 12.809 & 0.562 & 9.251 & 0.235 & 4.432 & 0.244 & 343.523\\
HELP\_J095845.278& 149.689 & 2.261 & 2.868 & 10.704 & 0.258 & 12.642 & 0.406 & 9.153 & 0.154 & 4.509 & 0.310 & 343.533\\
HELP\_J100041.578& 150.173 & 2.464 & 2.306 & 11.486 & 0.276 & 12.754 & 0.366 & 9.141 & 0.359 & 4.516 & 0.365 & 343.528\\
HELP\_J100012.929& 150.054 & 2.203 & 2.917 & 11.375 & 0.235 & 12.797 & 0.374 & 8.737 & 0.239 & 4.640 & 0.453 & 344.772\\
HELP\_J100158.959& 150.496 & 2.116 & 2.176 & 11.436 & 0.353 & 12.512 & 0.352 & 9.170 & 0.218 & 4.714 & 0.407 & 343.520\\
HELP\_J100120.835& 150.337 & 2.440 & 2.268 & 10.693 & 0.257 & 12.263 & 0.477 & 9.288 & 0.188 & 4.835 & 0.312 & 343.532\\
HELP\_J095854.192& 149.726 & 2.279 & 2.964 & 11.114 & 0.317 & 12.694 & 0.361 & 8.910 & 0.312 & 4.863 & 0.389 & 343.474\\
HELP\_J095933.431& 149.889 & 2.396 & 2.142 & 11.115 & 0.288 & 12.450 & 0.406 & 9.360 & 0.231 & 4.966 & 0.397 & 343.519\\
HELP\_J100117.738& 150.324 & 2.752 & 2.719 & 11.400 & 0.344 & 12.564 & 0.434 & 9.014 & 0.102 & 5.034 & 0.316 & 343.528\\
HELP\_J095941.266& 149.922 & 2.290 & 2.334 & 11.291 & 0.187 & 12.535 & 0.306 & 9.375 & 0.221 & 5.063 & 0.279 & 343.524\\
HELP\_J100022.824& 150.095 & 1.861 & 1.989 & 11.280 & 0.306 & 12.242 & 0.238 & 9.355 & 0.317 & 5.109 & 0.286 & 343.523\\
HELP\_J100124.470& 150.352 & 1.938 & 2.584 & 11.436 & 0.309 & 12.579 & 0.320 & 9.053 & 0.205 & 5.109 & 0.609 & 343.485\\
HELP\_J100003.861& 150.016 & 2.792 & 1.760 & 11.350 & 0.314 & 12.772 & 0.392 & 9.445 & 0.214 & 5.198 & 0.309 & 343.524\\
HELP\_J100122.000 & 150.342 & 2.729 & 2.672 & 10.914 & 0.227 & 12.217 & 0.498 & 9.304 & 0.216 & 5.345 & 0.289 & 343.532\\
HELP\_J100251.632& 150.715 & 2.545 & 2.920 & 11.427 & 0.257 & 12.762 & 0.365 & 9.191 & 0.304 & 5.473 & 0.271 & 343.532\\
HELP\_J100119.533 & 150.331 & 2.162 & 3.139 & 10.796 & 0.310 & 12.713 & 0.352 & 8.923 & 0.072 & 5.504 & 0.350 & 343.532\\
HELP\_J095942.853 & 149.929 & 2.494 & 4.287 & 10.832 & 0.358 & 12.326 & 0.325 & 9.382 & 0.296 & 5.568 & 0.239 & 239.984\\
HELP\_J100031.099& 150.130 & 2.621 & 1.407 & 11.387 & 0.467 & 12.150 & 0.215 & 9.240 & 0.329 & 5.668 & 0.680 & 343.496\\
HELP\_J100024.947 & 150.104 & 2.186 & 2.209 & 11.434 & 0.298 & 12.429 & 0.279 & 9.536 & 0.422 & 5.720 & 0.281 & 343.528\\
HELP\_J100132.304 & 150.385 & 2.536 & 4.742 & 10.392 & 0.307 & 12.334 & 0.332 & 9.279 & 0.197 & 5.927 & 0.490 & 343.473\\
HELP\_J100043.177 & 150.180 & 2.089 & 2.435 & 11.801 & 0.348 & 12.963 & 0.417 & 9.297 & 0.289 & 6.320 & 0.645 & 343.495\\
HELP\_J100215.646 & 150.565 & 2.786 & 2.174 & 11.409 & 0.282 & 12.485 & 0.447 & 9.305 & 0.135 & 6.367 & 0.337 & 343.527\\
HELP\_J100016.253& 150.068 & 2.791 & 2.725 & 11.475 & 0.334 & 12.913 & 0.403 & 9.366 & 0.368 & 6.446 & 0.294 & 343.524\\
HELP\_J100224.785 & 150.603 & 2.537 & 2.194 & 10.850 & 0.319 & 12.783 & 0.588 & 9.300 & 0.132 & 6.738 & 0.315 & 343.532\\
HELP\_J100142.547& 150.427 & 2.004 & 2.601 & 10.840 & 0.140 & 12.647 & 0.391 & 9.391 & 0.230 & 6.981 & 0.306 & 343.532\\
HELP\_J100031.833& 150.133 & 2.212 & 1.988 & 11.323 & 0.197 & 12.886 & 0.524 & 9.310& 0.122 & 7.609 & 1.829 & 343.531\\
HELP\_J100023.654 & 150.099 & 2.365 & 2.378 & 11.301 & 0.295 & 12.636 & 0.463 & 9.091 & 0.099 & 7.627 & 0.614 & 343.489\\
HELP\_J095912.209 & 149.801 & 2.166 & 2.354 & 11.505 & 0.207 & 12.648 & 0.336 & 9.480 & 0.151 & 7.646 & 0.330 & 343.528\\
HELP\_J095953.305 & 149.972 & 1.714 & 2.864 & 11.566 & 0.354 & 12.744 & 0.363 & 9.519 & 0.312 & 7.877 & 0.332 & 343.528\\
HELP\_J095837.347 & 149.656 & 2.716 & 2.114 & 11.393 & 0.264 & 12.631 & 0.334 & 9.478 & 0.147 & 8.209 & 0.903 & 343.495\\
HELP\_J100232.097 & 150.634 & 2.578 & 2.684 & 11.041 & 0.279 & 12.640 & 0.356 & 9.575 & 0.329 & 8.365 & 0.311 & 343.532\\
HELP\_J100145.219 & 150.438 & 1.856 & 2.599 & 11.398 & 0.359 & 12.701 & 0.350 & 9.502 & 0.262 & 8.422 & 0.346 & 343.471\\
HELP\_J100224.008 & 150.600 & 2.640 & 2.556 & 11.401 & 0.257 & 12.839 & 0.450 & 9.543 & 0.216 & 8.718 & 0.293 & 343.532\\
HELP\_J100103.571& 150.265 & 1.803 & 2.306 & 10.888 & 0.237 & 12.204 & 0.351 & 9.417 & 0.272 & 8.839 & 0.429 & 343.471\\
HELP\_J100026.438 & 150.110 & 2.752 & 2.242 & 10.671 & 0.266 & 12.637 & 0.335 & 9.516 & 0.203 & 9.038 & 0.565 & 343.528\\

\end{longtable}\vspace*{.4\baselineskip}  
	\end{small}

\textbf{Column descriptions:} Column 1: Source ID as in HELP database; Columns 2-3: Coordinates of sources (RA, Dec) expressed in degrees; Column 4: Observed redshift (see \hyperref[sec:2]{Section 2}); Columns 5-10: Main SED derived properties with \texttt{CIGALE} given in the form of base-10 logarithms (from left to right: stellar mass, IR luminosity and dust mass with accompanied uncertainties). All physical properties and their corresponding uncertainties are estimated as the likelihood-weighted means and standard deviations (see \hyperref[sec:3]{Section 3} for the detailed SED modelling procedure). Columns 11-13: ALMA flux estimation, corresponding uncertainty and observed frequency as in $\mathrm{A^{3}COSMOS}$ database (see \citealt{liu19a} for details).
%\end{thebibliography}
%\end{appendix}

\begin{appendix} 
			   \twocolumn
%First online appendix
	%%%%%%%%%%%%%%%%%%%%%
		\section{SED fitting systematics}
		\label{sec:appA}

In order to better evaluate our SED	method and explore the eventual biases, we made simulated data set and fit it using the exact same method that we applied to our observed galaxies. The purpose of using simulations is to analyse eventual observational effects on SED fitting results.  To achieve this goal, we follow the methods presented in \citealt{ciesla15} and \cite{malek19} who use CIGALE to create mock catalogue of objects for each galaxy for which the physical parameters are known. To build the simulated sample we adopt the best-fit SED model for each fitted object which gives one artificial model per galaxy. Input fluxes obtained from the best SEDs are then perturbed following a Gaussian distribution, with $\sigma$ corresponding to the observed uncertainty per each photometric band. The fitting of mock galaxies is further performed with the exact same choice of physical models and their input parameters as for our real data. 

\begin{figure}[ht]
			\label{Fig:app1}
			%	\vspace{-0.2cm}
			\centering
%			\hspace{-1.0cm}
			%	\includegraphics [width=13.89cm]{dsfg-tracks.pdf}
			\includegraphics [width=7.33cm]{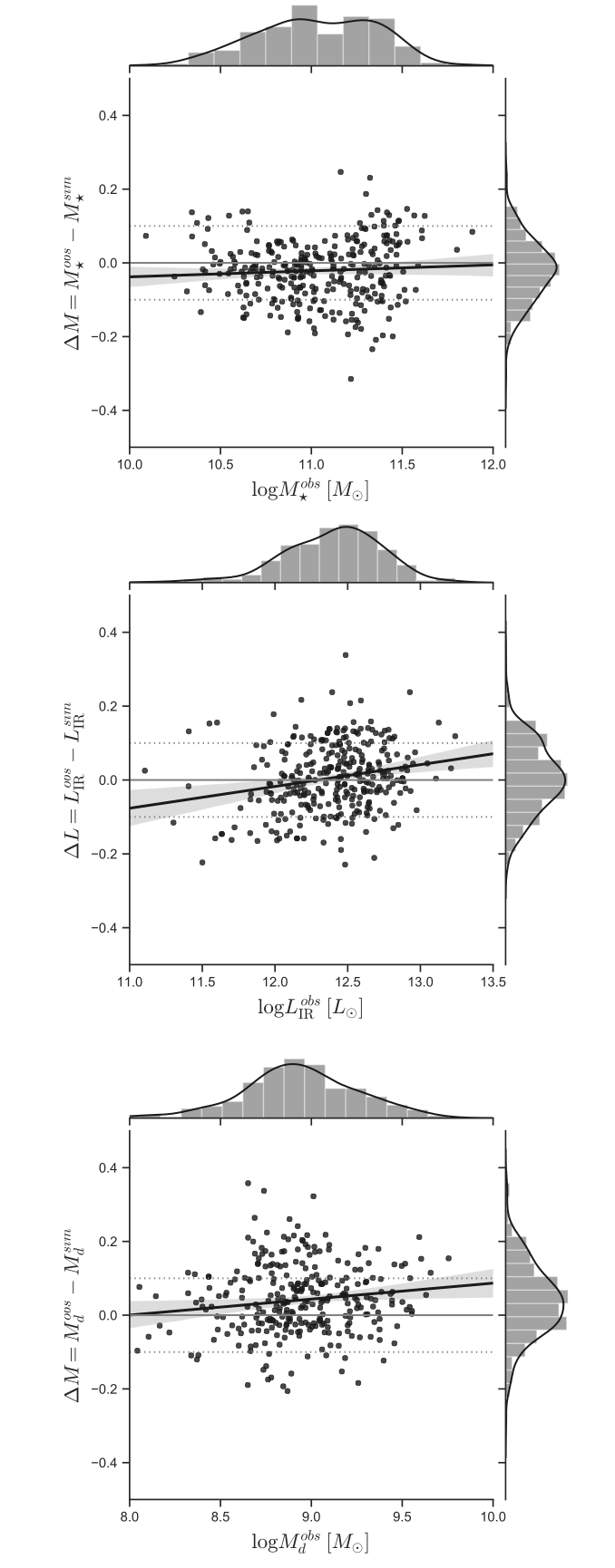}
			\caption{For each panel: offset between the estimated and simulated value. From top to bottom: stellar mass, dust mass and dust luminosity. The black line and the corresponding shaded region is the best linear regression fit to the parameter offset.}
					\end{figure}

The \hyperref[Fig:app1]{Fig. A.1} illustrates the $\log$ difference between the observed physical quantities and the best output parameters of the simulated catalogue. For the stellar mass, dust mass and IR luminosity, such dispersion is expressed as $\Delta M_{\star}$, $\Delta M_{\rm dust}$ and $\Delta L_{\rm IR}$, respectively. We find that for all main physical quantities analysed here, dispersion follows normal distributions with very small offset ($\lesssim0.1$)  Namely, for each physical parameter we find that more than $75\%$ of sources lie within the mean offset of $\pm0.1$. Therefore, we conclude that our SED  fitting procedure does not introduce any significant systematics to derived quantities. %We thus conclude there is no significant systematics attributed to our choice of SED models described in Section 3. 

\end{appendix}

%%%%%%%%%%%%%%%%%%%%%%%%%%%%%%%%%%%%%%%%%%%%%%%555

%%%%%%%%%%%%%%%%%%%%%%%%
\end{document}